\documentclass[aps,english,preprint,nofootinbib]{revtex4}
\pdfoutput=1

\usepackage{graphicx}
\usepackage{hyperref}
\usepackage{amsmath, amssymb}
\usepackage{babel}
\usepackage{color}
\usepackage{mathrsfs}
\usepackage{slashed}

\begin{document}

\title{A New Theory Framework for the Electroweak Radiative Corrections in $K_{l3}$ Decays}

\author{Chien-Yeah Seng$^{1}$, Daniel Galviz$^{1}$ and Ulf-G. Mei\ss{}ner$^{1,2,3}$}

\affiliation{$^1$Helmholtz-Institut f\"ur Strahlen- und Kernphysik and Bethe Center for Theoretical Physics,\\ Universit\"at Bonn, 53115 Bonn, Germany \\
	$^{2}$Institute for Advanced Simulation, Institut f\"ur Kernphysik and J\"ulich Center for Hadron Physics, Forschungszentrum J\"ulich, 52425 J\"ulich, Germany\\
	$^{3}$Tbilisi State  University,  0186 Tbilisi, Georgia }

\date{\today}

 \begin{abstract}

We propose a new theory framework to study the electroweak radiative corrections in $K_{l3}$ decays by combining the classic current algebra approach with the modern effective field theory. Under this framework, the most important $\mathcal{O}(G_F\alpha)$ radiative corrections are described by a single tensor $T^{\mu\nu}$ involving the time-ordered product between the charged weak current and the electromagnetic current, and all remaining pieces are calculable order-by-order in Chiral Perturbation Theory. We further point out a special advantage in the $K_{l3}^{0}$ channel that it suffers the least impact from the poorly-constrained low-energy constants. This finding may serve as a basis for a more precise extraction of the matrix element $V_{us}$ in the future.
 \end{abstract}

\maketitle


\section{Introduction}

One of the high-precision tests of the Standard Model (SM) is the test of the unitarity of the Cabibbo-Kobayashi-Maskawa (CKM) matrix \cite{Cabibbo:1963yz,Kobayashi:1973fv}. In particular, its first-row matrix elements are required to satisfy the following relation:
\begin{equation}
\Delta_{\mathrm{CKM}}=|V_{ud}|^2+|V_{us}|^2+|V_{ub}|^2-1 = 0.\label{eq:unitarity}
\end{equation}
The contribution from $V_{ub}$ is negligible, so the test only concerns $|V_{ud}|$ and $|V_{us}|$. The quoted values of these two matrix elements in the ``CKM quark-mixing matrix" section of the 2018 Particle Data Group (PDG) read \cite{Tanabashi:2018oca}:
\begin{equation}
|V_{ud}|=0.97420(21),\:\:\:|V_{us}|=0.2243(5).
\end{equation}
With the numbers above, the deviation from the unitarity reads:
$\Delta_{\mathrm{CKM}}=-0.0006(5)$,
so unitarity requirement is well-satisfied, and it turns into stringent bounds on parameters of possible Beyond Standard Model (BSM) physics, e.g. the non-standard couplings $\{\epsilon_{i},\tilde{\epsilon}_{j}\}$ between quarks and leptons, and sets constraints to BSM physics at the scale $5-10$ TeV \cite{Bauman:2012fx,Gonzalez-Alonso:2018omy,Cirgiliano:2019nyn}.

The situation described above has changed since last year, following a series of re-evaluation of the electroweak radiative corrections in superallowed beta decays, from where $V_{ud}$ is most precisely extracted. Making use of a dispersion relation with existing neutrino/anti-neutrino scattering data, Ref.~\cite{Seng:2018yzq} reduces the existing theoretical uncertainty in the so-called single-nucleon $\gamma W$-box diagram by a half~\cite{Marciano:2005ec}, but at the same time also shifts its central value significantly. That calculation alone leads to an updated determination of $V_{ud}$:
\begin{equation}
|V_{ud}|=0.97366(15),
\end{equation}
and $\Delta_{\mathrm{CKM}}=-0.0017(4)$, a 4$\sigma$ violation of unitarity, raising new interests within the precision community. In follow-up works \cite{Seng:2018qru,Gorchtein:2018fxl}, several previously unconsidered nuclear effects are also pointed out. Preliminary investigations indicate that they largely cancel each other, and cause only a slight increase in the total uncertainty: $\Delta_{\mathrm{CKM}}=-0.0016(6)$~\cite{Gorchtein:2018fxl}. This is, however, not yet conclusive and needs to be further scrutinized. Further discussions along this line have also led to the identification of well-defined steps toward the further reduction of the $|V_{ud}|$ uncertainty: first, next-generation neutrino experiments at the Long-Baseline Neutrino Facility (LBNF) at Fermilab are expected to provide much more precise neutrino data input to the dispersion integral~\cite{POETIC2019}, second, a direct lattice approach to the $\gamma W$-box diagram is recently suggested~\cite{Seng:2019plg}, alternative computational methods of the box diagram are also proposed in comparison to the dispersive method~\cite{Czarnecki:2019mwq}, and finally, there are plans to study the above-mentioned new nuclear effects with ab-initio methods~\cite{Cirgiliano:2019nyn}. In short, one can be optimistic that the theoretical uncertainty in the $|V_{ud}|$ extraction will be further reduced in the near future.

As a result of the increased precision in $|V_{ud}|$, the value of $|V_{us}|$ now plays a central role in the first-row CKM unitarity. The quoted value in the PDG represents actually an average over the results from two types of experiments: the $K\rightarrow l\nu$ ($K_{l2}$) and $K\rightarrow\pi l\nu$ ($K_{l3}$) decays, which disagree among each other at a $\sim$2$\sigma$ level:
\begin{eqnarray}
K_{l2}&:&|V_{us}|=0.2253(7)\nonumber\\
K_{l3}&:&|V_{us}|=0.2231(8).
\end{eqnarray}
This seemingly small difference, however, makes a huge impact on $\Delta_{\mathrm{CKM}}$ when combined with the updated determination of $|V_{ud}|$ in Ref.~\cite{Seng:2018yzq}: 
\begin{eqnarray}
K_{l2}&:&\Delta_{\mathrm{CKM}}=-0.0012(4)\nonumber\\
K_{l3}&:&\Delta_{\mathrm{CKM}}=-0.0022(5).
\end{eqnarray}
The first line in this equation exhibits an interesting hint for possible BSM physics, but might be loosened
after accounting for possible missing nuclear effects in the $V_{ud}$ extraction. The second line,
on the other hand, represents almost a confirmed signal.
Furthermore, a recent lattice study of the $K\pi$ form factor at zero momentum transfer improves the $K_{l3}$-extraction : $|V_{us}|=0.22333(61)$, leading to $\Delta_{\mathrm{CKM}}=-0.0021(4)$, a 5$\sigma$ violation of unitarity \cite{Bazavov:2018kjg}. All of these call for an immediate re-examination of the theory inputs that enter the $K_{l3}$ extraction of $|V_{us}|$, in particular the treatment of electroweak radiative corrections in $K_{l3}$ which is the purpose of this paper. 

In standard treatments of $K_{l3}$ decays, the total decay rate is given by the following master formula (see, e.g. Ref.~\cite{Cirigliano:2011ny}):
\begin{equation}
\Gamma_{K_{l3}}=\frac{C_{K}^2G_{F}^{2}M_{K}^{5}}{128\pi^{3}}S_{\mathrm{EW}}|V_{us}|^{2}|f_{+}^{K^{0}\pi^{-}}(0)|^{2}I_{Kl}^{(0)}(\lambda_{i})\left(1+\delta_{\mathrm{em}}^{Kl}+\delta_{\mathrm{SU}(2)}^{K\pi}\right)~,\label{eq:EFTmaster}
\end{equation}
with $M_K$ the kaon mass.
The main sources of theoretical input, according to the formula above, are: (1) the $K\pi$ form factor and its momentum-dependence encoded in $|f_{+}^{K^{0}\pi^{-}}(0)|^2I_{Kl}^{(0)}(\lambda_{i})$, (2) the strong isospin breaking effect encoded in $\delta_{\mathrm{SU}(2)}^{K\pi}$, and (3) the electroweak radiative corrections, encoded in $S_{\mathrm{EW}}$ and $\delta_{\mathrm{em}}^{Kl}$. In particular, the constant logarithmic multiplicative factor:\footnote{Throughout, $\alpha$ denotes the Quantum Electrodynamics (QED) fine-structure constant and $\alpha_s$ the strong coupling constant.}
\begin{equation}
S_{\mathrm{EW}}=1+\frac{\alpha}{\pi}\ln\frac{M_{Z}^2}{M_{\rho}^{2}}+\mathcal{O}(\alpha\alpha_s),\label{eq:SEW}
\end{equation}
which is associated to the QED anomalous dimension of the effective four-fermion operators. It contains all universal, short-distance electroweak corrections that are not reabsorbed into the Fermi's constant $G_F$ and is common to all semileptonic transitions. Here, $M_Z$ and $M_\rho$ denote the mass of the $Z$-boson and of the $\rho$-meson, respectively. 
In the meantime, $\delta_{\mathrm{em}}^{Kl}$ represents the model-dependent, long-distance electromagnetic correction. The latter is calculated within the framework of Chiral Perturbation Theory (ChPT) with dynamical photonic and leptonic degrees of freedom~\cite{Cirigliano:2001mk,Cirigliano:2008wn}, currently up to the order $\mathcal{O}(e^{2}p^{2})$.

In this paper, we argue that there exist several conceptual and practical issues in the existing calculations which could be improved upon. Firstly, we demonstrate that the existence of the ``universal factor" $S_{\mathrm{EW}}$ implies an incomplete resummation of $\mathcal{O}(e^{2}p^{2n})$ ($n\geq 2$) terms that leads to a relative systematic error of the order $\mathcal{O}\left(10^{-3}\right)$. In the current treatment their effects are part of the quoted error budget, but we will show that this can be significantly improved. Second, ChPT provides a model-independent framework. However, within such calculations, all the ultraviolet (UV) physics from electromagnetic corrections that are not fixed by chiral symmetry are contained in the low-energy constants (LECs), which can only be estimated by models and contain somewhat uncontrolled uncertainties, that turn into one of the main sources of theoretical uncertainties in the $K_{l3}$ decay rate, and consequently to $|V_{us}|$.
In view of the problems above, it is necessary to switch from the pure ChPT treatment to a new formalism that: (1) allows for an inclusive treatment of higher-order terms, and (2) minimizes the model-dependence, which is equivalent to effects of the LECs in the traditional ChPT calculation.

We realize that such a method actually exists since the late 70s~\cite{Sirlin:1977sv}. Making use of current algebra, one is able to express all the short-distanced-enhanced terms in terms of the momentum integral of a single hadronic tensor involving the time-ordered product between the electromagnetic and the charged weak current. This allows for a more systematic treatment of such terms, e.g. using the recently-developed dispersive method, that enables a better account of the smooth transition between physics at different scales. Meanwhile, for the non-enhanced terms that are suppressed in the neutron and superallowed beta decays but not in $K_{l3}$, we develop a diagrammatic approach to calculate them order-by-order in ChPT, and point out along this process that the $K_{l3}^{0}$ channel possesses a natural advantage of having minimal dependence on the poorly-constrained LECs. 

The outline of this paper is as follows. In Sec.~\ref{sec:tree} we define our basic notations and review the kinematics of $K_{l3}$ decay at tree level. In Sec.~\ref{sec:Semileptonic} we present some important universal features for electroweak radiative corrections in general semi-leptonic beta decays, in particular how it is matched to a Fermi theory with definite multiplicative coefficients. In Sec.~\ref{sec:ChPT} we switch temporarily to an effective field theory (EFT) description, and calculate the $\mathcal{O}(e^{2}p^{2})$ corrections to the $K\pi$ form factors, which serve as important references in the latter discussions. The next four sections are the core of this paper. In Sec.~\ref{sec:OMS} we review the current algebra method developed in the late 70s that divides the radiative corrections to the $K\pi$ form factor into a ``three-point function" and a ``two-point function". The latter, as well as the $\gamma W$-box diagram, contain all the hadronic short-distance enhancements and are elegantly expressed in terms of the momentum integral of a single hadronic tensor $T^{\mu\nu}$. Based on this framework, we derive in Sec.~\ref{sec:logs} the well-known short-distance logarithmic factor, and demonstrate the advantage of our proposed method in the matching of physics at different scales. In Sec.~\ref{sec:diagrammatic} we develop a novel diagrammatic approach to calculate the non-enhanced three-point function order-by-order in EFT, and use it in Sec.~\ref{sec:matching} to split the $\mathcal{O}(e^{2}p^{2})$ corrections in Sec.~\ref{sec:ChPT} into two- and three-point functions. We demonstrate that the unknown LECs in the three-point function makes a minimal impact on the decay rate for the case of $K_{l3}^{0}$, hence the latter is a preferred channel for the $V_{us}$ extraction. Our conclusions are summarized in Sec.~\ref{sec:conclusion}.

\section{\label{sec:tree}$K_{l3}$ decay at tree level}

We start by defining the electromagnetic and charged weak current in the hadron sector that are responsible for generic semi-leptonic beta decays:
\begin{eqnarray}
J_{\mathrm{em}}^{\mu} &=& \frac{2}{3}\bar{u}\gamma^{\mu} u-\frac{1}{3}\bar{d}\gamma^{\mu} d-\frac{1}{3}\bar{s}\gamma^{\mu} s\nonumber\\
J_{W}^{\mu} &=& V_{ud}\bar{u}\gamma^{\mu}(1-\gamma_{5})d+V_{us}\bar{u}\gamma^{\mu}(1-\gamma_{5})s.\label{eq:currents}
\end{eqnarray}
For the $K_{l3}$ decay, the involved charged weak current is $J_{W}^{\mu\dagger}$, and it satisfies the following equal-time commutation relation:
\begin{equation}
\left[J_{W}^{0\dagger}(\vec{x},t),J_{\mathrm{em}}^{\mu}(\vec{y},t)\right]=J_{W}^{\mu\dagger}(\vec{x},t)\delta^{(3)}(\vec{x}-\vec{y}),\label{eq:CA}
\end{equation}
which is the well-known current algebra relation, and is protected from perturbative Quantum Chromodynamics (pQCD) corrections.

Next we summarize some basic results from the tree-level $K_{l3}$ phenomenology, which are quite standard and are available in a number of references (e.g. Ref.~\cite{Cirigliano:2001mk,Cirigliano:2008wn}). There are two types of $K_{l3}$ decays:
\begin{eqnarray}
K_{l3}^{0}&:&K^{0}(p)\rightarrow \pi^{-}(p')l^{+}(p_{l})\nu_{l}(p_{\nu})\nonumber\\
K_{l3}^{+}&:&K^{+}(p)\rightarrow \pi^{0}(p')l^{+}(p_{l})\nu_{l}(p_{\nu}),
\end{eqnarray}
where $l=e,\mu$. We may define the momentum transfer $q=p-p'$.
At low energy, the charged weak interaction between leptons and quarks is described by Fermi's theory, with the following effective Lagrangian:
\begin{equation}
\mathcal{L}_{4f}=-\frac{G_{F}}{\sqrt{2}}J_{W}^{\mu} \bar{l}\gamma_{\mu}(1-\gamma_{5})\nu_{l}+ {\rm h.c.}~.
\end{equation}
Therefore, the decay amplitude at tree level is given by:
\begin{equation}
\mathfrak{M}_{(0)}=-\frac{G_{F}}{\sqrt{2}}F_{K\pi}^{\mu}(p',p) \bar{u}_{\nu}(p_{\nu})\gamma_{\mu}(1-\gamma_{5})v_{l}(p_{l}),
\end{equation}
where the charged weak form factor is defined as $F_{K\pi}^{\mu}(p',p)=\left\langle \pi(p')\right|J_{W}^{\mu\dagger}(0)\left|K(p)\right\rangle$. One could in general parameterize it in terms of two invariant functions\footnote{It is customary to isolate an isospin factor $C_K$ out of the invariant functions $f_{\pm}^{K\pi}$, but we do not do it here because in the current algebra formalism, the most important pieces of the radiative corrections to $\delta f_{\pm}^{K\pi}$ will be expressed as momentum integrals, and it would look weird to have a factor $C_{K}^{-1}$ in front.}:
\begin{equation}
F_{K\pi}^{\mu}(p',p)=V_{us}^{*}\left[f_{+}^{K\pi}(q^{2})(p+p')^{\mu}+f_{-}^{K\pi}(q^{2})(p-p')^{\mu}\right].\label{eq:FFpara}
\end{equation}
By Lorentz covariance, it is obvious that only the vector component of $J_{W}^{\mu\dagger}$ contributes to the form factor. It is also a standard procedure to explicitly factor out the $q^{2}=0$ component of $f_{+}^{K\pi}$ by defining:
\begin{equation}
\bar{f}_{+}^{K\pi}(q^{2})=\frac{f_{+}^{K\pi}(q^{2})}{f_{+}^{K\pi}(0)},\:\:\:\:\bar{f}_{-}^{K\pi}(q^{2})=\frac{f_{-}^{K\pi}(q^{2})}{f_{+}^{K\pi}(0)},
\end{equation}
as the overall factor $f_{+}^{K\pi}(0)$ can then be studied with non-perturbative methods such as lattice QCD.

We further define the following kinematic quantities:
\begin{equation}
  y=\frac{2p.p_{l}}{M_{K}^{2}},\:\:z=\frac{2p.p'}{M_{K}^{2}},\:\:r_{\pi}=\frac{M_{\pi}^{2}}{M_{K}^{2}},\:\:r_{l}=\frac{m_{l}^{2}}{M_{K}^{2}}~,
\end{equation}
with $M_\pi$ and $m_l$ the pion and the lepton mass, in order.
With these, the $K_{l3}$ decay rate at tree level is given by:
\begin{equation}
\Gamma^{(0)}_{K\pi}=\frac{G_{F}^{2}|V_{us}^{2}|M_{K}^{5}}{128\pi^{3}}\left|f_{+}^{K\pi}(0)\right|^{2}\int_{\mathcal{D}_{3}}dydz\bar{\rho}^{(0)}_{K\pi}(y,z),
\end{equation}
where 
\begin{eqnarray}
\bar{\rho}^{(0)}_{K\pi}(y,z)&=&A_{1}^{(0)}(y,z)\left|\bar{f}_{+}^{K\pi}(q^{2})\right|^{2}+A_{2}^{(0)}(y,z)\bar{f}_{+}^{K\pi}(q^{2})\bar{f}_{-}^{K\pi}(q^{2})+A_{3}^{(0)}(y,z)\left|\bar{f}_{-}^{K\pi}(q^{2})\right|^{2}\nonumber\\
A_{1}^{(0)}(y,z)&=&4(1-y)(y+z-1)+r_{l}(4y+3z-3)-4r_{\pi}+r_{l}(r_{\pi}-r_{l})\nonumber\\
A_{2}^{(0)}(y,z)&=&2r_{l}(3-2y-z+r_{l}-r_{\pi})\nonumber\\
A_{3}^{(0)}(y,z)&=&r_{l}(1-z+r_{\pi}-r_{l}).
\end{eqnarray}
The physical domain $\mathcal{D}_{3}$ of the three-body final state can be expressed in two equivalent ways:
\begin{eqnarray}
&&c(z)-d(z)<y<c(z)+d(z),\:\:\:2\sqrt{r_{\pi}}<z<1+r_{\pi}-r_{l}\nonumber\\
&&c(z)=\frac{(2-z)(1+r_{l}+r_{\pi}-z)}{2(1+r_{\pi}-z)},\:\:\:d(z)=\frac{\sqrt{z^{2}-4r_{\pi}}(1+r_{\pi}-r_{l}-z)}{2(1+r_{\pi}-z)}
\end{eqnarray}
or
\begin{eqnarray}
&&a(y)-b(y)<z<a(y)+b(y),\:\:\:2\sqrt{r_{l}}<y<1+r_{l}-r_{\pi}\nonumber\\
&&a(y)=\frac{(2-y)(1+r_{\pi}+r_{l}-y)}{2(1+r_{l}-y)},\:\:\:b(y)=\frac{\sqrt{y^{2}-4r_{l}}(1+r_{l}-r_{\pi}-y)}{2(1+r_{l}-y)}
\end{eqnarray}
depending on whether $y$ or $z$ is first integrated.

An important observation from this analysis is that the contribution from $f_{-}^{K\pi}(q^{2})$ to the decay rate, which is always attached to $A_{2,3}^{(0)}$, is suppressed by $r_{l}=m_{l}^{2}/M_{K}^{2}$, even for the muon, this factor still reads $\sim 0.04\ll1$ so the suppression is significant. This also implies the following: the effect of any theoretical uncertainty in the electromagnetic radiative correction to $f_{-}^{K\pi}$ is suppressed in the total error budget of the $K_{l3}$ decay rate.

\section{\label{sec:Semileptonic}General formalism for radiative corrections in semi-leptonic beta decays, and its application to $K_{l3}$}

\begin{figure}[tb]
	\begin{centering}
		\includegraphics[scale=0.35]{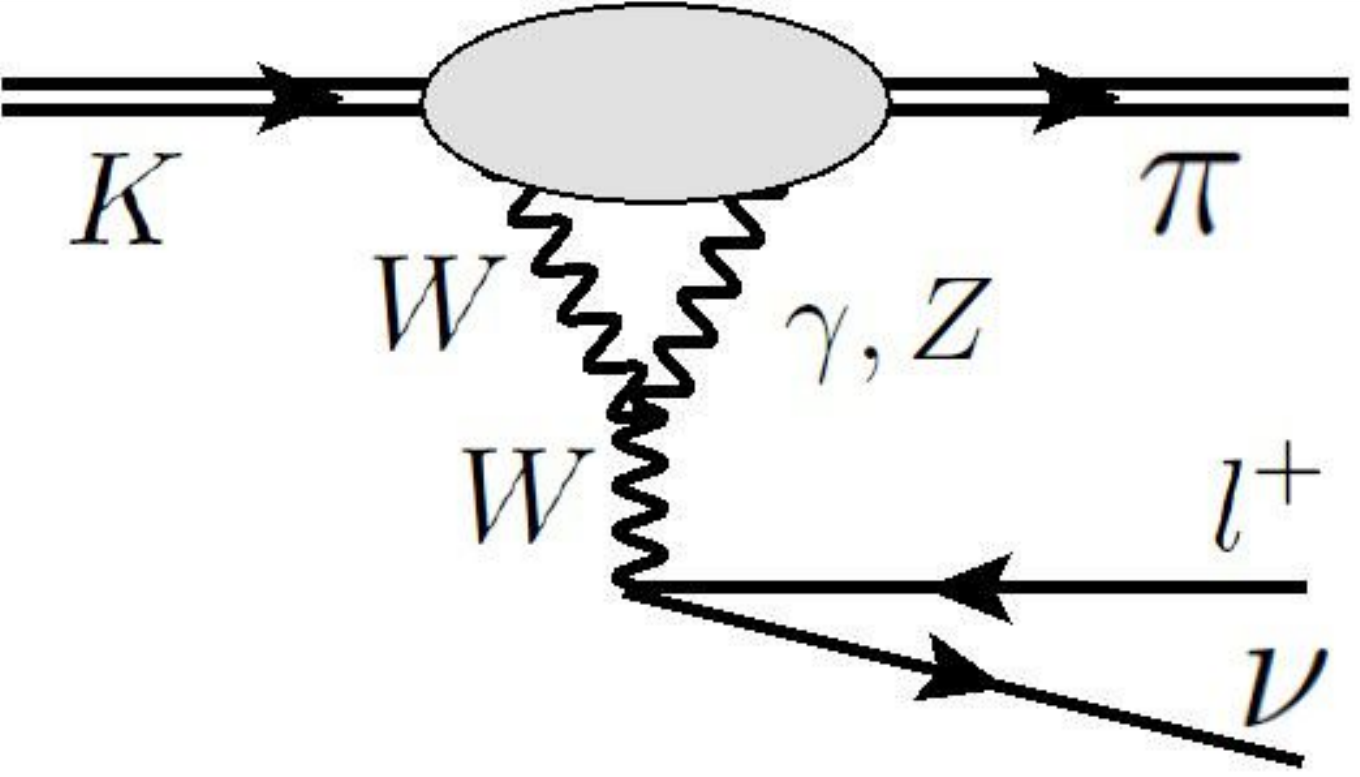}\hfill
		\includegraphics[scale=0.35]{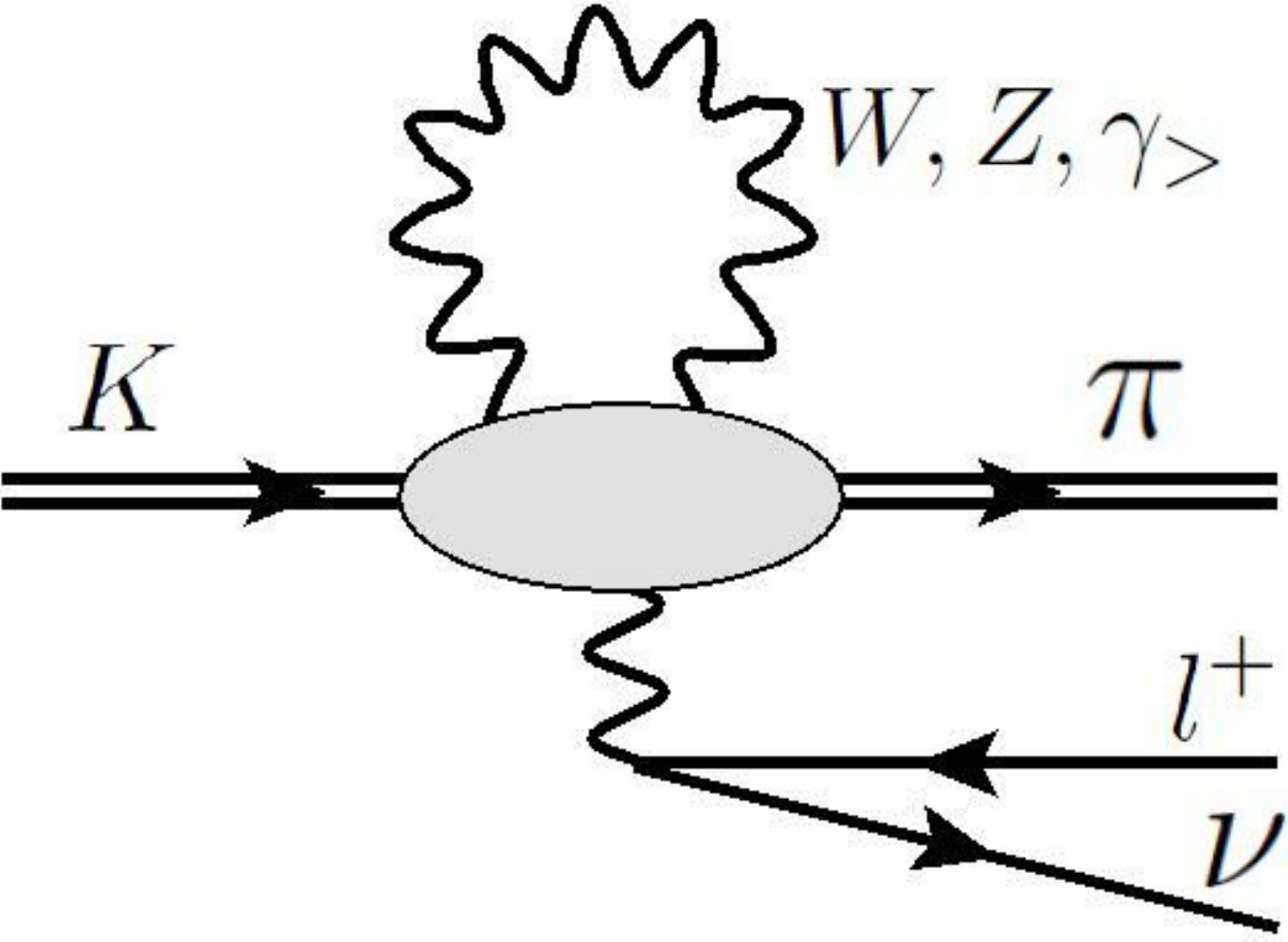}\hfill
		\includegraphics[scale=0.35]{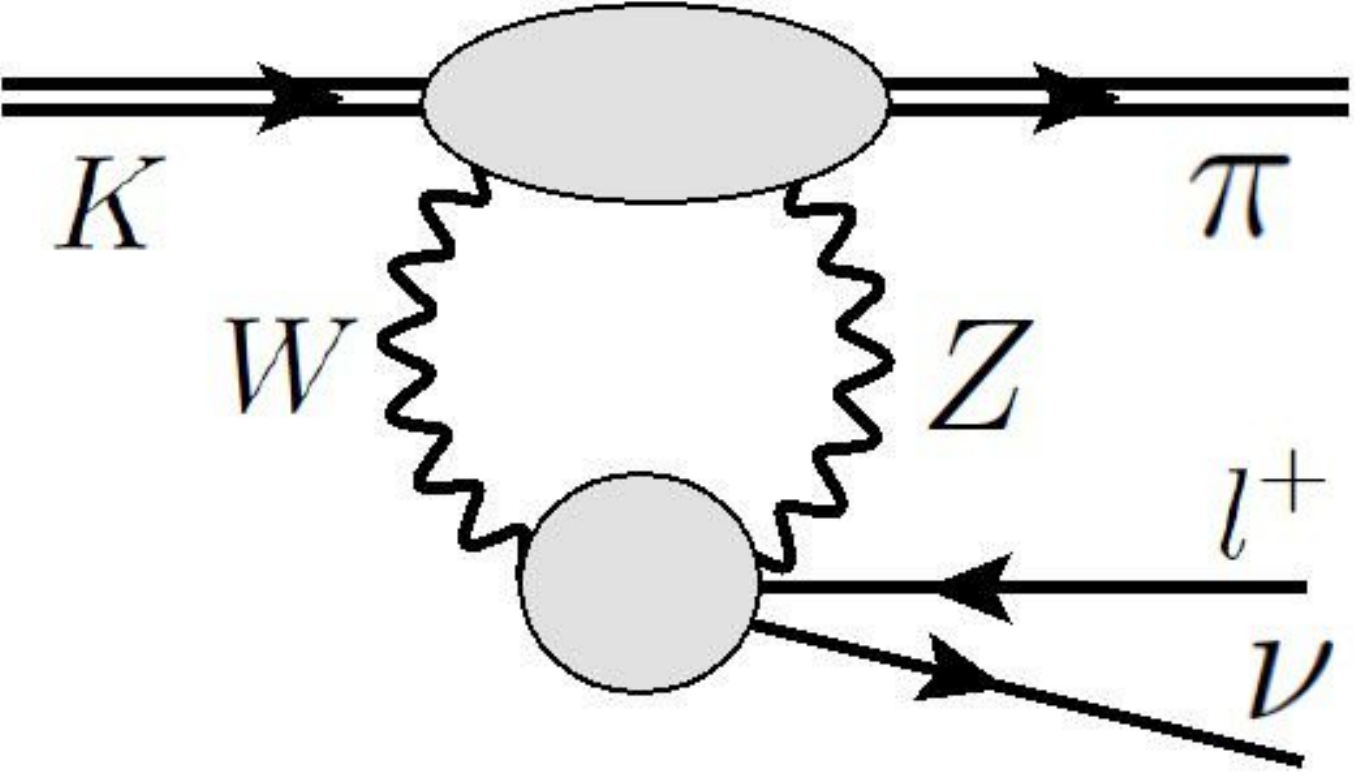}
		\par\end{centering}
	\caption{\label{fig:EWdiagrams}All one-loop weak corrections involving the hadronic piece in $K_{l3}$ decay amplitude in SM. $\gamma_{>}$ denotes photon with a ``mass" $M_{W}$.}
\end{figure}

To start our discussions on electroweak radiative corrections, we first review the known results of muon decay which serves as an important reference point. The Fermi constant $G_{F}=1.1663787(6)\times 10^{-5}$ GeV$^{-2}$ \cite{Tanabashi:2018oca} is defined through the muon lifetime after the inclusion of $\mathcal{O}(G_F\alpha)$ electromagnetic radiative corrections calculated in the Fermi theory~\cite{Kinoshita:1958ru}:
\begin{equation}
\frac{1}{\tau_{\mu}}=\frac{G_{F}^{2}m_{\mu}^{5}}{192\pi^{3}}F(x)\left[1+\frac{\alpha}{2\pi}\left(\frac{25}{4}-\pi^{2}+\mathcal{O}(m_{e})\right)\right],\label{eq:Fermi}
\end{equation} 
where $F(x)=1-8x-12x^{2}\ln x+8x^{3}-x^{4}$ and $x=m_{e}^{2}/m_{\mu}^{2}$. 

In connection to the full SM calculation, it is useful to follow Sirlin's approach that splits the full photon propagator in the fermion self-energy diagrams into two pieces~\cite{Sirlin:1977sv}:
\begin{equation}
\frac{1}{k^{2}-M_{\gamma}^{2}}=\frac{1}{k^{2}-M_{W}^{2}}+\frac{M_{W}^{2}}{M_{W}^{2}-k^{2}}\frac{1}{k^{2}-M_{\gamma}^{2}}.\label{eq:photon}
\end{equation}
Here we have introduced a small photon mass $M_{\gamma}$ as an infrared (IR) regulator. The first term at the right hand side (RHS) of Eq.~\eqref{eq:photon} describes a ``massive photon", whereas the second term describes a massless photon with an extra Pauli-Villars (PV) regulator~\cite{Pauli:1949zm} with $\Lambda=M_{W}$. The benefit of such a separation is evident: fermion self-energy corrections involving the PV-regulated photon propagator, together with the $\gamma W$ box diagram, give exactly the PV-regulated electromagnetic radiative corrections to the Fermi theory as described in Eq.~\eqref{eq:Fermi}, up to $\mathcal{O}(G_{F}^{2})$ discrepancies which may be safely neglected. Meanwhile, all remaining $\mathcal{O}(G_F\alpha)$ SM corrections, including both one-loop and counterterm contributions shall be known, with a slight abuse of terminology, as ``weak radiative corrections" that are sensitive only to physics at $k\sim M_{W}$ and give rise only to a constant renormalization factor to the differential decay rate which is absorbed into the definition of $G_{F}$ in Eq.~\eqref{eq:Fermi}. Notice also that, since a PV-regulator preserves Abelian gauge symmetry, the PV-regulated Fermi theory description above is manifestly $U(1)_{\mathrm{em}}$-invariant.

One may now proceed to discuss the general semi-leptonic beta decay of hadrons. Through a sophisticated current algebra analysis~\cite{Sirlin:1977sv}, it is shown that most of the weak radiative corrections (depicted in Fig.~\ref{fig:EWdiagrams}, using $K_{l3}$ as example) are identical to that in the muon decay, which means that they are simply reabsorbed into the definition of $G_{F}$, the only exceptions are a few definite, calculable pQCD corrections stemming from Fig.~\ref{fig:EWdiagrams}, as well as the $WZ$ box diagram from which the electroweak currents probe not only the difference in the electric charge of the SU(2)$_{L}$ doublet but also their average. These small discrepancies are encoded as a constant multiplicative factor to the Fermi constant. As a result, the semi-leptonic beta decay of a generic hadron is now fully described by the following Lagrangian\footnote{Under this formalism, the baryon masses (or squared masses for mesons) are given consistently by $m_{B}=m_{B,\mathrm{QCD}}+\delta m_{B,\gamma_<}$.}:
\begin{equation}
\mathcal{L}=\mathcal{L}_{\mathrm{QCD}}+\mathcal{L}_{\mathrm{QED},\gamma_{<}}+\mathcal{L}_{4f}^{\prime}\label{eq:Lsemileptonic}
\end{equation}
where $\mathcal{L}_{\mathrm{QCD}}$ is the ordinary QCD Lagrangian, $\mathcal{L}_{\mathrm{QED},\gamma_{<}}$ is the QED Lagrangian with a PV-regulated photon propagator at $\Lambda=M_{W}$, and $\mathcal{L}_{4f}^{\prime}$ is the Fermi theory Lagrangian with the above-mentioned multiplicative factor:
\begin{equation}
\mathcal{L}_{4f}^{\prime}=-\left\{1-\frac{3\alpha}{8\pi}\left[\left(2\bar{Q}+1\right)
\ln\frac{M_{W}^{2}}{M_{Z}^{2}}+a_{\mathrm{pQCD}}\right]\right\}\frac{G_{F}}{\sqrt{2}}J_{W}^{\mu} \bar{l}\gamma_{\mu}(1-\gamma_{5})\nu+ {\rm h.c.}~.\label{eq:SMtoFermi}
\end{equation} 
Here, $G_{F}$ is the experimentally-measured Fermi constant from the muon decay, $\bar{Q}=1/6$ is the average charge of the quark doublet, and $a_{\mathrm{pQCD}}$ describes the pQCD correction (up to $\mathcal{O}(\alpha_{s})$) to Fig.~\ref{fig:EWdiagrams}:
\begin{eqnarray}
a_{\mathrm{pQCD}}&=&\frac{1}{3}\int d\kappa^{2}\left\{\frac{2M_{W}^{2}}{(\kappa^{2}+M_{W}^{2})^{2}}-\frac{M_{W}^{2}(M_{Z}^{2}-M_{W}^{2})\kappa^{2}}{(\kappa^{2}+M_{W}^{2})^{2}(\kappa^{2}+M_{Z}^{2})^{2}}\right.\nonumber\\
&&\left.+\frac{M_{W}^{2}}{(\kappa^{2}+M_{W}^{2})(\kappa^{2}+M_{Z}^{2})}\left[\frac{c_w^2}{s_w^2}+6\bar{Q}\frac{s_w^2}{c_w^2}\right]\right\}\frac{\alpha_s(\kappa^{2})}{\pi},\label{eq:aPQCD}
\end{eqnarray}
where $c_{w}^{2}=1-s_{w}^{2}=M_{W}^{2}/M_{Z}^{2}$. The first two terms at the RHS in Eq.~\eqref{eq:aPQCD} come from the first two diagrams in Fig.~\ref{fig:EWdiagrams}, and the third term comes from the third diagram.

Before proceeding further, we have two comments on the Lagrangian in Eq.~\eqref{eq:Lsemileptonic}. First, the same result can be derived through a one-loop analysis with massless quarks as in Section~2 of Ref.~\cite{DescotesGenon:2005pw}, although the latter does not explicitly spell out the multiplicative factor in Eq.\eqref{eq:SMtoFermi}. Second, we want to stress that Eq.~\eqref{eq:Lsemileptonic} is strictly speaking not a low-energy EFT of the SM; it \textit{is}, up to $\mathcal{O}(G_F\alpha)$, the SM itself in terms of describing  semi-leptonic beta decays. This is evident, as in Eq.~\eqref{eq:Lsemileptonic} the active energy scale still goes all the way to $q\sim M_W$, and yet all the one-loop corrections are UV-finite. No extra counterterms are needed.

\begin{figure}[tb]
	\begin{centering}
		\includegraphics[scale=0.2]{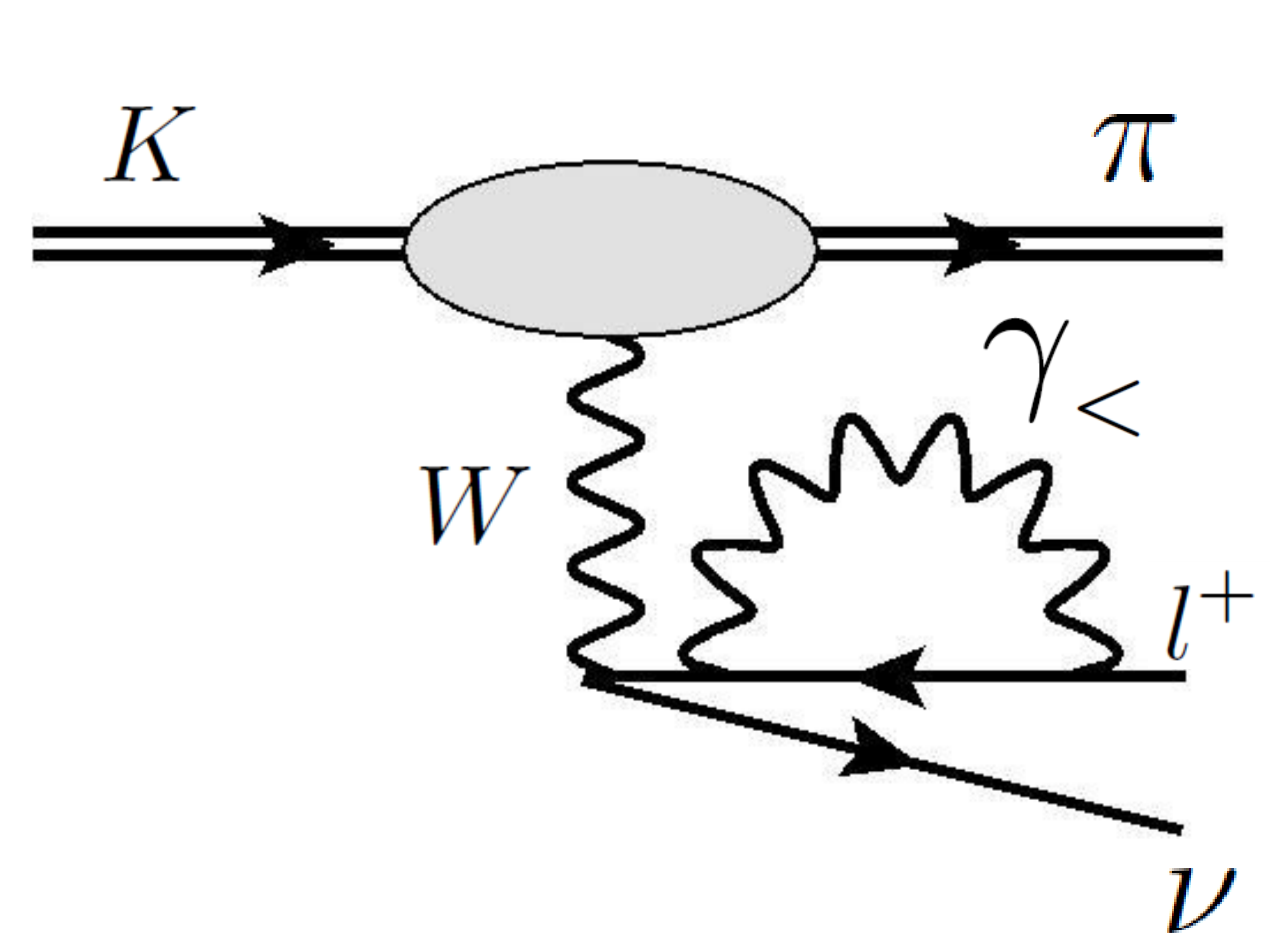}\hfill
		\includegraphics[scale=0.2]{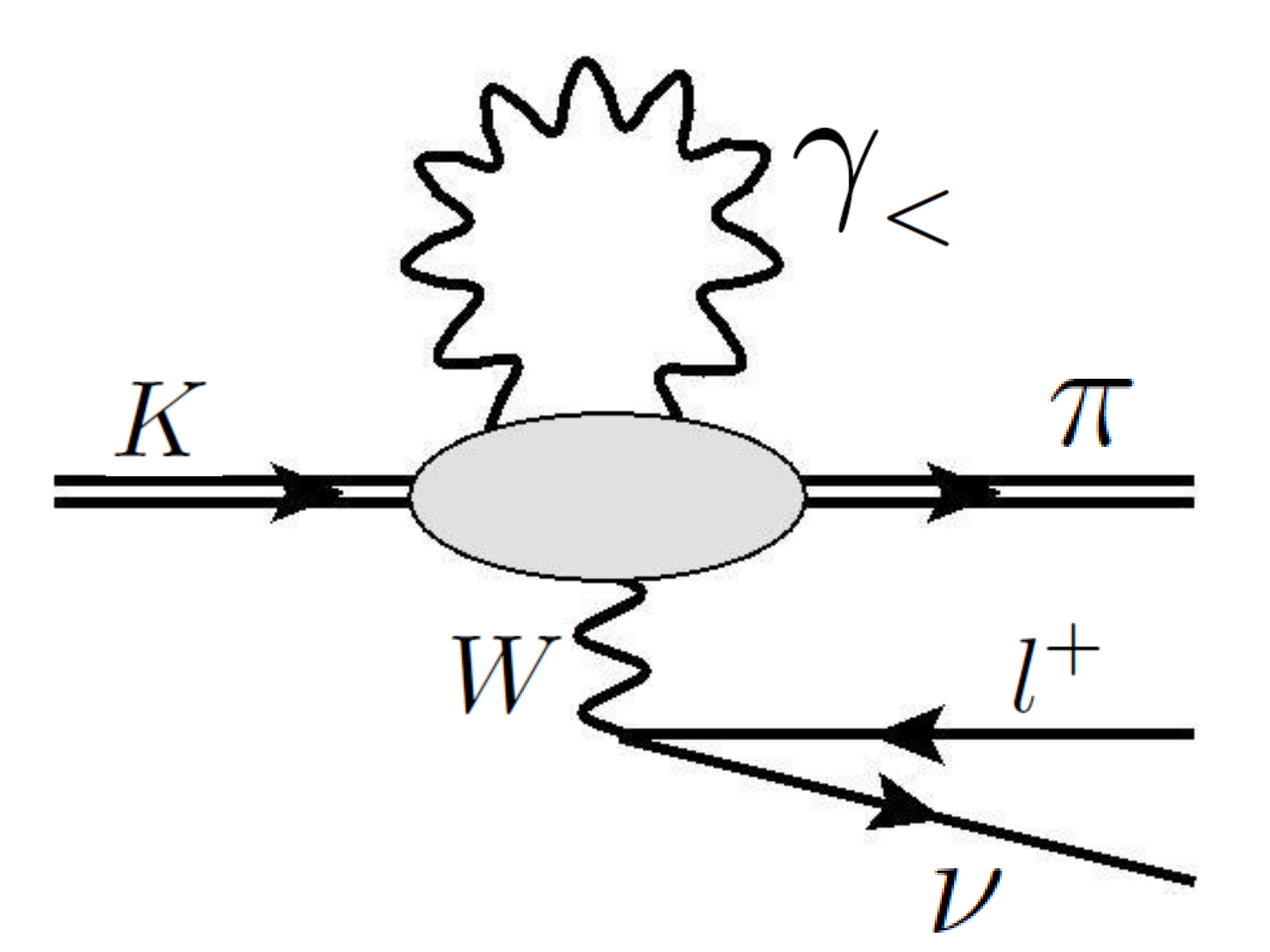}\hfill
		\includegraphics[scale=0.2]{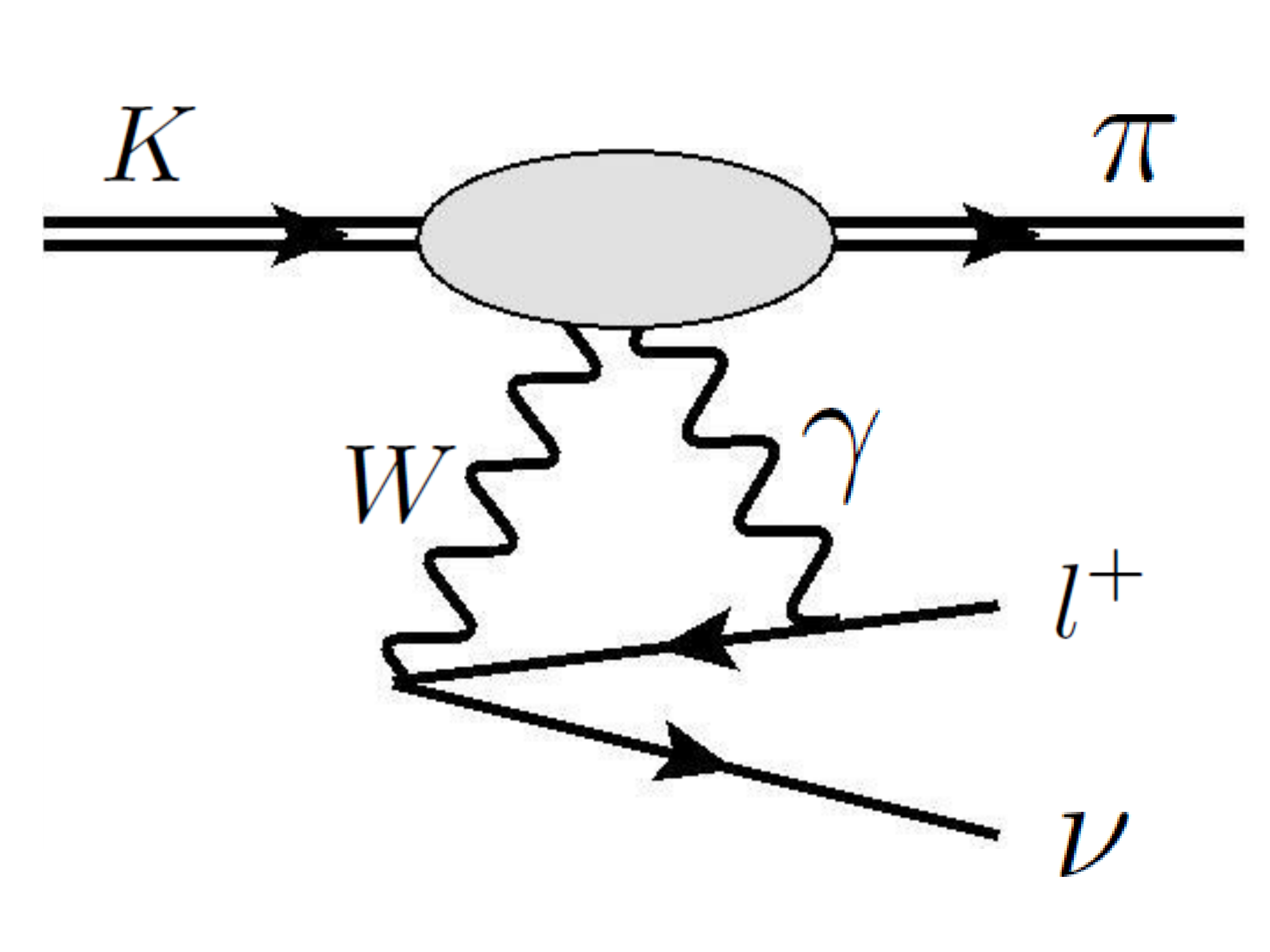}
		\par\end{centering}
	\caption{\label{fig:EMdiagrams}All one-loop electromagnetic corrections to the $K_{l3}$ decay amplitude in SM. $\gamma_<$ denotes a PV-regulated photon propagator with $\Lambda=M_{W}$.}
\end{figure}

One can now apply the formalism above to $K_{l3}$. Up to $\mathcal{O}(G_{F}\alpha)$, the decay amplitude is given by (after substituting $\bar{Q}=1/6$):
\begin{eqnarray}
  \mathfrak{M}&=&-\sqrt{Z_{l}}\left[1-\frac{\alpha}{2\pi}\left(\ln\frac{M_{W}^{2}}{M_{Z}^{2}}+\frac{3}{4}a_{\mathrm{pQCD}}\right)\right]\frac{G_{F}}{\sqrt{2}}F_{K\pi}^{\mu}\bar{u}_{\nu}\gamma_{\mu}(1-\gamma_{5})\nu_{l}\nonumber\\
  && -\frac{G_{F}}{\sqrt{2}}\delta F_{K\pi}^{\mu}\bar{u}_{\nu}\gamma_{\mu}(1-\gamma_{5})\nu_{l}+\Box_{\gamma W}.\label{eq:basicM}
\end{eqnarray}
The first term at the RHS contains the tree-level contribution, the weak+pQCD multiplicative factor as well as the electromagnetic radiative correction to the external charged lepton  (depicted in the first diagram of Fig.~\ref{fig:EMdiagrams}), expressed in terms of the $K\pi$ form factor and the charged lepton wavefunction renormalization (in Feynman gauge, which we will adopt throughout this paper):
\begin{equation}
 Z_{l}=1-\frac{\alpha}{4\pi}\left[\ln\frac{M_{W}^{2}}{m_{l}^{2}}+\frac{9}{2}-2\ln\frac{m_{l}^{2}}{M_{\gamma}^{2}}\right].\label{eq:Zl}
\end{equation}
The second term comes from the electromagnetic radiative correction to the $K\pi$ form factor (the second diagram of Fig.~\ref{fig:EMdiagrams}):
\begin{equation}
F_{K\pi}^{\mu}(p',p)\rightarrow F_{K\pi}^{\mu}(p',p)+\delta F_{K\pi}^{\mu}(p',p).
\end{equation}
Finally, $\Box_{\gamma W}$ denotes the $\gamma W$ box diagram contribution (the third diagram in Fig.~\ref{fig:EMdiagrams}):
\begin{equation}
\Box_{\gamma W}=-\frac{G_{F}e^{2}}{\sqrt{2}}\int\frac{d^{4}k}{(2\pi)^{4}}\frac{\bar{u}_{\nu}\gamma^{\nu}(1-\gamma_{5})(\slashed{k}-\slashed{p}_{l}+m_{l})\gamma^{\mu} v_{l}}{(p_{l}-k)^{2}-m_{l}^{2}}\frac{1}{k^{2}-M_{\gamma}^{2}}\frac{M_{W}^{2}}{M_{W}^{2}-k^{2}}T_{\mu\nu}(k;p',p),\label{eq:box}
\end{equation}
where 
\begin{equation}
T^{\mu\nu}(k;p',p)=\int d^{4}xe^{ik\cdot x}\left\langle \pi(p')\right|T\left\{J_{\mathrm{em}}^{\mu}(x)J_{W}^{\nu\dagger}(0)\right\}\left|K(p)\right\rangle\label{eq:GCompton}
\end{equation}
which is known as the  ``generalized Compton tensor''. 
The factor $M_{W}^{2}/(M_{W}^{2}-k^{2})$ in Eq.~\eqref{eq:box} comes from the W-propagator, except that we have neglected its dependence on external momenta which brings only $\mathcal{O}(G_{F}^{2})$ corrections. By doing so, Eq.~\eqref{eq:box} is completely equivalent to the box diagram calculated in the Fermi theory with a PV-regulator, again confirming our previous claim. 

Up to this point we have identified all elements needed to fully understand the $\mathcal{O}(G_{F}\alpha)$ electroweak radiative corrections to the $K_{l3}$ amplitude: they are just $\delta F_{K\pi}^{\mu}$ and $\Box_{\gamma W}$, i.e. the last two diagrams in Fig.~\ref{fig:EMdiagrams}. Of course in principle one also needs the real photon emission diagrams to cancel the IR divergences, but the latter involve just standard tree-level calculations so we shall not discuss them any further in this paper.

\section{\label{sec:ChPT}ChPT calculation of the electromagnetic radiative corrections to the $K\pi$ form factors}

The current standard treatment of radiative corrections in $K_{l3}$ is equivalent to calculating $\delta Z_{l}$, $\delta F_{K\pi}^{\mu}$ and $\Box_{\gamma W}$ using ChPT, the low-energy EFT of QCD in the light quark sector.
The first systematic calculation to the order $\mathcal{O}(e^{2}p^{2})$ appeared in Ref.~\cite{Cirigliano:2001mk}, and was subsequently updated in Ref.~\cite{Cirigliano:2008wn} making use of more updated LECs obtained in Ref.~\cite{Ananthanarayan:2004qk,DescotesGenon:2005pw}. A comprehensive review of general kaon decays in SM is also available in Ref.~\cite{Cirigliano:2011ny}. Among these results, the $\mathcal{O}(e^{2}p^{2})$ corrections to $F_{K\pi}^{\mu}$ are most relevant to our work so we shall devote this section to it. 

We start from a short review of the basic ChPT formalism. Consider a three-flavor QCD Lagrangian coupled with the photon field $A_{\mu}$ as well as an auxiliary, complex vector field $\mathcal{V}_{\mu}$ which will be used later to derive the charged weak current:
\begin{equation}
\mathcal{L}_{\mathrm{QCD}}^{\mathrm{ext}}=\bar{q}i\slashed{D}q-\bar{q}M_q q-e\bar{q}Q_{\mathrm{em}}\gamma^{\mu} qA_{\mu}+\bar{q}\gamma^{\mu} Q_{W}q \mathcal{V}_{\mu}+\bar{q}\gamma^{\mu} Q_{W}^{\dagger} q \mathcal{V}_{\mu}^{\dagger}-\frac{1}{4}G_{\mu\nu}^{a} G^{a\mu\nu}-\frac{1}{4}F_{\mu\nu}F^{\mu\nu},
\end{equation}
where $q=(u\:d\:s)^{T}$, and the matrices read:
\begin{equation}
M_{q}=\left(\begin{array}{ccc}
m_{u} & 0 & 0\\
0 & m_{d} & 0\\
0 & 0 & m_{s}
\end{array}\right),\:\:Q_{\mathrm{em}}=\left(\begin{array}{ccc}
\frac{2}{3} & 0 & 0\\
0 & -\frac{1}{3} & 0\\
0 & 0 & -\frac{1}{3}
\end{array}\right),\:\:Q_{W}=\left(\begin{array}{ccc}
0 & 0 & 0\\
V_{ud}^{*} & 0 & 0\\
V_{us}^{*} & 0 & 0
\end{array}\right).
\end{equation}
Throughout this work we will further assume isospin symmetry, i.e. $m_u=m_d=\hat{m}$, which turns off the leading order (LO) $\pi^{0}-\eta$ mixing. The vector component of the charged weak current responsible for the kaon decays can then be obtained from $\mathcal{L}_{\mathrm{QCD}}^{\mathrm{ext}}$ through:
\begin{equation}
(J_{W}^{\mu\dagger})_V=\left(\frac{\partial\mathcal{L}_{\mathrm{QCD}}^{\mathrm{ext}}}{\partial \mathcal{V}_{\mu}}\right)_{\mathcal{V},\mathcal{V}^{\dagger}=0}.
\end{equation}
One could also re-express the Lagrangian above in terms of left- and right-handed quark fields:
\begin{equation}
\mathcal{L}_{\mathrm{QCD}}^{\mathrm{ext}}=\mathcal{L}_{\mathrm{QCD}}^{0}+\bar{q}_R\gamma^{\mu} r_{\mu} q_R+\bar{q}_L\gamma^{\mu} l_{\mu} q_L-\bar{q}_R M_q q_L-\bar{q}_LM_q^{\dagger} q_R-\frac{1}{4}F_{\mu\nu}F^{\mu\nu}, 
\end{equation}
where $\mathcal{L}_{\mathrm{QCD}}^{0}$ is the massless QCD Lagrangian, and the external fields
\begin{equation}
l_{\mu}=r_{\mu}=Q_{W} \mathcal{V}_{\mu}+Q_{W}^{\dagger} \mathcal{V}_{\mu}^{\dagger}-eQ_{\mathrm{em}}A_{\mu}
\end{equation}
are vector spurions that are traceless in the flavor space.

With these one  writes down the LO chiral Lagrangian with dynamical photons:
\begin{equation}
\mathcal{L}_{p^{2}+e^{2}}=\frac{F_0^{2}}{4}\left\langle D^{\mu} U(D_{\mu} U)^{\dagger}+U\chi^{\dagger}+U^{\dagger} \chi\right\rangle+Ze^{2}F_0^{4}\left\langle Q_{\mathrm{em}}UQ_{\mathrm{em}}U^{\dagger}\right\rangle-\frac{1}{4}F_{\mu\nu}F^{\mu\nu}\label{eq:ChPTLO}
\end{equation}
where $D_{\mu} U=\partial_{\mu} U-ir_{\mu} U+iUl_{\mu}$, $\chi=2B_0 M_q$ and $\left\langle...\right\rangle$ denotes the trace in flavor space. Further, $F_0$ is the pion decay constant in the chiral limit. In what follows,
we will equate it with its physical value. 
The first term in Eq.~\eqref{eq:ChPTLO} is the ordinary $\mathcal{O}(p^{2})$ ChPT Lagrangian, while the second term encodes the effect of short-distance virtual photons at $\mathcal{O}(e^{2})$ \cite{Ecker:1988te}. The parameter $Z$ can be related to the squared-mass difference of the pions:
\begin{equation}
Z=\frac{M_{\pi^{\pm}}^{2}-M_{\pi^{0}}^{2}}{8\pi\alpha F_0^{2}}\approx 0.8.
\end{equation}
At the same time, the squared masses of the neutral mesons that are not affected by $Z$, satisfy the Gell-Mann-Okubo relation:
$3M_{\eta}^2=4M_{K^{0}}^{2}-M_{\pi^{0}}^{2}$.

The ChPT representation of the vector charged weak current at LO is:
\begin{equation}
(J_{W}^{\mu\dagger})_{V,\mathrm{LO}}=\frac{iF_0^{2}}{4}\left\langle Q_{W}\left(\left[D^{\mu} U^{\dagger},U\right]+\left[D^{\mu} U,U^{\dagger}\right]\right)\right\rangle_{\mathcal{V},\mathcal{V}^{\dagger}=0}.\label{eq:JWEFT}
\end{equation}
It is important to notice that the current above contains not only the meson fields but also photon fields. Also, with this convention of charged weak current, one obtains $f_{+}^{K\pi}=-1$ at tree level, which has a sign difference with some other literature. Of course such a difference does not affect any physical result as long as one sticks with a given convention consistently throughout the whole calculation. 

Applying the Lagrangian in Eq.~\eqref{eq:ChPTLO} to one loop yields UV-divergences which must be absorbed by the $\mathcal{O}(p^{4})$ and $\mathcal{O}(e^{2}p^{2})$ counterterms. The only $\mathcal{O}(p^{4})$ counterterm needed in this work is \cite{Gasser:1984gg}:
\begin{eqnarray}
\mathcal{L}_{p^{4}}=-iL_{9}\left\langle f_{\mu\nu}^{R} D^{\mu} U(D^{\nu} U)^{\dagger}+f_{\mu\nu}^{L}(D^{\mu} U)^{\dagger} D^{\nu} U\right\rangle,
\end{eqnarray}
where the field-strength tensors are defined as:
\begin{equation}
f^{R}_{\mu\nu}=\partial_{\mu} r_{\nu}-\partial_{\nu} r_{\mu}-i[r_{\mu},r_{\nu}],\:\:f^{L}_{\mu\nu}=\partial_{\mu} l_{\nu}-\partial_{\nu} l_{\mu}-i[l_{\mu},l_{\nu}].
\end{equation}
At the same time, the part of the $\mathcal{O}(e^{2}p^{2})$ counterterms needed in this work is given by \cite{Urech:1994hd}:
\begin{eqnarray}
\mathcal{L}_{e^{2}p^{2}}&=&e^{2}F_0^{2}\Bigl\{K_{3}\left[\left\langle (D^{\mu} U)^{\dagger} Q_{\mathrm{em}} U\right\rangle\left\langle (D_{\mu} U)^{\dagger} Q_{\mathrm{em}} U\right\rangle+\left\langle D^{\mu} U Q_{\mathrm{em}} U^{\dagger}\right\rangle\left\langle D_{\mu} U Q_{\mathrm{em}} U^{\dagger}\right\rangle\right]\Bigr.\nonumber\\
&&+K_{4}\left\langle (D^{\mu} U)^{\dagger} Q_{\mathrm{em}} U\right\rangle\left\langle D_{\mu} U Q_{\mathrm{em}} U^{\dagger}\right\rangle+K_{5}\left\langle Q_{\mathrm{em}}^{2}\left[(D^{\mu} U)^{\dagger} D_{\mu} U+D^{\mu} U(D_{\mu} U)^{\dagger}\right]\right\rangle\nonumber\\
&&+K_{6}\left\langle (D^{\mu} U)^{\dagger} D_{\mu} U Q_{\mathrm{em}} U^{\dagger} Q_{\mathrm{em}}U+D^{\mu} U(D_{\mu} U)^{\dagger} Q_{\mathrm{em}}UQ_{\mathrm{em}}U^{\dagger}\right\rangle\nonumber\\
&&+K_{9}\left\langle\left(\chi^{\dagger} U+U^{\dagger}\chi+\chi U^{\dagger}+U\chi^{\dagger}\right)Q_{\mathrm{em}}^{2}\right\rangle\nonumber\\
&&+K_{10}\left\langle\left(\chi^{\dagger} U+U^{\dagger}\chi\right)Q_{\mathrm{em}}U^{\dagger} Q_{\mathrm{em}}U+\left(\chi U^{\dagger}+U\chi^{\dagger}\right)Q_{\mathrm{em}}UQ_{\mathrm{em}}U^{\dagger}\right\rangle\nonumber\\
&&\Bigl.+K_{12}\left\langle (D^{\mu} U)^{\dagger}[c_{\mu}^{R}Q_{\mathrm{em}},Q_{\mathrm{em}}]U+D^{\mu} U[c_{\mu}^{L} Q_{\mathrm{em}},Q_{\mathrm{em}}]U^{\dagger}\right\rangle\Bigr\},\label{eq:Oe2p2}
\end{eqnarray}
where $c_{\mu}^{R}Q_{\mathrm{em}}=-i[r_{\mu},Q_{\mathrm{em}}],c_{\mu}^{L}Q_{\mathrm{em}}=-i[l_{\mu},Q_{\mathrm{em}}]$. The definition of renormalized LECs is as follows: 
\begin{eqnarray}
L_{i}^{r}(\mu)&=&L_{i}-\Gamma_i\lambda\nonumber\\
K_{i}^{r}(\mu)&=&K_{i}-\Sigma_i\lambda,
\end{eqnarray}
where
\begin{equation}
\lambda=\frac{\mu^{d-4}}{16\pi^{2}}\left(\frac{1}{d-4}-\frac{1}{2}\left[\ln 4\pi-\gamma_E+1\right]\right)~,
\end{equation}
with $\mu$ the scale of dimensional regularization, $d$ the number of space-time dimensions,
and $\gamma_E \simeq 0.5772$ the Euler-Mascheroni constant.
The coefficients $\{\Gamma_{i},\Sigma_{i}\}$ we need to know are \cite{Gasser:1984gg,Urech:1994hd}:
\begin{equation}
\Gamma_{9}=\frac{1}{4},\:\:\Sigma_{3}=-\frac{3}{4},\:\:\Sigma_{4}=2Z,\:\:\Sigma_{5}=-\frac{9}{4},\:\:\Sigma_{6}=\frac{3}{2}Z,\:\:\Sigma_{9}=-\frac{1}{4},\:\:\Sigma_{10}=\frac{1}{4}+\frac{3}{2}Z,\:\:\Sigma_{12}=\frac{1}{4}.
\end{equation}

In a full ChPT treatment of $K_{l3}$ decays, it is also necessary to introduce dynamical leptons in the chiral Lagrangian; this leads to extra $\mathcal{O}(e^{2}p^{2})$ counterterms with LECs denoted as $\{X_i\}$ \cite{Knecht:1999ag}. In our approach, however, Feynman diagrams with dynamical leptons (such as the lepton wavefunction renormalization and the $\gamma W$-box diagram) are not studied using ChPT, so such terms are not needed. Finally, it is beneficial at this point to compare the treatment of UV-physics between the representation in Section~\ref{sec:Semileptonic} and in the EFT language. In the former, all the physics at the scale $k\sim M_{W}$ are explicitly evaluated, and all the remaining loop integrals are finite due to the existence of the PV-regulator. On the other hand, in an EFT which works intrinsically at a scale $k\ll M_{W}$, one does not need a PV-regulator and the UV-divergences are regulated by dimensional regularization. All finite multiplicative factors in Section~\ref{sec:Semileptonic} that come from the UV-end, such as the $\ln (M_{W}^{2}/M_{Z}^{2})$ factor and the $\mathcal{O}(\alpha_s)$ corrections, are contained in the renormalized LECs $K_{12}^{r}$ and $\{X_i^{r}\}$. 

\begin{figure}[tb]
	\begin{centering}
		\includegraphics[scale=0.3]{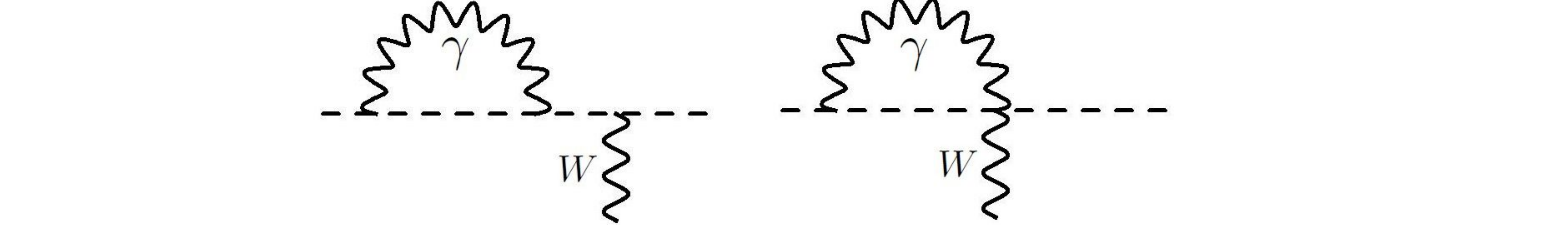}\hfill
		\includegraphics[scale=0.3]{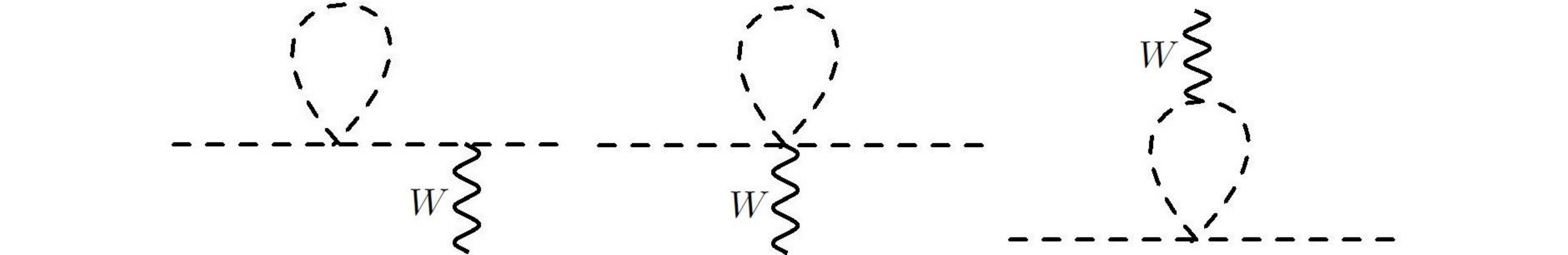}
		\par\end{centering}
	\caption{\label{fig:EMandChloops}The electromagnetic (first two) and chiral (last three) loop diagrams that contribute to $\delta F_{K\pi}^{\mu}$ in ChPT.}
\end{figure}

Now let us focus on $\delta F_{K\pi}^{\mu}$. There are two types of one-loop diagrams that give $\mathcal{O}(e^{2}p^{2})$ corrections to $F_{K\pi}^{\mu}$: the electromagnetic loops and chiral loops, see Fig.~\ref{fig:EMandChloops}, and the full analytical results are available in Ref.~\cite{Cirigliano:2001mk}. In there, however, all these diagrams are treated in the same footing, and in particular, the chiral loop diagrams contain simultaneously the $\mathcal{O}(p^{4})$ and $\mathcal{O}(e^{2}p^{2})$ corrections to $F_{K\pi}^{\mu}$. We find this not the best representation because the former, including strong isospin-breaking effects, is a part of the pure-QCD dressings to the $K\pi$ form factor, which could be studied more systematically with other methods, such as in a dispersive representation  \cite{Bernard:2006gy,Bernard:2007tk,Bernard:2009zm,Bernard:2011ae,Bernard:2013jxa} or in lattice QCD. It is the inclusion of dynamical photons that greatly complicates the problem, for instance, it was not until very recently that the first direct lattice calculation of the electromagnetic effects to $K_{l2}$ and $\pi_{l2}$ decay rates was available \cite{Giusti:2017dwk}, 
and the same method applied to semi-leptonic decays is expected to be extremely challenging \cite{Cirigliano:2019jig,DiCarlo:2019thl}. Therefore, it would sound more natural to regard the $\mathcal{O}(p^{4})$ and $\mathcal{O}(e^{2}p^{2})$ corrections as two separate problems, and only the latter belongs to the electromagnetic radiative correction which is the focus of this paper. For that purpose, it is necessary to explicitly isolate and retain only the pieces linear in $\alpha$.

Our results are given as follows: For $K_{l3}^{+}$,
\begin{eqnarray}
\delta f_{+}^{K^{+}\pi^{0}}(q^{2})&=&\frac{\alpha}{16\sqrt{2}\pi}\left[8+\frac{M_{K}^{2}}{\mu^{2}}-4\ln\frac{M_{K}^{2}}{M_{\gamma}^{2}}\right]+\frac{Z\alpha}{2\sqrt{2}\pi}\frac{M_{K}^{2}}{M_{\eta}^{2}-M_{\pi}^{2}}\left[1+\ln\frac{M_{K}^{2}}{\mu^{2}}\right]\nonumber\\
&&-\frac{8\pi Z\alpha}{\sqrt{2}}\left[\frac{1}{2}\bar{h}_{K^{+}\pi^{0}}(q^{2})+\bar{h}_{K^{0}\pi^{-}}(q^{2})+\frac{3}{2}\bar{h}_{K^{+}\eta}(q^{2})\right]\nonumber\\
&&-\frac{4\pi\alpha}{\sqrt{2}}\left[-2K_{3}^{r}+K_{4}^{r}+\frac{2}{3}K_{5}^{r}+\frac{2}{3}K_{6}^{r}+2K_{12}^{r}\right.\nonumber\\
&&\left.+\frac{2}{3}\frac{M_{\pi}^{2}}{M_{\eta}^{2}-M_{\pi}^{2}}\left(-6K_{3}^{r}+3K_{4}^{r}+2K_{5}^{r}+2K_{6}^{r}-2K_{9}^{r}-2K_{10}^{r}\right)\right]\nonumber\\
\delta f_{-}^{K^{+}\pi^{0}}&=&-\frac{\alpha}{16\sqrt{2}\pi}\left[3\ln\frac{M_{K}^{2}}{\mu^{2}}-4\right]-\frac{Z\alpha}{4\sqrt{2}\pi}\left[1+\frac{M_{K}^{2}}{\mu^{2}}\right]-\frac{3Z\alpha}{8\sqrt{2}\pi}\left[1+\frac{M_{\pi}^{2}}{\mu^{2}}\right]\nonumber\\
&&+\frac{8\pi Z\alpha}{\sqrt{2}}\left[\frac{M_{K}^{2}-M_{\pi}^{2}+3q^{2}}{2q^{2}}K_{K\pi}(q^{2})+\frac{M_{K}^{2}-M_{\pi}^{2}-3q^{2}}{2q^{2}}K_{K\eta}(q^{2})\right.\nonumber\\
&&\left.\left.+\frac{3F_{0}^{2}}{2q^{2}}\left(H_{K\pi}(q^{2})+H_{K\eta}(q^{2})\right)\right]-\frac{8\pi Z\alpha F_{0}^{2}}{\sqrt{2}}\sum_{PQ}\left[\left(a_{PQ}+\frac{M_{P}^{2}-M_{Q}^{2}}{2q^{2}}b_{PQ}\right)\right.\right.\nonumber\\
&&\left. \bar{K}_{PQ}(q^{2})+b_{PQ}\frac{1}{q^{2}}\bar{h}_{PQ}(q^{2})\right]-\frac{4\pi\alpha}{\sqrt{2}}\left[4K_{3}^{r}-2K_{4}^{r}-\frac{2}{3}K_{5}^{r}-\frac{2}{3}K_{6}^{r}\right],
\end{eqnarray}
and for $K_{l3}^{0}$,
\begin{eqnarray}
\delta f_{+}^{K^{0}\pi^{-}}(q^{2})&=&\frac{\alpha}{16\pi}\left[\ln\frac{M_{\pi}^{2}}{\mu^{2}}+8-4\ln\frac{M_{\pi}^{2}}{M_{\gamma}^{2}}\right]-8\pi Z\alpha\left[\frac{1}{2}\bar{h}_{K^{+}\pi^{0}}(q^{2})+\bar{h}_{K^{0}\pi^{-}}(q^{2})+\frac{3}{2}\bar{h}_{K^{+}\eta}(q^{2})\right]\nonumber\\
&&-8\pi\alpha K_{12}^{r}\nonumber\\
\delta f_{-}^{K^{0}\pi^{-}}(q^{2})&=&\frac{\alpha}{16\pi}\left[3\ln\frac{M_{\pi}^{2}}{\mu^{2}}-4\right]-\frac{Z\alpha}{8\pi}\left[1+\ln\frac{M_{\pi}^{2}}{\mu^{2}}\right]-8\pi Z\alpha\left[\frac{M_{K}^{2}-M_{\pi}^{2}-q^{2}}{q^{2}}K_{K\pi}(q^{2})\right.\nonumber\\
&&\left.-\frac{M_{K}^{2}-M_{\pi}^{2}}{q^{2}}K_{K\eta}(q^{2})+\frac{3F_0^{2}}{2q^{2}}\left(H_{K\pi}(q^{2})+H_{K\eta}(q^{2})\right)\right]\nonumber\\
&&-8\pi Z\alpha F_0^{2}\sum_{PQ}\left[\left(c_{PQ}+d_{PQ}\frac{M_P^{2}-M_Q^{2}}{2q^{2}}\right)\bar{K}_{PQ}(q^{2})+\frac{1}{q^{2}}d_{PQ}\bar{h}_{PQ}(q^{2})\right]\nonumber\\
&&+\frac{8\pi\alpha}{3}(K_{5}^{r}+K_{6}^{r}).
\end{eqnarray} 
Here $\{PQ\}=\{K^{+}\pi^{0},K^{0}\pi^{-},K^{+}\eta\}$. The coefficients $\{a_{PQ},b_{PQ},c_{PQ},d_{PQ}\}$ and the functions $\{H_{PQ},K_{PQ}\}$ are defined in Appendix A,C of Ref.~\cite{Cirigliano:2001mk} (in particular, $L_{9}^{r}$ is absorbed into $H_{PQ}$), except that here one have already explicitly factored out $Z$ so one should take $Z\rightarrow 0$ in all these quantities (which also means that $M_{\pi^{\pm}}=M_{\pi^{0}}=M_{\pi}$ and $M_{K^{+}}=M_{K^{0}}=M_{K}$). The extra loop functions $\{\bar{K}_{PQ},\bar{h}_{PQ}\}$ that are required for the isolation of $\mathcal{O}(e^2p^2)$ pieces are defined in Appendix.~\ref{sec:loopfunction}. One sees that each of the invariant functions above is UV-finite and scale-independent. 

Let us pause and comment on the accuracy of the ChPT results above. There are two main sources of uncertainties at a superficial level: (1) the neglected higher-order terms that scale as $\mathcal{O}(e^{2}p^{4})$, and (2) the uncertainties in the renormalized LECs. According to the analysis in Ref.~\cite{Cirigliano:2008wn}, the former is assigned a universal uncertainty of 0.19\% based on chiral power counting, while the latter carries an uncertainty of (0.11-0.16)\%. Let us first consider the LECs. We see that the results above depend on $\{K_i^{r}\}$; if the $\gamma W$-box diagram is also included, then they depend also on $\{X_i^{r}\}$. In present literature, they are expressed in terms of a set of QCD correlation functions, which are then simply estimated within a resonance model \cite{Baur:1996ya,Bijnens:1996kk,Ananthanarayan:2004qk,DescotesGenon:2005pw}. Quantitative analysis of their corresponding theoretical uncertainties are not available, and for most practical purposes they are simply taken as a na\"{\i}ve loop factor $1/(4\pi)^{2}\approx 6.3\times10^{-3}$ \cite{Cirigliano:2001mk} or adopt an arbitrarily-assigned value, say 50\% of the total counterterm contribution \cite{DescotesGenon:2005pw}. Such arbitrariness limits the possibility of a future reduction of the error bars. Furthermore, as we advertised in the Introduction and will prove in Sec.~\ref{sec:logs}, there is also the issue of an incomplete resummation following the introduction of $S_{\mathrm{EW}}$ as a multiplicative factor. All the above-mentioned issues motivate us to consider an alternative approach to the problem in contrast to the pure ChPT treatment, which will be described in the following sections.

\section{\label{sec:OMS}On-Mass-Shell Formula, Ward Identity, and Current Algebra}

The method we propose in this paper is based on a representation of the first-order perturbation correction to strong-interaction amplitudes known as the on-mass-shell (OMS) formula, first introduced in the late 60s \cite{Brown:1970dd}. The statement is as follows. Suppose a form factor $F^{\mu}(p',p)$ is defined through the matrix element:
\begin{equation}
F^{\mu}(p',p)=\left\langle B(p')\right|J^{\mu}(0)\left|A(p)\right\rangle
\end{equation}
where $A,B$ are strongly-interacting scalar particles (just for simplicity), and $p,p'$ are on-shell momenta: $p^{2}=M_A^{2}$, $p^{\prime 2}=M_B^{2}$, where $M_{A,B}^{2}$ are the unperturbed squared masses. If there exists a perturbative Lagrangian $\delta\mathcal{L}$ that causes a first-order perturbation to the form factor, $F^{\mu}(p',p)\rightarrow F^{\mu}(p',p)+\delta F^{\mu}(p',p)$, then the OMS formula states that $\delta F^{\mu}$ can be written totally in terms of on-shell matrix elements as follows:
\begin{equation}
\delta F^{\mu}(p',p)=\lim_{\bar{q}\rightarrow q}iT^{\mu}(\bar{q};p',p)=
\lim_{\bar{q}\rightarrow q}\left\{i\bar{T}^{\mu}(\bar{q};p',p)-iB^{\mu}(\bar{q};p',p)\right\},\label{eq:OMS}
\end{equation}
where
\begin{eqnarray}
\bar{T}^{\mu}(\bar{q};p',p)&=&\int d^{4} xe^{i\bar{q}\cdot x}\left\langle B(p')\right|T\left\{J^{\mu}(x)\delta\mathcal{L}(0)\right\}\left|A(p)\right\rangle\nonumber\\
B^{\mu}(\bar{q};p',p)&=&-\frac{i\delta M_B^{2}}{(p-\bar{q})^{2}-M_B^{2}}F^{\mu}(p-\bar{q},p)-F^{\mu}(p',p'+\bar{q})\frac{i\delta M_A^{2}}{(p'+\bar{q})^{2}-M_A^{2}}.\label{eq:TbarB}
\end{eqnarray}
Here, $\delta M_{A,B}^{2}$ are the first-order perturbation to the squared mass of $A,B$ due to the effect of $\delta\mathcal{L}$: $M_{A,B}^{2}\rightarrow M_{A,B}^{2}+\delta M_{A,B}^{2}$, which can be written as:
\begin{equation}
\delta M_A^{2}=-\left\langle A(p)\right|\delta\mathcal{L}(0)\left|A(p)\right\rangle,\:\:\delta M_B^{2}=-\left\langle B(p')\right|\delta\mathcal{L}(0)\left|B(p')\right\rangle.
\end{equation}
Notice that both $\bar{T}^{\mu}$ and $B^{\mu}$ possess the same single-particle poles at $(p-\bar{q})^{2}=M_B^{2}$ and $(p'+\bar{q})^{2}=M_A^{2}$ which cancel each other, making $T^{\mu}$ pole-free (more discussions are given in Sec.~\ref{sec:diagrammatic}). Simply speaking, the validity of Eq.~\eqref{eq:OMS} stems from the fact that the difference between $iB^{\mu}$ and the one-particle-reducible piece in $i\bar{T}^{\mu}$ in the $\bar{q}\rightarrow q$ limit reproduces the effect of the wavefunction renormalization as well as the mass perturbation of the external states. Interested readers shall refer to Ref.~\cite{Brown:1970dd} for more details. 

Sirlin further developed a Ward identity (WI) treatment on top of the OMS formula \cite{Sirlin:1977sv}. One starts from the following mathematical identity:
\begin{equation}
iT^{\mu}(\bar{q};p',p)=-\bar{q}_{\nu}\frac{\partial}{\partial\bar{q}_{\mu}}iT^{\nu}(\bar{q};p',p)+\frac{\partial}{\partial\bar{q}_{\mu}}\left(\bar{q}_{\nu} iT^{\nu}(\bar{q};p',p)\right).\label{eq:separate}
\end{equation}
Furthermore, since $T^{\nu}$ involves a time-ordered product, one could use the following identity:
\begin{equation}
\partial_{\mu} T\left\{J^{\mu}(x)\delta\mathcal{L}(0)\right\}=T\left\{\partial_{\mu} J^{\mu}(x)\delta\mathcal{L}(0)\right\}+\delta(x_0)\left[J^{0}(x),\delta\mathcal{L}(0)\right],
\end{equation}
to rewrite the second term at the RHS of Eq.~\eqref{eq:separate} as:
\begin{equation}
\bar{q}_{\nu} iT^{\nu}(\bar{q};p',p)=i(D-\bar{q}\cdot B)-\int d^{4}xe^{i\bar{q}\cdot x}\delta(x_0)\left\langle B(p')\right|\left[J^{0}(x),\delta\mathcal{L}(0)\right]\left|A(p)\right\rangle,\label{eq:WI}
\end{equation}
where 
\begin{equation}
D(\bar{q};p',p)=i\int d^{4}xe^{i\bar{q}\cdot x}\left\langle B(p')\right|T\left\{\partial_{\mu} J^{\mu}(x)\delta\mathcal{L}(0)\right\}\left|A(p)\right\rangle.\label{eq:D}
\end{equation}
After such rewriting, we can take the $\bar{q}\rightarrow q$ limit on both sides of Eq.~\eqref{eq:separate}. Using the OMS formula, Eq.~\eqref{eq:separate} now tells us that $\delta F^{\mu}$ can be split into two terms:
\begin{equation}
\delta F^{\mu}(p',p)=\delta F^{\mu}_{2}(p',p)+\delta F^{\mu}_{3}(p',p),\label{eq:OMSWI}
\end{equation}
where:
\begin{eqnarray}
\delta F^{\mu}_{2}(p',p)&=&-\lim_{\bar{q}\rightarrow q}\frac{\partial}{\partial\bar{q}_{\mu}}\int d^{4}xe^{i\bar{q}\cdot x}\delta(x_0)\left\langle B(p')\right|\left[J^{0}(x),\delta\mathcal{L}(0)\right]\left|A(p)\right\rangle\nonumber\\
\delta F^{\mu}_{3}(p',p)&=&-\lim_{\bar{q}\rightarrow q}\bar{q}_{\nu}\frac{\partial}{\partial\bar{q}_{\mu}}iT^{\nu}(\bar{q};p',p)+\lim_{\bar{q}\rightarrow q}i\frac{\partial}{\partial\bar{q}_{\mu}}\left(D(\bar{q};p',p)-\bar{q}\cdot B(\bar{q};p',p)\right)\label{eq:gen2and3pt}
\end{eqnarray} 
shall be known as the ``two-point function'' and the ``three-point function'', respectively. 

The formalism above can be straightforwardly applied to study the electromagnetic radiative corrections to $\delta F_{K\pi}^{\mu}$ based on the setup in Section~\ref{sec:Semileptonic}. The only complication is that $\delta\mathcal{L}$ is not anymore a local interaction, but comes from one-loop QED corrections with a PV-regulator to the photon propagator. That gives~\cite{Brown:1970dd}:
\begin{equation}
\delta\mathcal{L}(0)=\frac{e^{2}}{2}\int\frac{d^{4}k}{(2\pi)^{4}}\int d^{4}x e^{ik\cdot x}\frac{1}{k^{2}-M_{\gamma}^{2}}\frac{M_{W}^{2}}{M_{W}^{2}-k^{2}}T\left\{J_{\mathrm{em}}^{\mu}(x)J^{\mathrm{em}}_{\mu}(0)\right\}.\label{eq:deltaLQED}
\end{equation}
That means $iT^{\mu}$ now involves a time-ordered product of three currents. Its corresponding WI can be derived using the following identity:
\begin{eqnarray}
\partial_{x,\mu}T\left\{J_{W}^{\mu\dagger}(x)J_{\mathrm{em}}^{\nu}(y)J^{\mathrm{em}}_{\nu}(0)\right\}&=&T\left\{\partial_{\mu} J_{W}^{\mu\dagger}(x)J_{\mathrm{em}}^{\nu}(y)J^{\mathrm{em}}_{\nu}(0)+\delta(x^{0}-y^{0})[J_{W}^{0\dagger}(x),J_{\mathrm{em}}^{\nu}(y)]J^{\mathrm{em}}_{\nu}(0)\right.\nonumber\\
&&\left.+\delta(x^{0})[J_{W}^{0\dagger}(x),J_{\mathrm{em}}^{\nu}(0)]J^{\mathrm{em}}_{\nu}(y)\right\}~.
\end{eqnarray}
Notice that the last two terms are equal-time commutators that satisfy the current algebra relation \eqref{eq:CA}. Therefore, through an identical derivation as before, one obtains the following splitting for the electromagnetic radiative correction to $F_{K\pi}^{\mu}$ into two-point and three-point functions:
\begin{equation}
\delta F_{K\pi}^{\mu}(p',p)=\delta F_{K\pi,2}^{\mu}(p',p)+\delta F_{K\pi,3}^{\mu}(p',p),
\end{equation}
where 
\begin{eqnarray}
\delta F^{\mu}_{K\pi,2}(p',p)&=&-\frac{e^{2}}{2}\int\frac{d^{4}k}{(2\pi)^{4}}T^{\lambda}_{\:\:\:\lambda}(k;p',p)\frac{\partial}{\partial k_{\mu}}\left(\frac{1}{k^{2}-M_{\gamma}^{2}}\frac{M_{W}^{2}}{M_{W}^{2}-k^{2}}\right)
\nonumber\\
\delta F^{\mu}_{K\pi,3}(p',p)&=&-\lim_{\bar{q}\rightarrow q}\bar{q}_{\nu}\frac{\partial}{\partial\bar{q}_{\mu}}iT^{\nu}(\bar{q};p',p)+\lim_{\bar{q}\rightarrow q}i\frac{\partial}{\partial\bar{q}_{\mu}}\left(D(\bar{q};p',p)-\bar{q}\cdot B(\bar{q};p',p)\right),\label{eq:twopt}
\end{eqnarray}
with $iT^{\mu}=i\bar{T}^{\mu}-iB^{\mu}$, and 
\begin{eqnarray}
\bar{T}^{\mu}(\bar{q};p',p)&=&\frac{e^{2}}{2}\int\frac{d^{4}k}{(2\pi)^{4}}\frac{1}{k^{2}-M_{\gamma}^{2}}\frac{M_{W}^{2}}{M_{W}^{2}-k^{2}}\nonumber\\ &\times& \int d^{4}xe^{i\bar{q}\cdot x}\int d^{4}ye^{ik\cdot y}
\left\langle \pi(p')\right|T\left\{J_{W}^{\mu\dagger}(x)J_{\mathrm{em}}^{\nu}(y)J^{\mathrm{em}}_{\nu}(0)\right\}\left|K(p)\right\rangle\nonumber\\
D(\bar{q};p',p)&=&\frac{ie^{2}}{2}\int\frac{d^{4}k}{(2\pi)^{4}}\frac{1}{k^{2}-M_{\gamma}^{2}}\frac{M_{W}^{2}}{M_{W}^{2}-k^{2}}\nonumber\\ &\times&\int d^{4}xe^{i\bar{q}\cdot x}\int d^{4}ye^{ik\cdot y}
\left\langle \pi(p')\right|T\left\{\partial_{\mu} J_{W}^{\mu\dagger}(x)J_{\mathrm{em}}^{\nu}(y)J^{\mathrm{em}}_{\nu}(0)\right\}\left|K(p)\right\rangle.
\end{eqnarray}
Obviously only the vector component of $J_{W}^{\mu\dagger}$ contributes to $\bar{T}^{\mu}$ and $D$. The definition of $B^{\mu}(\bar{q};p',p)$ is the same as in Eq.~\eqref{eq:TbarB}, except that we adopt the form of $\delta\mathcal{L}$ in Eq.~\eqref{eq:deltaLQED}. An important observation is that the two-point function depends on the generalized Compton tensor $T^{\mu\nu}(k;p',p)$ as defined in Eq.~\eqref{eq:GCompton}, which is exactly the same as that appeared in the $\gamma W$ box diagram (except that here the Lorentz indices are contracted among themselves).

The original motivations for the OMS+WI treatment of semi-leptonic beta decay form factors are of three-fold. First, it provides a convenient framework to study the matching between SM semileptonic beta decay and Fermi theory (i.e. the derivation of Eq.~\eqref{eq:SMtoFermi}). Second, for the electromagnetic radiative corrections to the form factor, all the $\ln M_{W}$-enhanced terms are contained in the two-point function, so such a separation makes it easy to discuss the large $M_{W},M_{Z}$ behavior of the generic $\mathcal{O}(G_F\alpha)$ correction to the decay rate \cite{Sirlin:1981ie}. Third, one observes that the three-point function would vanish if the vector charged weak current is exactly conserved, i.e. $(\partial_{\nu} J^{\nu\dagger}_{W})_V=0$, and $q^{\mu}\rightarrow 0$. This is particularly important in superallowed beta decays of nuclei, because the three-point function is then suppressed by $q/M,\hat{m}/M\sim 10^{-3}$, where $M$ is the nuclear mass, which makes it completely negligible. 

We argue that this formalism is equally useful in $K_{l3}$. First, the most important pieces in the $\mathcal{O}(G_F\alpha)$ corrections, namely $\delta F_{K\pi,2}^{\mu}$ and $\Box_{\gamma W}$, are now collectively represented by the momentum integral of a single hadronic quantity $T^{\mu\nu}(k;p',p)$. Therefore, through a systematic non-perturbative study of the latter, one could go beyond the $\mathcal{O}(e^{2}p^{2})$ result in ChPT and sum up terms to all orders in the chiral power counting. Second, although unlike the superallowed beta decay, the three-point function in $K_{l3}$ is not suppressed by small scales, we are still able to calculate it order-by-oder in ChPT as we will demonstrate in Sec.~\ref{sec:matching}, and we will find in certain channels that the unknown LECs in the three-point functions have negligible impact on the decay rate. Therefore, the combined OMS+WI+ChPT framework possesses both conceptual and practical advantages over the pure ChPT framework, and thus is a better starting point.

\section{\label{sec:logs}Obtaining the large-logs and comparing with the pure EFT language}

As the first application of the method, let us reproduce the short-distance logarithmic factor which was first derived in Ref.~\cite{Sirlin:1981ie}. For simplicity, let us neglect the $\mathcal{O}(\alpha_{s})$ pQCD corrections. The key to study the asymptotic behavior is the following leading-twist, free-field operator product expansion (OPE) for the generalized Compton tensor $T^{\mu\nu}(k;p',p)$ at large and negative $k^2$~\cite{Bjorken:1966jh,Johnson:1966se}:
\begin{equation}
T^{\mu\nu}(k;p',p)\rightarrow \frac{i}{k^2}\left\{g^{\mu\nu}k^{\lambda}-k^{\mu}g^{\nu\lambda}-k^{\nu}g^{\mu\lambda}-2i\bar{Q}\varepsilon^{\mu\nu\alpha\lambda}k_{\alpha}\right\}\left\langle \pi(p')\right|J_{W\lambda}^{\dagger}(0)\left|K(p)\right\rangle,\label{eq:OPE}
\end{equation}
where $\varepsilon^{0123}=-1$ in our convention. Eq.~\eqref{eq:OPE} can be obtained simply by taking the expressions of the electroweak currents in Eq.~\eqref{eq:currents} and contracting a pair of quark fields to obtain a massless quark propagator.

We can now split the two-point function in Eq.~\eqref{eq:twopt} into two pieces:
\begin{eqnarray}
\delta F^{\mu}_{K\pi,2}&=&-\frac{e^{2}}{2}\int_{|k|>\tilde{\mu}}\frac{d^{4}k}{(2\pi)^{4}}T^{\lambda}_{\:\:\:\lambda}(k;p',p)\frac{\partial}{\partial k_{\mu}}\left(\frac{1}{k^{2}}\frac{M_{W}^{2}}{M_{W}^{2}-k^{2}}\right)+\delta F^{\mathrm{res},\mu}_{K\pi,2}(\tilde{\mu})
\end{eqnarray}
where $\tilde{\mu}$ is a scale above which the leading-twist, free-field OPE is valid (but at the same time $\tilde{\mu}\ll M_{W}$, so a common educational guess is $\tilde{\mu}\approx 2$ GeV), and:
\begin{eqnarray}
\delta F^{\mathrm{res},\mu}_{K\pi,2}(\tilde{\mu})&=&-\frac{e^{2}}{2}\int_{|k|<\tilde{\mu}}\frac{d^{4}k}{(2\pi)^{4}}T^{\lambda}_{\:\:\:\lambda}(k;p',p)\frac{\partial}{\partial k_{\mu}}\left(\frac{1}{k^{2}-M_{\gamma}^{2}}\frac{M_{W}^{2}}{M_{W}^{2}-k^{2}}\right)
\end{eqnarray}
is the residual, non-asymptotic piece which is now $\tilde{\mu}$-dependent. Then, substituting the expression in Eq.~\eqref{eq:OPE} the $|k|>\tilde{\mu}$ integral can be performed analytically:
\begin{equation}
-\frac{e^{2}}{2}\int_{|k|>\tilde{\mu}}\frac{d^{4}k}{(2\pi)^{4}}T^{\lambda}_{\:\:\:\lambda}(k;p',p)\frac{\partial}{\partial k_{\mu}}\left(\frac{1}{k^{2}}\frac{M_{W}^{2}}{M_{W}^{2}-k^{2}}\right)=-\frac{\alpha}{8\pi}\left(\ln\frac{M_{W}^{2}}{\tilde{\mu}^{2}}\right)F_{K\pi}^{\mu}.
\end{equation}
Similarly, the $\gamma W$-box amplitude in Eq.~\eqref{eq:box} can be split into two pieces:
\begin{eqnarray}
\Box_{\gamma W}&=&-\frac{G_Fe^2}{\sqrt{2}}\int_{|k|>\tilde{\mu}}\frac{d^{4}k}{(2\pi)^{4}}\frac{\bar{u}_{\nu}\gamma^{\nu}(1-\gamma_{5})\slashed{k}\gamma^{\mu}v_{l}}{k^{2}}\frac{1}{k^{2}}\frac{M_{W}^{2}}{M_{W}^{2}-k^{2}}T_{\mu\nu}(k;p',p)+\Box_{\gamma W}^{\mathrm{res}}(\tilde{\mu})\nonumber\\
&&=\frac{3\alpha}{4\pi}\left(\ln\frac{M_{W}^2}{\tilde{\mu}^{2}}\right)\mathfrak{M}_{0}+\Box_{\gamma W}^{\mathrm{res}}(\tilde{\mu}),
\end{eqnarray}
where we have again evaluated the integral in the first term analytically using the OPE relation~\eqref{eq:OPE}. Finally, one isolates the large-logarithmic term in the charged-lepton wavefunction renormalization in Eq.~\eqref{eq:Zl} by writing:
\begin{equation}
\delta Z_{l}=-\frac{\alpha}{4\pi}\ln\frac{M_{W}^{2}}{\tilde{\mu}^{2}}+\delta Z_{l}^{\mathrm{res}}(\tilde{\mu}).
\end{equation}
After these treatments, one can now write the total $K_{l3}$ decay amplitude in Eq.~\eqref{eq:basicM} as:
\begin{eqnarray}
\mathfrak{M}&=&\left[1-\frac{\alpha}{2\pi}\ln\frac{M_{W}^{2}}{M_{Z}^{2}}-\frac{\alpha}{8\pi}\ln\frac{M_{W}^{2}}{\tilde{\mu}^{2}}+\frac{3\alpha}{4\pi}
\ln\frac{M_{W}^{2}}{\tilde{\mu}^2}-\frac{\alpha}{8\pi}\ln\frac{M_{W}^{2}}{\tilde{\mu}^{2}}\right]\frak{M}_{0}\nonumber\\
&&-\frac{G_F}{\sqrt{2}}\bar{u}_{\nu}\gamma_{\mu}(1-\gamma_{5})\nu_{l}\left[\frac{1}{2}\delta Z_{l}^{\mathrm{res}}(\tilde{\mu})F_{K\pi}^{\mu}+ \delta F_{K\pi,2}^{\mathrm{res},\mu}(\tilde{\mu})+\delta F_{K\pi,3}^{\mu}\right]+\Box_{\gamma W}^{\mathrm{res}}(\tilde{\mu})\nonumber\\
&&=\tilde{S}^{\frac{1}{2}}_{\mathrm{EW}}(\tilde{\mu})\mathfrak{M}_0-\frac{G_F}{\sqrt{2}}\bar{u}_{\nu}\gamma_{\mu}(1-\gamma_{5})\nu_{l}\left[\frac{1}{2}\delta Z_{l}^{\mathrm{res}}(\tilde{\mu})F_{K\pi}^{\mu}+ \delta F_{K\pi,2}^{\mathrm{res},\mu}(\tilde{\mu})+\delta F_{K\pi,3}^{\mu}\right]+\Box_{\gamma W}^{\mathrm{res}}(\tilde{\mu}),\nonumber\\
\label{eq:separatelog}
\end{eqnarray}
where 
\begin{equation}
\tilde{S}_{\mathrm{EW}}(\tilde{\mu})=1+\frac{\alpha}{\pi}\ln\frac{M_{Z}^{2}}{\tilde{\mu}^2}
\end{equation}
is the desired short-distance logarithmic factor. {One observes that $\tilde{S}_{\mathrm{EW}}(M_{\rho})=S_{\mathrm{EW}}$.

With the preparations above we are now able to compare the matching procedure of physics at different scales in our formalism and in the usual ChPT formalism. Suppose $\delta\mathfrak{M}$ is the correction to the $K_{l3}$ decay amplitude contributed from both the two-point function and the $\gamma W$-box, then we can always decompose it schematically as:
\begin{eqnarray}
\mathfrak{M}_0+\delta\mathfrak{M}=\mathfrak{M}_{0}+\frac{\alpha}{\pi}\int d^{4}k \mathfrak{A}(k)&=&\left[\mathfrak{M}_0+\frac{\alpha}{\pi}\int_{|k|>\tilde{\mu}}d^4k\mathfrak{A}(k)\right]+\frac{\alpha}{\pi}\int_{|k|<\tilde{\mu}}d^4k\mathfrak{A}(k)\nonumber\\
&\equiv&\tilde{S}_{\mathrm{EW}}^{\frac{1}{2}}(\tilde{\mu})\mathfrak{M}_{0}+\frac{\alpha}{\pi}\Delta(\tilde{\mu})\nonumber\\
&=&S_{\mathrm{EW}}^{\frac{1}{2}}\mathfrak{M}_0+\frac{\alpha}{\pi}\left[\left(\ln\frac{M_{\rho}}{\tilde{\mu}}\right)\mathfrak{M}_{0}+\Delta(\tilde{\mu})\right].\label{eq:correctresum}
\end{eqnarray}
The two terms in the square bracket in the third line must add up to be $\tilde{\mu}$-independent. In the meantime, in a standard ChPT calculation of $\delta\mathfrak{M}$ up to $\mathcal{O}(e^{2}p^{2})$, one obtains:
\begin{equation}
\delta\mathfrak{M}^{e^2p^2}=\frac{\alpha}{\pi}\left(\ln\frac{M_{Z}}{M_{\rho}}\right)\mathfrak{M}_{0}^{p^{2}}+\frac{\alpha}{\pi}\left[\left(\ln\frac{M_{\rho}}{\tilde{\mu}}\right)\mathfrak{M}_{0}^{p^{2}}+\Delta^{p^{2}}(\tilde{\mu})\right],
\end{equation}
where in fixed chiral order one is able to demonstrate explicitly that the terms in the square bracket is indeed $\tilde{\mu}$-independent (see, e.g. Ref.~\cite{DescotesGenon:2005pw}). The problem, however, comes from the application of the result above to the master formula~\eqref{eq:EFTmaster}, which implies:
\begin{equation}
\mathfrak{M}_{0}+\delta\mathfrak{M}_{0}\rightarrow S_{\mathrm{EW}}^{\frac{1}{2}}\mathfrak{M}_0+\frac{\alpha}{\pi}\left[\left(\ln\frac{M_{\rho}}{\tilde{\mu}}\right)\mathfrak{M}_{0}^{p^{2}}+\Delta^{p^{2}}(\tilde{\mu})\right].\label{eq:wrongresum}
\end{equation}
The first term in the RHS resums $\mathfrak{M}_{0}$ to all order in chiral power counting, and that arises naturally from Eq.~\eqref{eq:EFTmaster} due to the isolation of $S_{\mathrm{EW}}$ as a constant multiplicative factor. In contrast, the terms in the square bracket goes into $\delta_{\mathrm{em}}^{Kl}$ and are not resummed, unlike those in Eq.~\eqref{eq:correctresum}. Therefore, the existing ChPT treatment actually performs an incomplete resummation of $\mathcal{O}(e^{2}p^{2n})$ terms from the two-point function and the $\gamma W$-box. The effects of these neglected terms are in principle embedded in the estimated universal 0.19\% uncertainty in Ref.~\cite{Cirigliano:2008wn}, but it is now clear that in our formalism this part is automatically improved. The key is to study $\Box_{\gamma W}$ and $\delta F_{K\pi,2}^{\mu}$ directly from their integral form, Eq.~\eqref{eq:box} and~\eqref{eq:twopt}, without relying on chiral power expansion. Making use of a dispersion relation, one may re-express the $k$-integrals not with respect to $T^{\mu\nu}$ but its discontinuity\footnote{We display here only the right-hand discontinuity for illustration.}:
\begin{equation}
W^{\mu\nu}(k;p',p)=\sum_{X}(2\pi)^{4}\delta^{4}(k+p'-p_{X})\left\langle \pi(p')\right|J_{\mathrm{em}}^{\mu}(0)\left|X\right\rangle \left\langle X\right|J_{W}^{\nu\dagger}(0)\left|K(p)\right\rangle,
\end{equation}
which consists of products of single-current on-shell matrix elements. With that one may more easily identify the dominant intermediate states at different kinematical regions and perform better matching on boundaries. One should, however, be aware of several extra complications comparing to the case of neutron and superallowed beta decay~\cite{Seng:2018qru}. For instance, the tensor $T^{\mu\nu}(k;p',p)$ is highly non-forward, so one expects its corresponding dispersion relation to be more involved. Detailed research along this direction will be performed in a follow-up work.

\section{\label{sec:diagrammatic}Diagrammatic Approach to the Three-point function}

\begin{figure}[tb]
	\begin{centering}
		\includegraphics[scale=0.2]{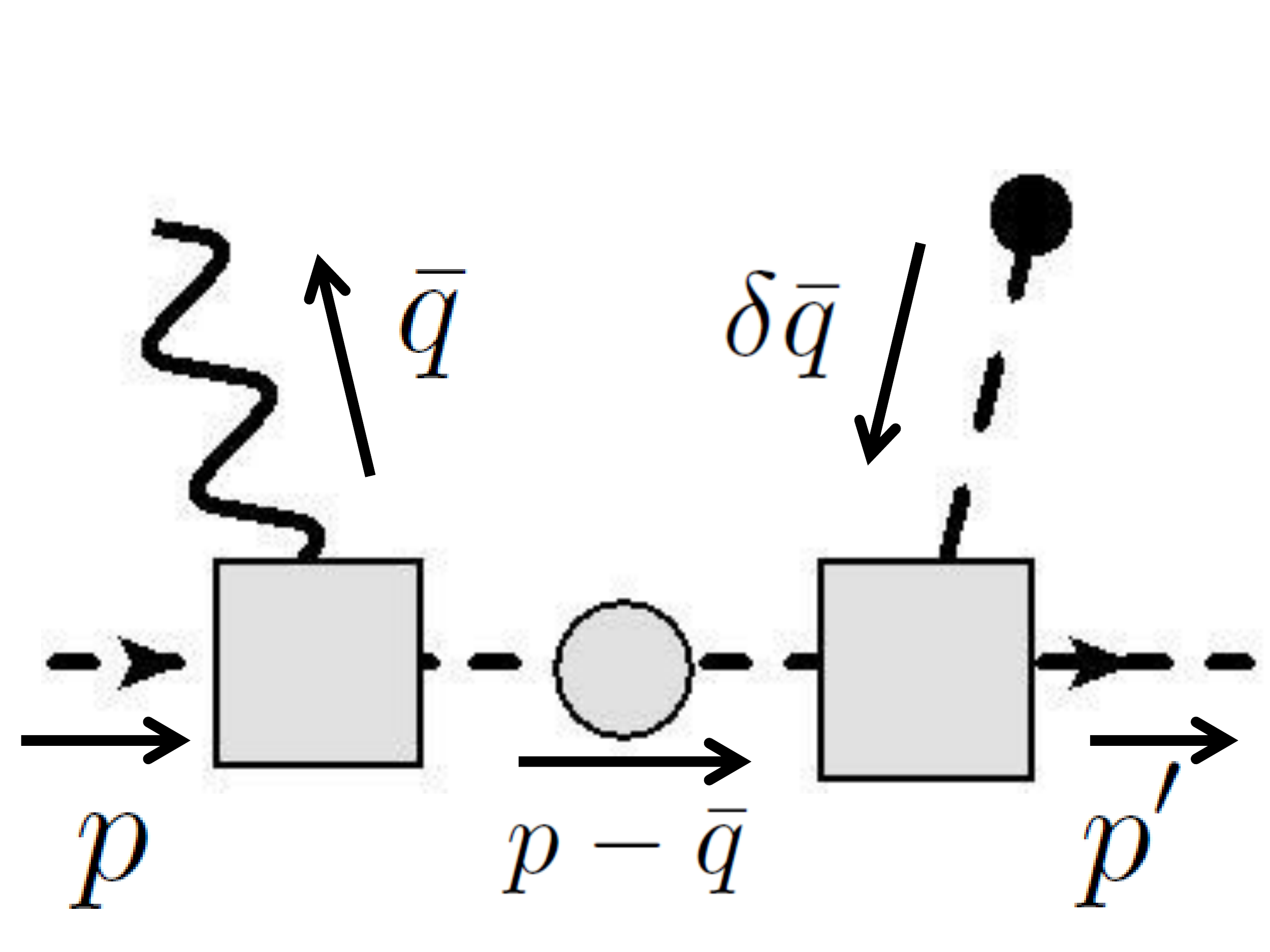}\hfill
		\includegraphics[scale=0.2]{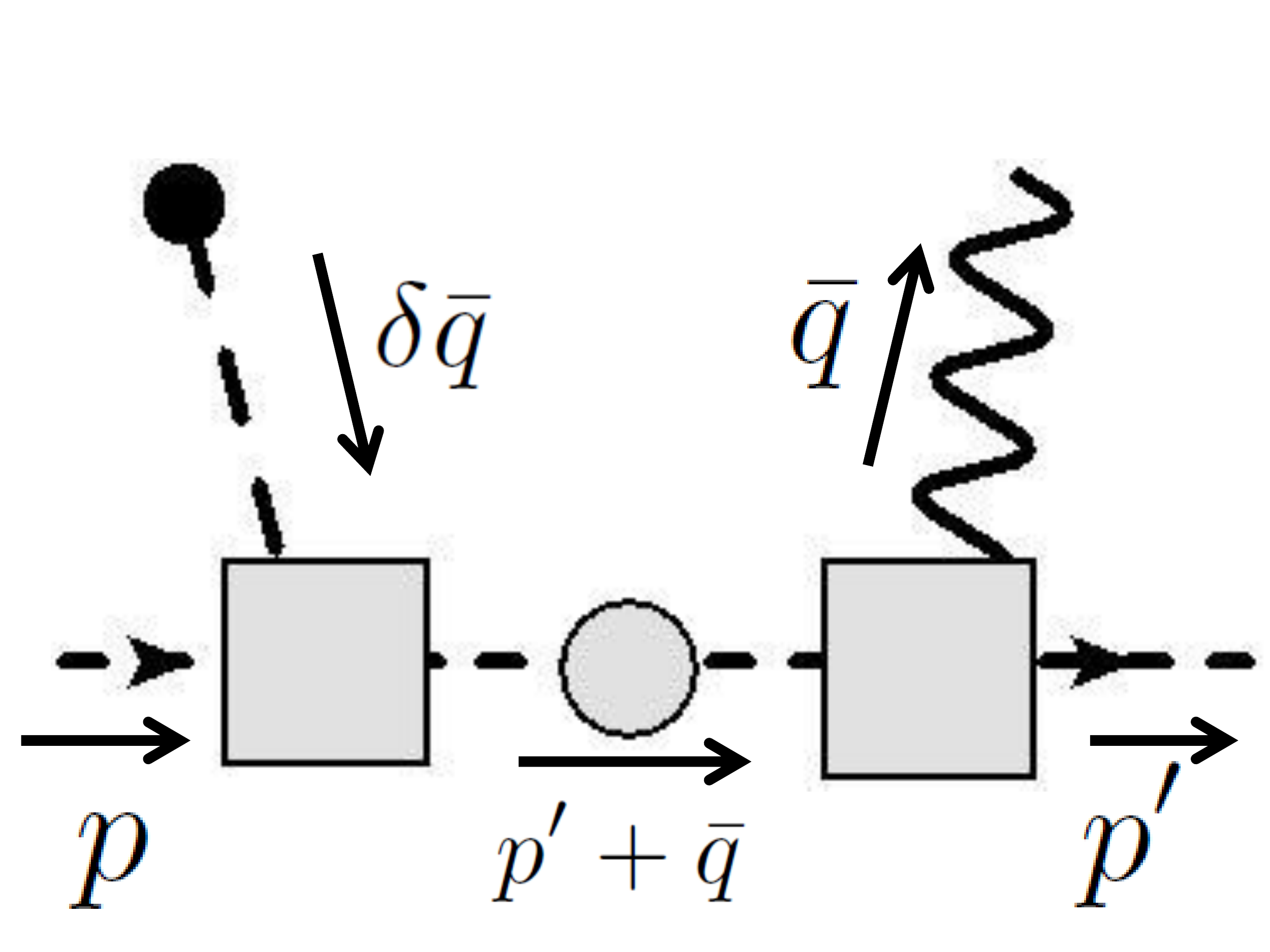}\hfill
		\includegraphics[scale=0.2]{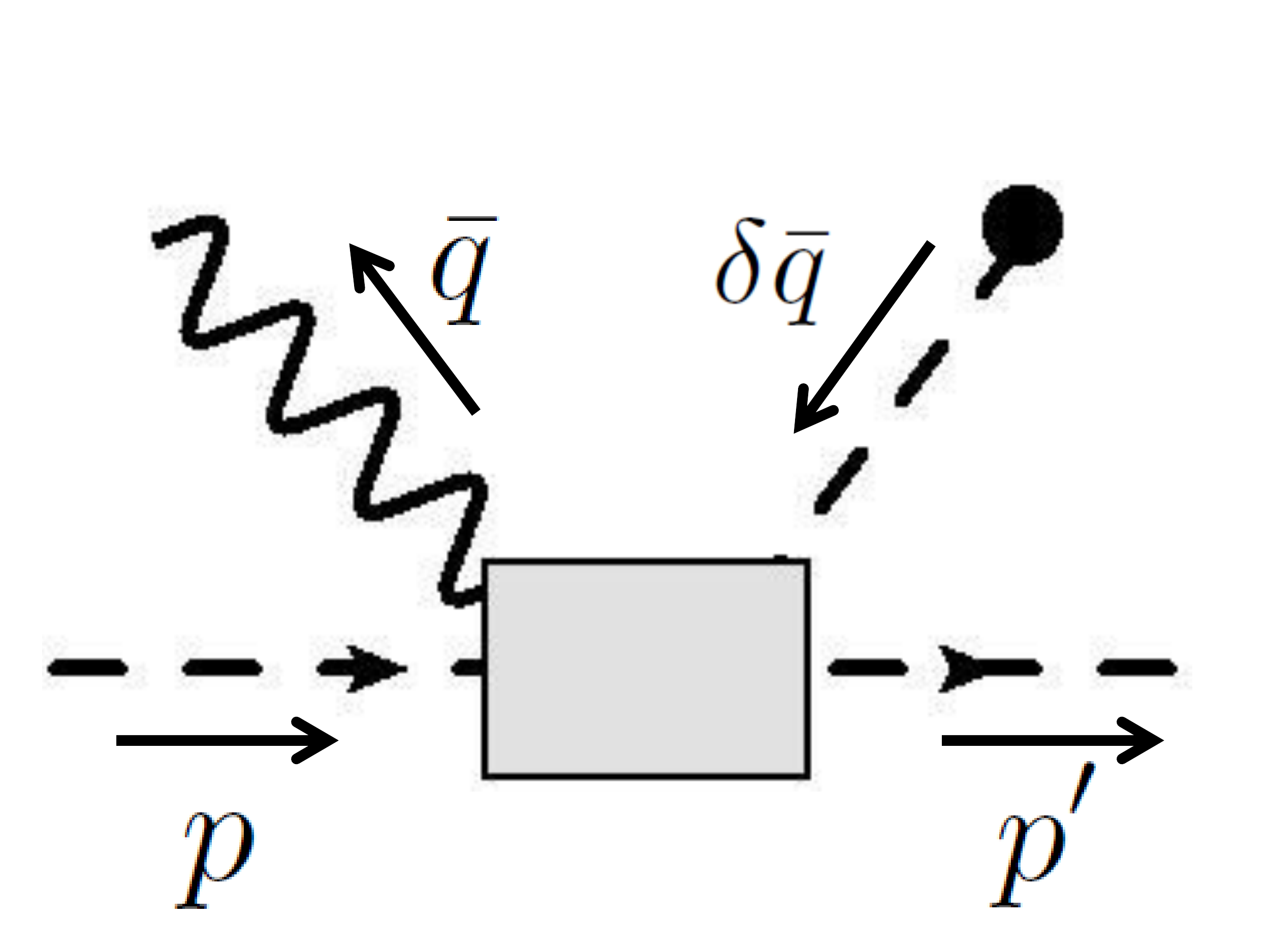}
		\par\end{centering}
	        \caption{\label{fig:Tbarmu}Diagrammatic representation of $i\bar{T}^{\mu}$ (or $D$). Grey boxes represent 1PI correlation functions containing the insertion of $J^{\mu}$ (or $\partial_{\nu} J^{\nu}$), $i\delta\mathcal{L}$, or both, grey circles denote full single-particle propagator, and dashed lines with black circle at the end depict the external momentum insertion to the $i\delta\mathcal{L}$ vertex.}
\end{figure}

To really make practical use of our new method, we must be able to calculate not just $\Box_{\gamma W}$, $\delta F_{K\pi,2}^{\mu}$ but also the three-point function $\delta F_{K\pi,3}^{\mu}$. The latter consists of time-ordered products of three currents and it is more difficult to be studied in a completely non-perturbative approach. Fortunately, since it is not short-distance-enhanced, one is able to calculate it safely in an EFT. Below we describe a diagrammatic approach that permits an order-by-order calculation of the three-point function. We shall begin with a generic argument, and then apply it to $K_{l3}$. 
	
The general definition of a three-point function, as in Eq.~\eqref{eq:gen2and3pt}, consists of two terms. To obtain the first term one needs to compute the quantity $i\bar{T}^{\mu}(\bar{q};p',p)$ based on its definition in Eq.~\eqref{eq:TbarB}. From a simple analysis based on Wick's theorem, one sees that $i\bar{T}^{\mu}$ is simply given by the three types of diagrams in Fig.~\ref{fig:Tbarmu}, multiplied with the wavefunction renormalization factor $(Z_A Z_B)^{1/2}$ due the on-shell external states $A$ and $B$. In each diagram, there is a momentum $\bar{q}$ flowing out of the $J^{\mu}$ vertex, and a small external momentum $\delta\bar{q}=\bar{q}-q$ flowing into the $i\delta\mathcal{L}$ vertex. All these diagrams can be calculated using ordinary Feynman rules. One also observes that the first two types of diagrams have a single-particle pole in the $\bar{q}\rightarrow q$ limit whereas the third type does not. These poles are, however, exactly canceled by those in $iB^{\mu}(\bar{q};p',p)$, making $iT^{\mu}(\bar{q};p',p)$ pole-free. Such a cancellation can be seen as follows: let us consider the first diagram in Fig.~\ref{fig:Tbarmu} as an example. We shall name the one-particle-irreducible (1PI) correlation function due to the insertion of $J^{\mu}$ and $i\delta\mathcal{L}$ as $(Z_A Z_B)^{-1/2}\mathscr{F}^{\mu}(p-\bar{q},p)$ and
$-iZ_B^{-1}\Gamma_B(p',p-\bar{q})$, respectively. They obviously satisfy the following relations:
\begin{equation}
\lim_{(p-\bar{q})^{2}\rightarrow M_B^{2}}\mathscr{F}^{\mu}(p-\bar{q},p)=F^{\mu}(p-\bar{q},p),\:\:\:\lim_{(p-\bar{q})^{2}\rightarrow M_B^{2}}\Gamma_B(p',p-\bar{q})=\delta M_B^{2}.\label{eq:limits}
\end{equation}
In fact one can be more specific about the form factor: if we parameterize $\mathscr{F}^{\mu}(p_{2},p_{1})$ as:
\begin{equation}
\mathscr{F}^{\mu}(p_{2},p_{1})=F_{1}\left(p_{2}^{2},p_{1}^{2},(p_{2}-p_{1})^{2}\right)p_{1}^{\mu}+F_{2}\left(p_{2}^{2},p_{1}^{2},(p_{2}-p_{1})^{2}\right)p_{2}^{\mu},
\end{equation}
where $p_{1},p_{2}$ are generic off-shell momenta, then the on-shell form factor is simply defined as:
\begin{equation}
F^{\mu}(p_{2},p_{1})=F_{1}\left(M_B^{2},M_A^{2},(p_{2}-p_{1})^{2}\right)p_{1}^{\mu}+F_{2}\left(M_B^{2},M_A^{2},(p_{2}-p_{1})^{2}\right)p_{2}^{\mu}.
\end{equation}
On the other hand, the full propagator of the scalar particle $B$ can be written as:
\begin{equation}
S_B(p-\bar{q})=\frac{iZ_B}{(p-\bar{q})^{2}-M_B^{2}}+...~,
\end{equation}
where ``+..." are the remaining, pole-free terms. Therefore, adding this diagram to the first term of $-iB^{\mu}(\bar{q};p',p)$ as in Eq.~\eqref{eq:TbarB} gives:
\begin{equation}
iT^{\mu}(\bar{q};p',p)=\frac{1}{(p-\bar{q})^{2}-M_B^{2}}\left[\Gamma_B(p',p-\bar{q})\mathscr{F}^{\mu}(p-\bar{q},p)-\delta M_B^{2}F^{\mu}(p-\bar{q},p)\right]+...
\end{equation}
which is obviously pole-free because the numerator vanishes as $(p-\bar{q})^{2}\rightarrow M_B^{2}$. The second diagram in Fig.~\ref{fig:Tbarmu} follows the same argument, so we have proven that $iT^{\mu}(\bar{q};p',p)$ is pole-free. 

To obtain the second term in the three-point function, one needs to similarly compute the quantity $D(\bar{q};p',p)$ diagrammatically.
Through its definition in Eq.~\eqref{eq:D}, it appears to be exactly the same as $i\bar{T}^{\mu}(\bar{q};p',p)$ upon the replacement $J^{\mu}(x)\rightarrow \partial_{\nu} J^{\nu}(x)$ at the latter; therefore the corresponding Feynman diagrams are also those in Fig.~\ref{fig:Tbarmu}. However, one would obtain erroneous results if one calculated it based on na\"{\i}ve Feynman rules of the operator $\partial_{\nu} J^{\nu}$ including the derivative. In fact, if one did so then one would obtain $D=\bar{q}_{\mu} T^{\mu}$ which would invalidate the WI in Eq.~\eqref{eq:WI}. Let us understand the origin of this apparent paradox: in deriving Eq.~\eqref{eq:WI} one assumes the following definition of the time-ordered product:
\begin{equation}
T[A(t)B(0)]=\Theta(t)A(t)B(0)+\Theta(-t)B(0)A(t).\label{eq:timeordered}
\end{equation}
Notice, however, that this is not necessarily the same time-ordered product as one calculates using covariant perturbation theory (i.e. with ordinary Feynman rules), as the latter could lead to extra additive terms proportional to $
\delta(t)$ at the RHS in order to maintain the Lorentz covariance of the final result. Since $T^{\mu}$ is defined in covariant formalism, a more rigorous way to write Eq.~\eqref{eq:WI} is as follows:
\begin{equation}
\bar{q_{\nu}}iT^{\nu}=i(D-\bar{q}\cdot B)-\int d^{4}xe^{i\bar{q}\cdot x}\delta(x_0)\left\langle B(p')\right|\left[J^{0}(x),\delta\mathcal{L}(0)\right]\left|A(p)\right\rangle+\bar{q}_{\nu} i\delta \bar{T}^{\nu},
\end{equation}
where $i\delta\bar{T}^{\nu}$ denotes the difference between $i\bar{T}^{\nu}$ calculated with the covariant formalism and with the definition \eqref{eq:timeordered}; this term must be added to the commutator term to form a complete, covariant two-point function $\delta F_{2}^{\mu}$. 

Back to our discussion on $D$. We now know that the time-ordered product in $D$ is defined as in Eq.~\eqref{eq:timeordered}, nevertheless there is still a way to calculate it using covariant perturbation theory. The key is to realize that the discrepancy between these two methods stems from the existence of derivative operator in $\partial_{\nu} J^{\nu}$, which on the one hand picks up the off-shellness of the intermediate-state momenta in the covariant formalism. On the other hand, based on Eq.~\eqref{eq:timeordered} one could always insert a complete set of on-shell states between the two operators, and now we are dealing with on-shell matrix elements of the form $\left\langle i\right|\partial_{\nu} J^{\nu}(x)\left|j\right\rangle$, which can be simplified using the equation of motion (EOM). Suppose the current $J^{\mu}(x)$ satisfies the following EOM:
\begin{equation}
\partial_{\nu} J^{\nu}(x)=s(x)
\end{equation}
where $s(x)$ is a scalar (or pseudo-scalar) current, then the identity $\left\langle i\right|\partial_{\nu} J^{\nu}(x)\left|j\right\rangle=\left\langle i\right|s(x)\left|j\right\rangle$ holds. Therefore one may first replace $\partial_{\nu} J^{\nu}\rightarrow s$ and then remove the complete set of states. This is equivalent to rewriting $D$ as:
\begin{equation}
D(\bar{q};p',p)=i\int d^{4}xe^{i\bar{q}\cdot x}\left\langle B(p')\right|T\left\{s(x)\delta\mathcal{L}(0)\right\}\left|A(p)\right\rangle.\label{eq:Dmod}
\end{equation}
Now since the derivative operator is removed, one could safely evaluate the expression above with ordinary Feynman rules. 

To end our generic discussion, let us demonstrate that the combination $D-\bar{q}\cdot B$ appears in the three-point function is pole-free. Again we take the first diagram in Fig.~\ref{fig:Tbarmu} as an example. We name the 1PI correlation function due to the insertion of $s(x)$ as $(Z_AZ_B)^{-1/2}\mathfrak{S}(p-\bar{q},p)$. Then the following identity obviously holds due to EOM:
\begin{equation}
\lim_{(p-\bar{q})^{2}\rightarrow M_B^{2}}i\mathfrak{S}(p-\bar{q},p)=\lim_{(p-\bar{q})^{2}\rightarrow M_B^{2}}\bar{q}_{\mu} F^{\mu}(p-\bar{q},p).
\end{equation}
Therefore, combining with the first term of $-\bar{q}\cdot B$, we obtain:
\begin{equation}
D-\bar{q}\cdot B=\frac{-i}{(p-\bar{q})^{2}-M_B^{2}}\left[i\Gamma_B(p',p-\bar{q})\mathfrak{S}(p-\bar{q},p)-\delta M_B^{2}\bar{q}_{\mu} F^{\mu}(p-\bar{q},p)\right]+...
\end{equation}
which is pole-free because the numerator vanishes as $(p-\bar{q})^{2}\rightarrow M_B^{2}$. Hence we have shown that the entire three-point function is indeed pole-free.

We shall now discuss the application of the generic method above in the calculation of the three-point function of $K_{l3}$ using ChPT, where there are a few details to be considered. First: we notice that the electromagnetic perturbation Lagrangian $\delta\mathcal{L}$ can be either local or non-local. The local terms include the $Z$-dependent term in Eq.~\eqref{eq:ChPTLO} and the $\mathcal{O}(e^{2}p^{2})$ counterterms in Eq.~\eqref{eq:Oe2p2}. The non-local terms, on the other hand, comes from one-loop diagrams with virtual photons. The latter then possesses two vertices, and the external momentum $\delta\bar{q}$ can flow into either one of them, provided that one divides each of such diagram by 2. Next: we wish to calculate $\delta F_{K\pi,3}^{\mu}$ to $\mathcal{O}(e^{2}p^{2})$, consistent with the existing ChPT result. Given that a meson propagator scales as $\mathcal{O}(p^{-2})$, the requirement above implies that one should calculate the 1PI correlation function with a $(J^{\mu\dagger}_{W})_V$ (or $(\partial_{\mu} J^{\mu\dagger}_{W})_V$) insertion to $\mathcal{O}(p^{4})$, that with an $i\delta\mathcal{L}$ insertion to $\mathcal{O}(e^{2}p^{2})$, and that with simultaneous insertions of $(J^{\mu\dagger}_{W})_V$ (or $(\partial_{\mu} J^{\mu\dagger}_{W})_V$) and $i\delta\mathcal{L}$ to $\mathcal{O}(e^{2}p^{2})$. The corresponding diagrams are depicted in Fig.~\ref{fig:J1PI},\ref{fig:dL1PI} and \ref{fig:JdL1PI}. Third: to calculate $D$ we need the EOM-simplified ChPT representation of $(\partial_{\mu} J^{\mu\dagger}_{W})_V$. It is a simple matter to show that its LO expression reads:
\begin{equation}
(\partial_{\mu} J^{\mu\dagger}_{W})_{V,\mathrm{LO}}=\frac{iF_0^{2}}{4}\left\langle Q_{W}\left([\chi,U^{\dagger}]+[\chi^{\dagger},U]\right)\right\rangle.\label{eq:JWEOM}
\end{equation} 
And finally, we provide in Appendix~\ref{sec:Example} a simple example to demonstrate all the steps in the calculation of the three-point function, and show that it indeed adds up with the two-point function to give the full $\delta F_{K\pi}^{\mu}$.

\begin{figure}[tb]
	\begin{centering}
		\includegraphics[scale=0.3]{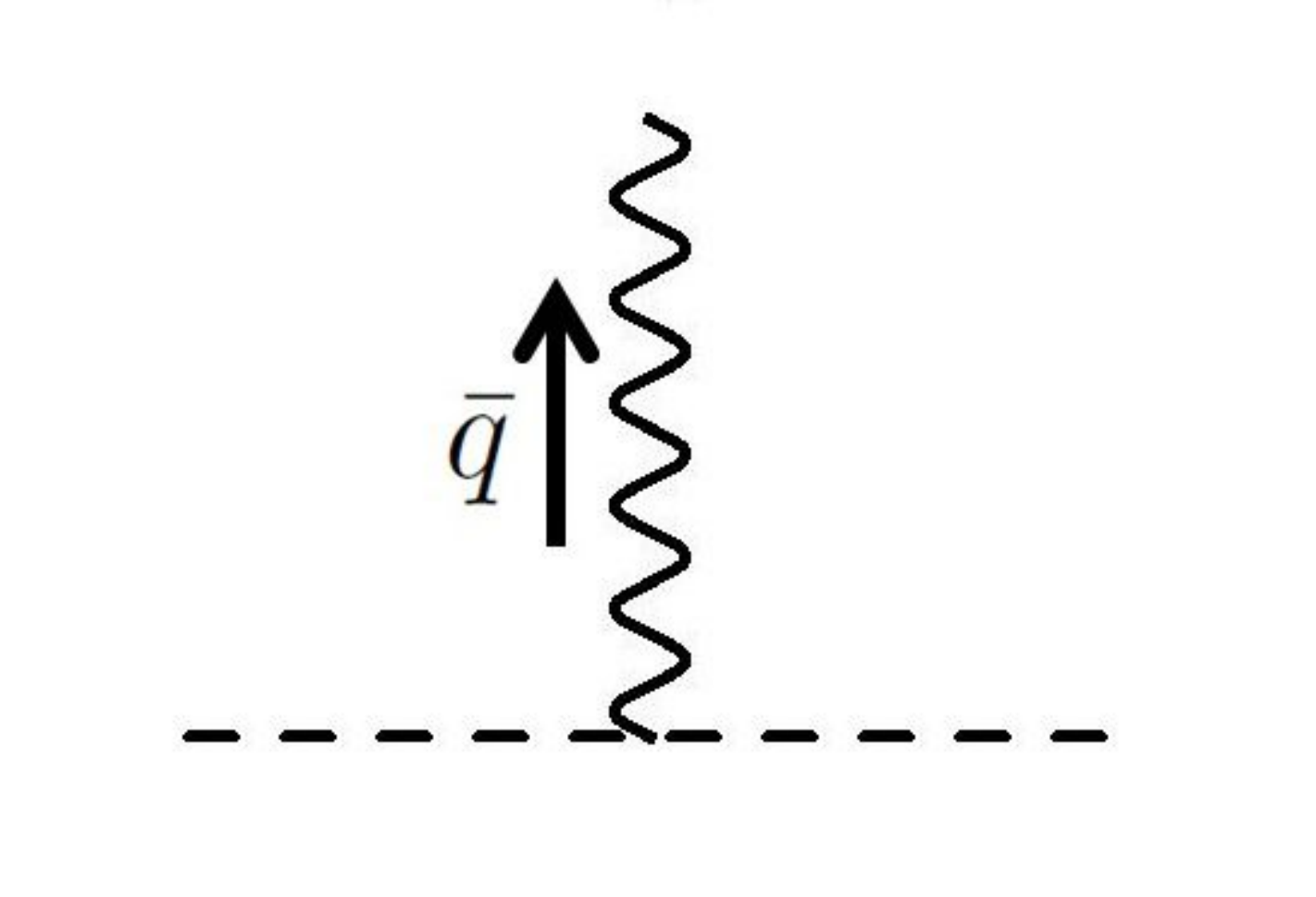}\hfill
		\includegraphics[scale=0.3]{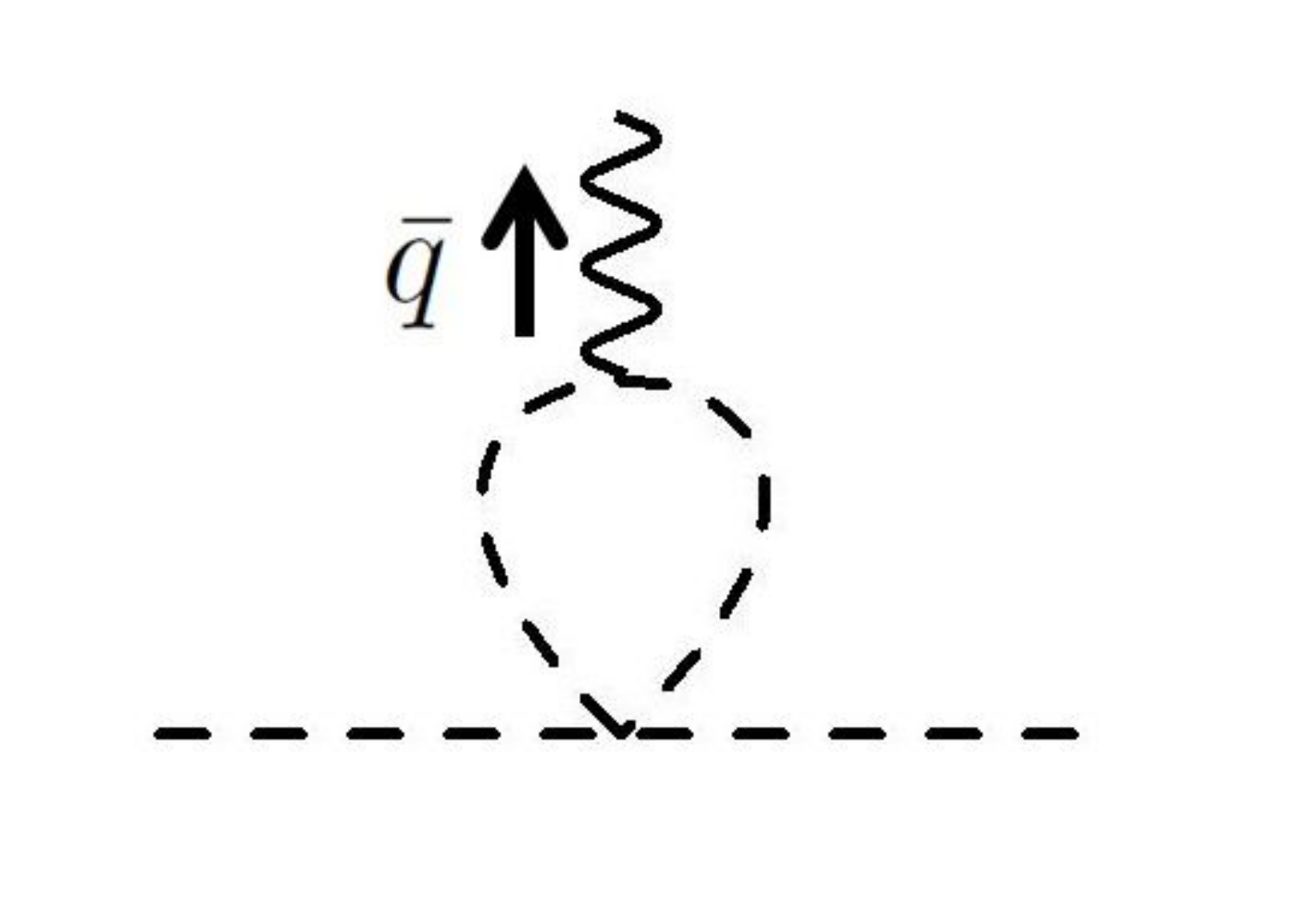}\hfill
		\includegraphics[scale=0.3]{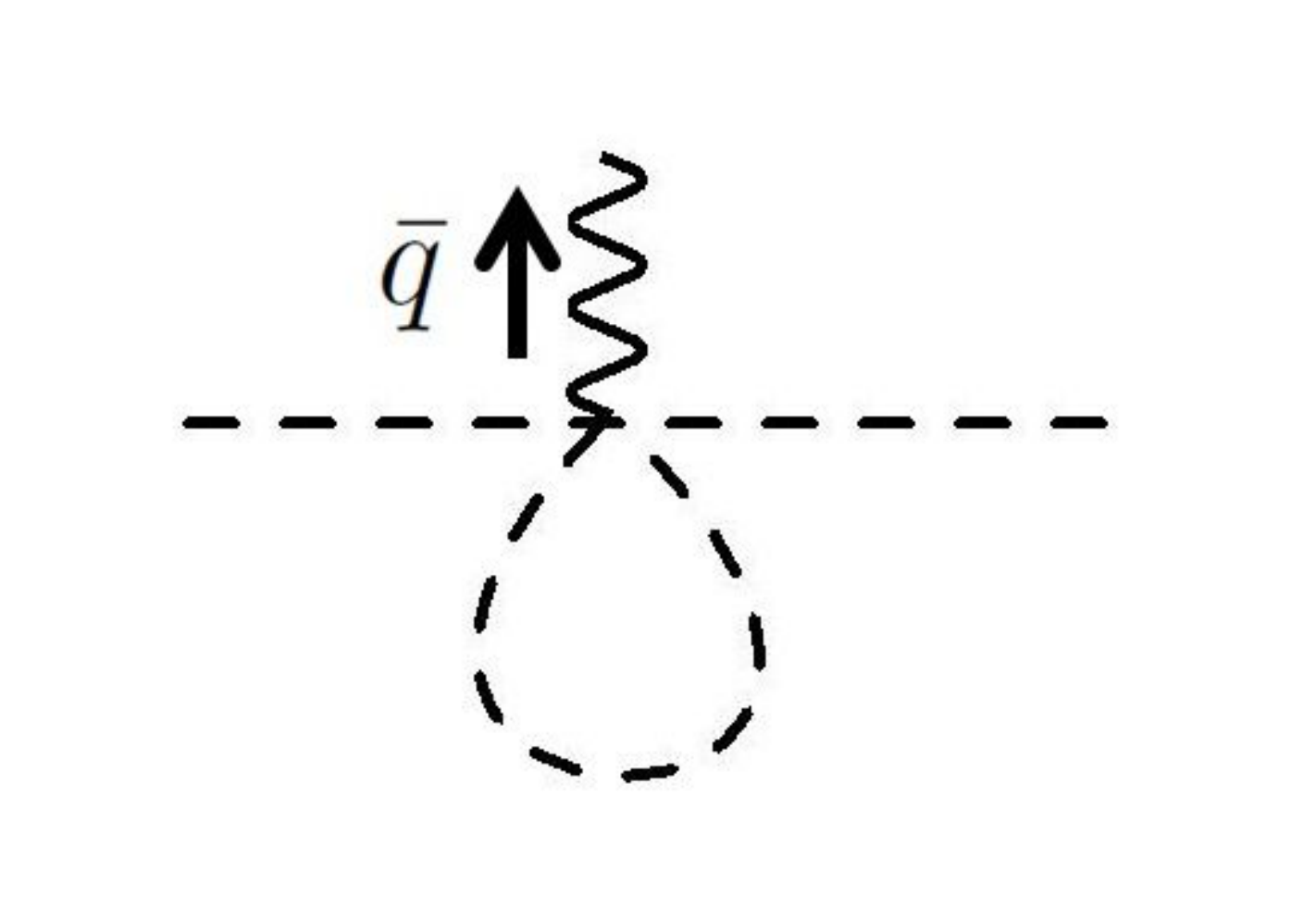}\hfill
		\includegraphics[scale=0.3]{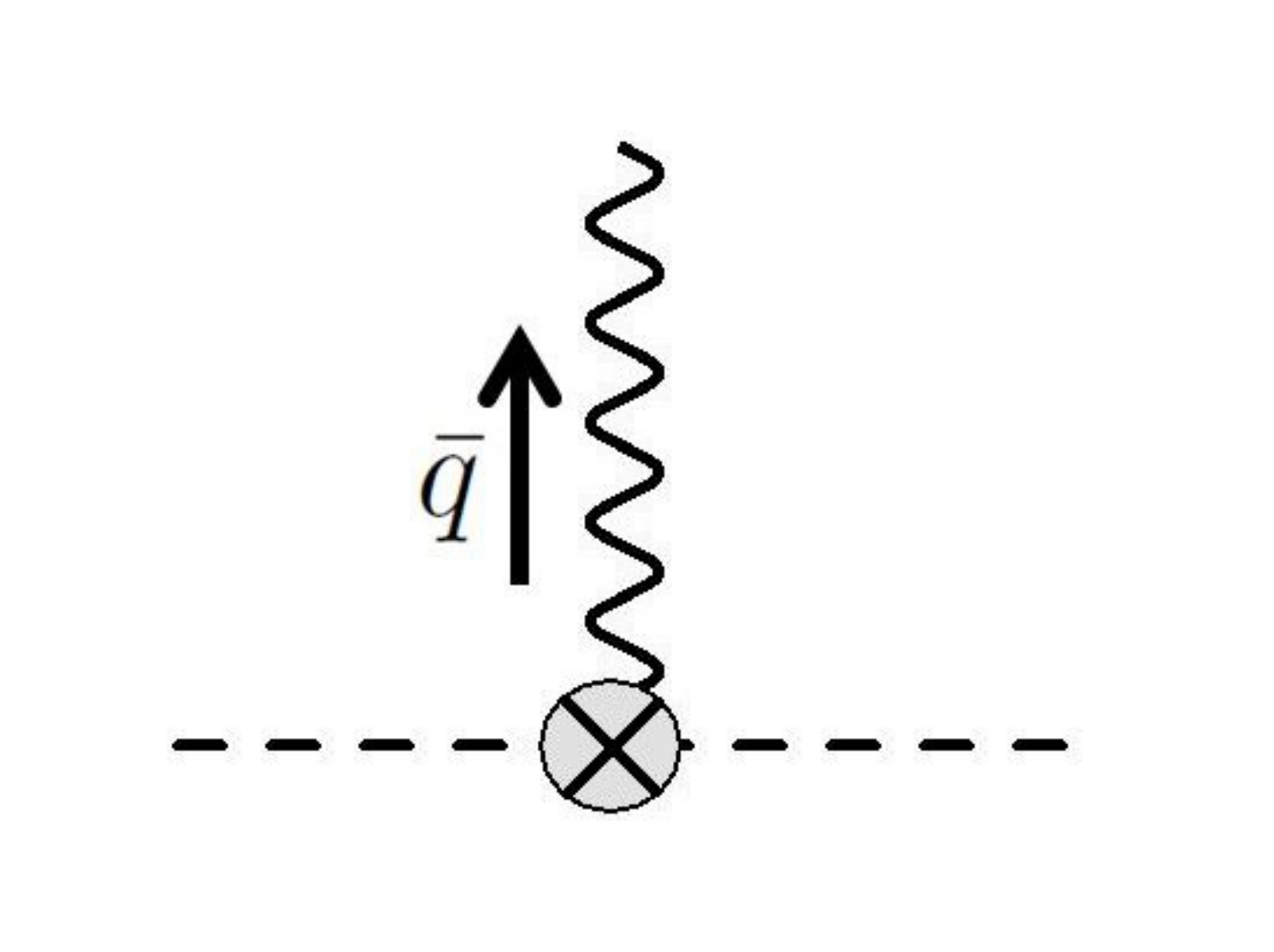}
		\par\end{centering}
	\caption{\label{fig:J1PI}Feynman diagrams up to $\mathcal{O}(p^{4})$ for the 1PI correlation function with the $(J^{\mu\dagger}_{W})_V$ (or $(\partial_{\mu} J_{W}^{\mu\dagger})_V$) insertion. The grey crossed circle represents $\mathcal{O}(p^{4})$ counterterms. }
\end{figure}

\begin{figure}[tb]
	\begin{centering}
		\includegraphics[scale=0.3]{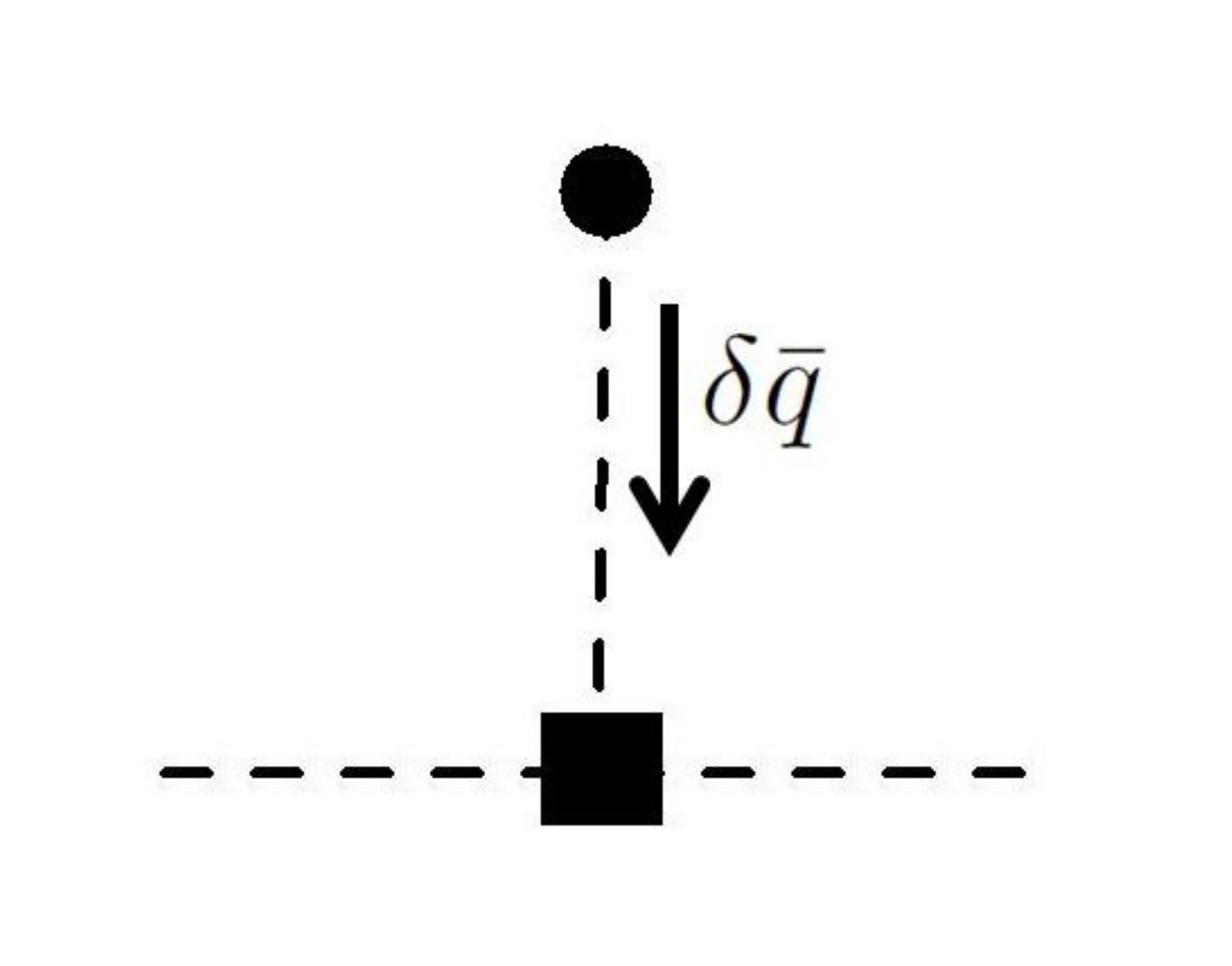}\hfill
		\includegraphics[scale=0.3]{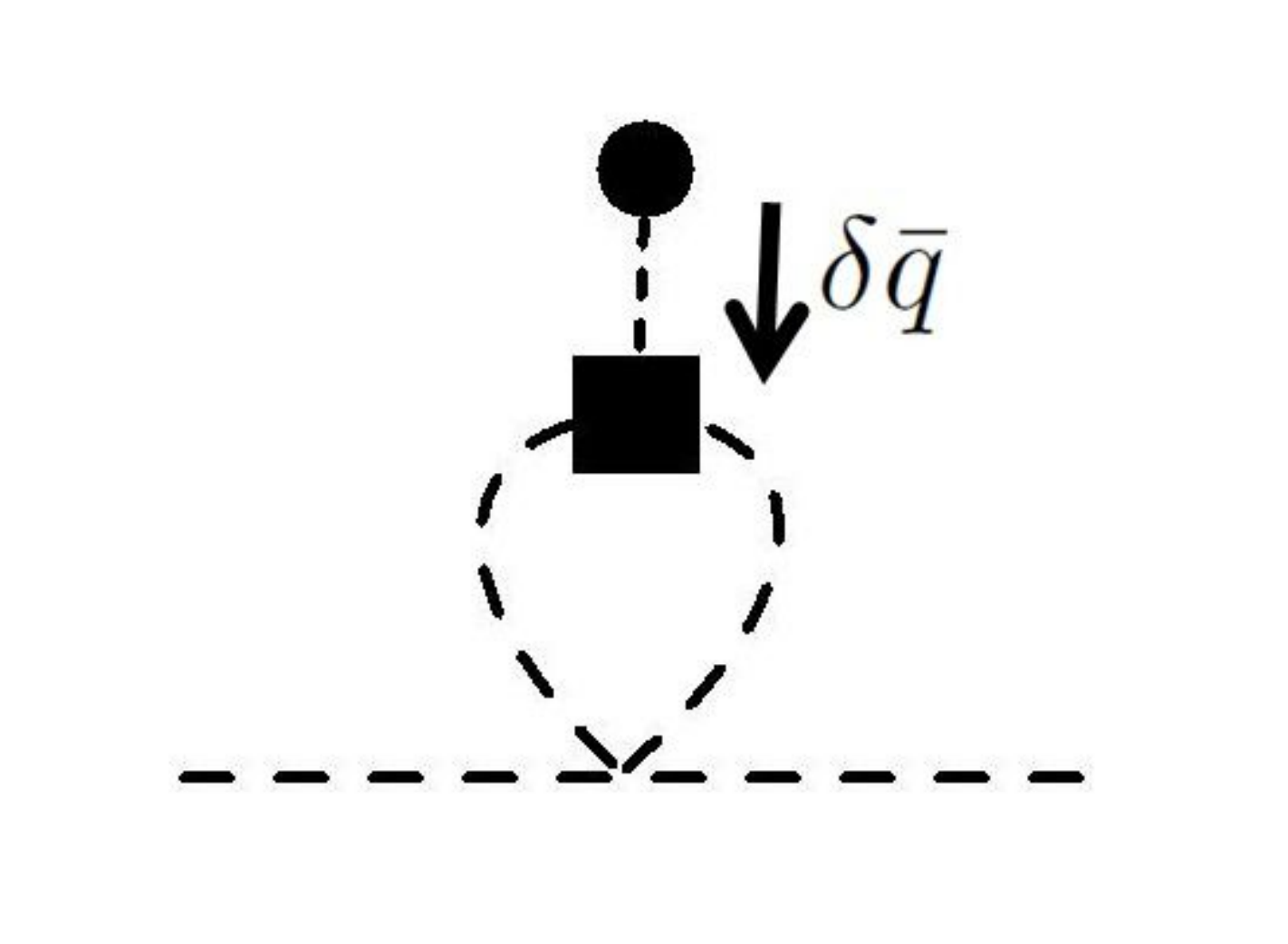}\hfill
		\includegraphics[scale=0.3]{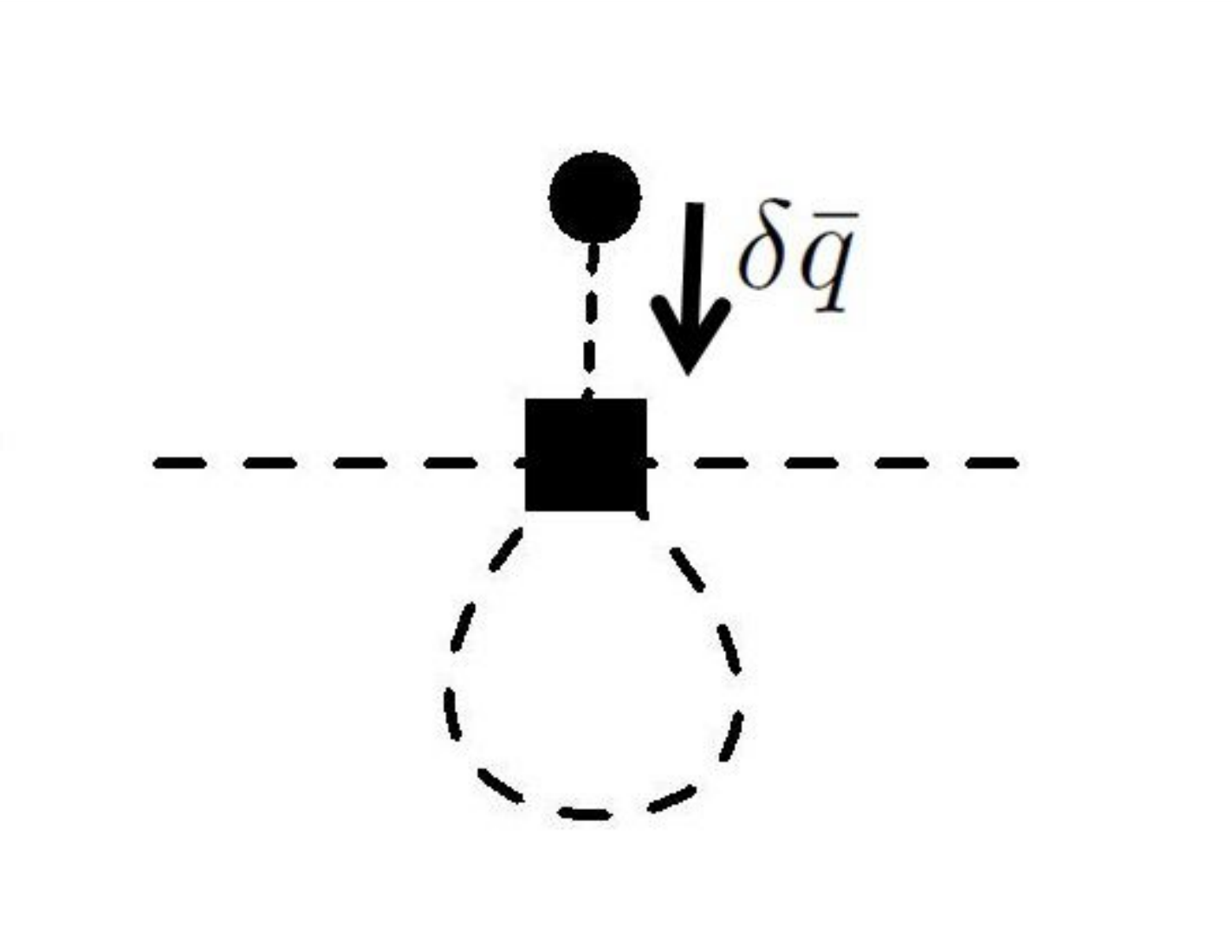}\hfill
		\includegraphics[scale=0.3]{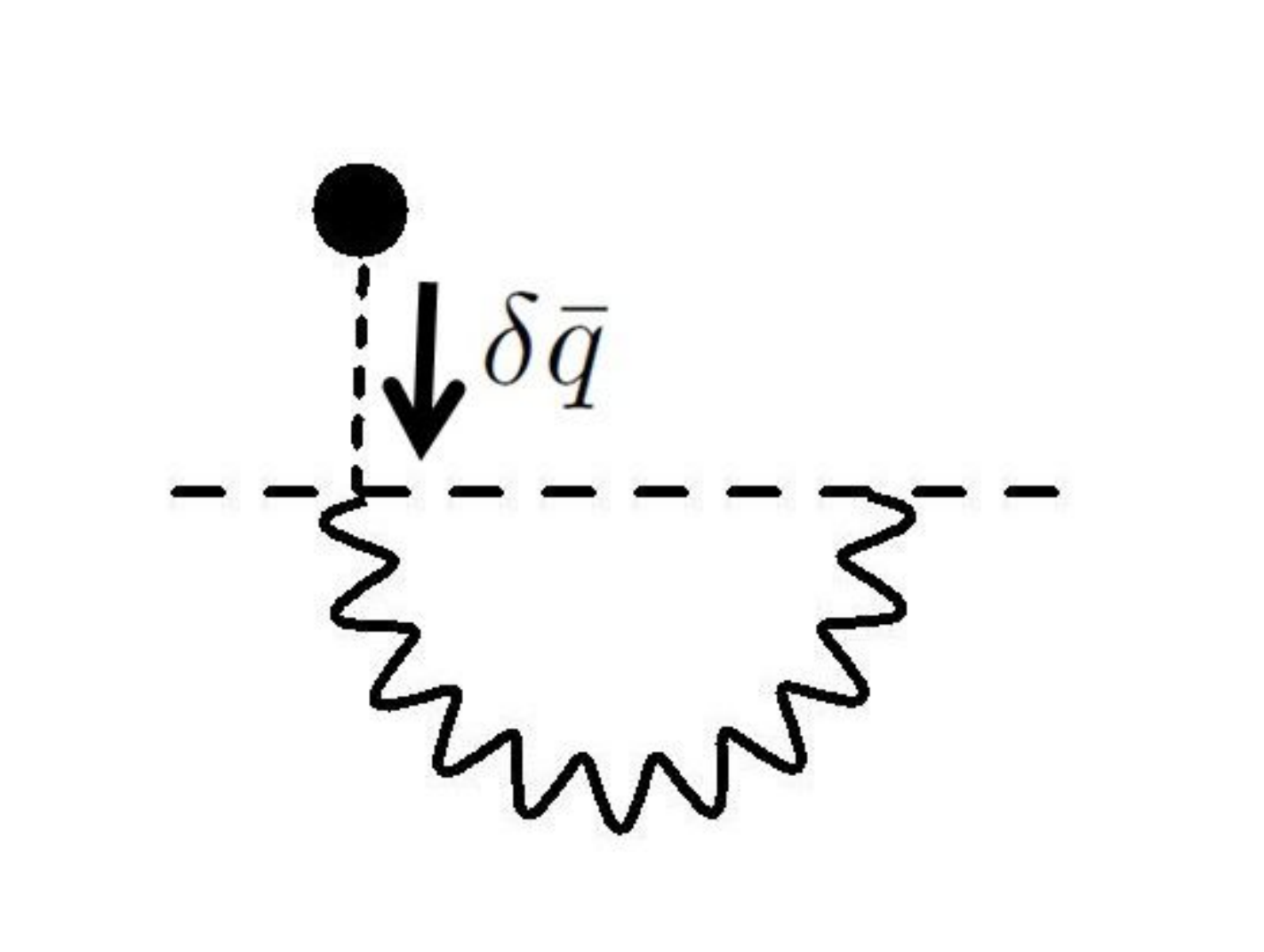}\hfill
		\includegraphics[scale=0.3]{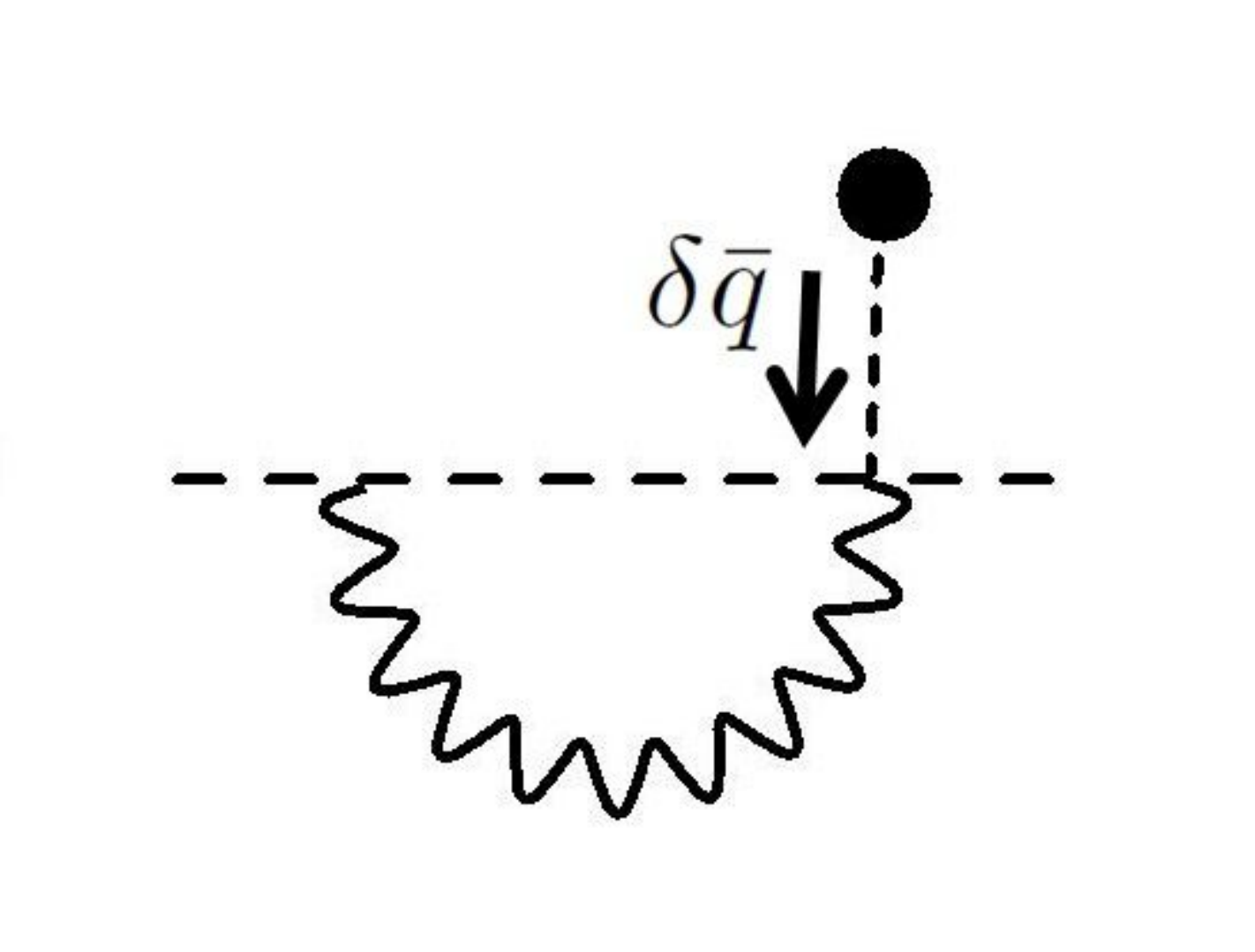}\hfill
		\includegraphics[scale=0.3]{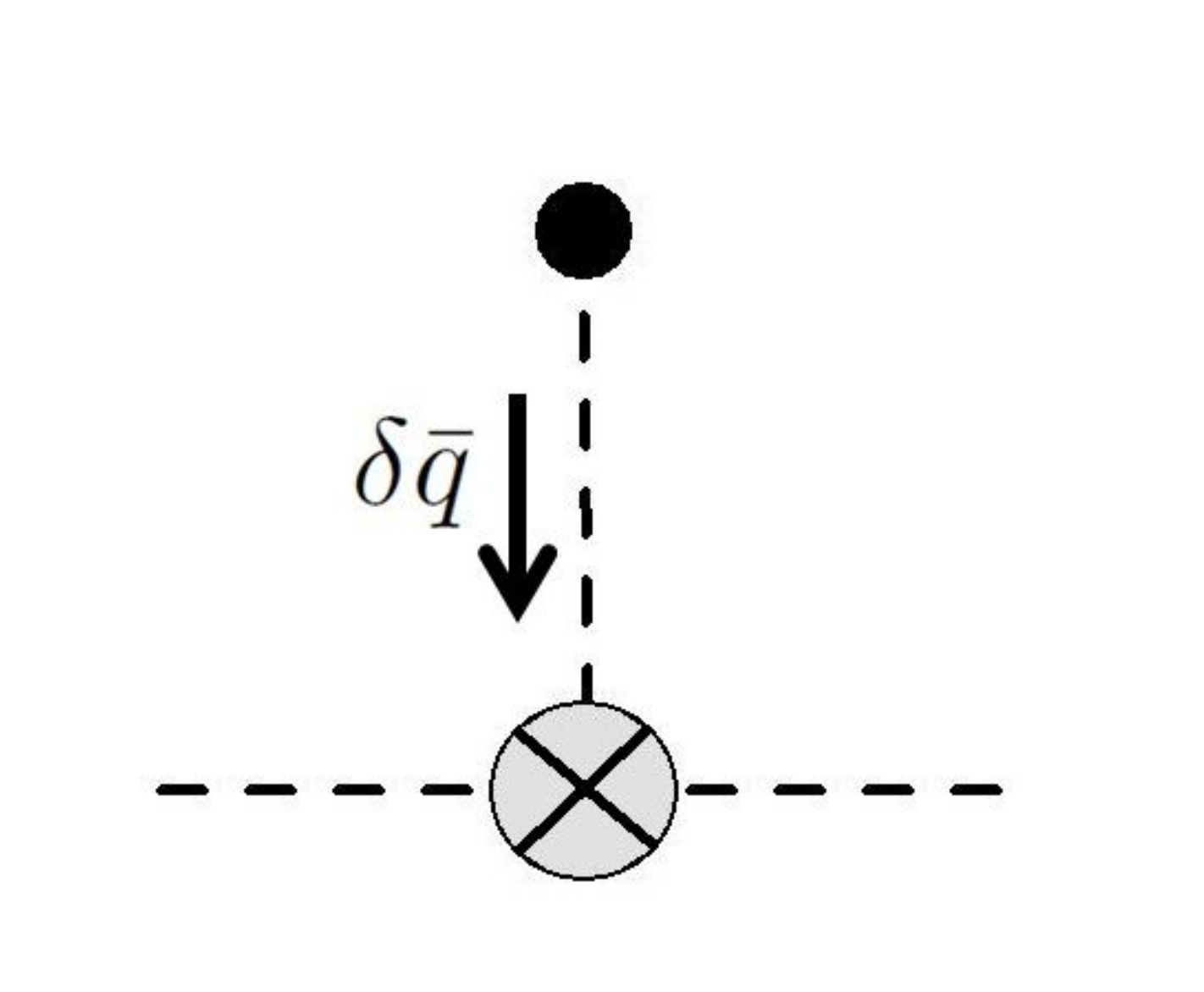}
		\par\end{centering}
	\caption{\label{fig:dL1PI}Feynman diagrams up to $\mathcal{O}(e^{2}p^{2})$ for the 1PI correlation function with the $i\delta\mathcal{L}$ insertion. The black box represent the vertex proportional to $Z$ whereas the grey crossed vertex represents $\mathcal{O}(e^{2}p^{2})$ counterterms. }
\end{figure}

\begin{figure}[tb]
	\begin{centering}
		\includegraphics[scale=0.3]{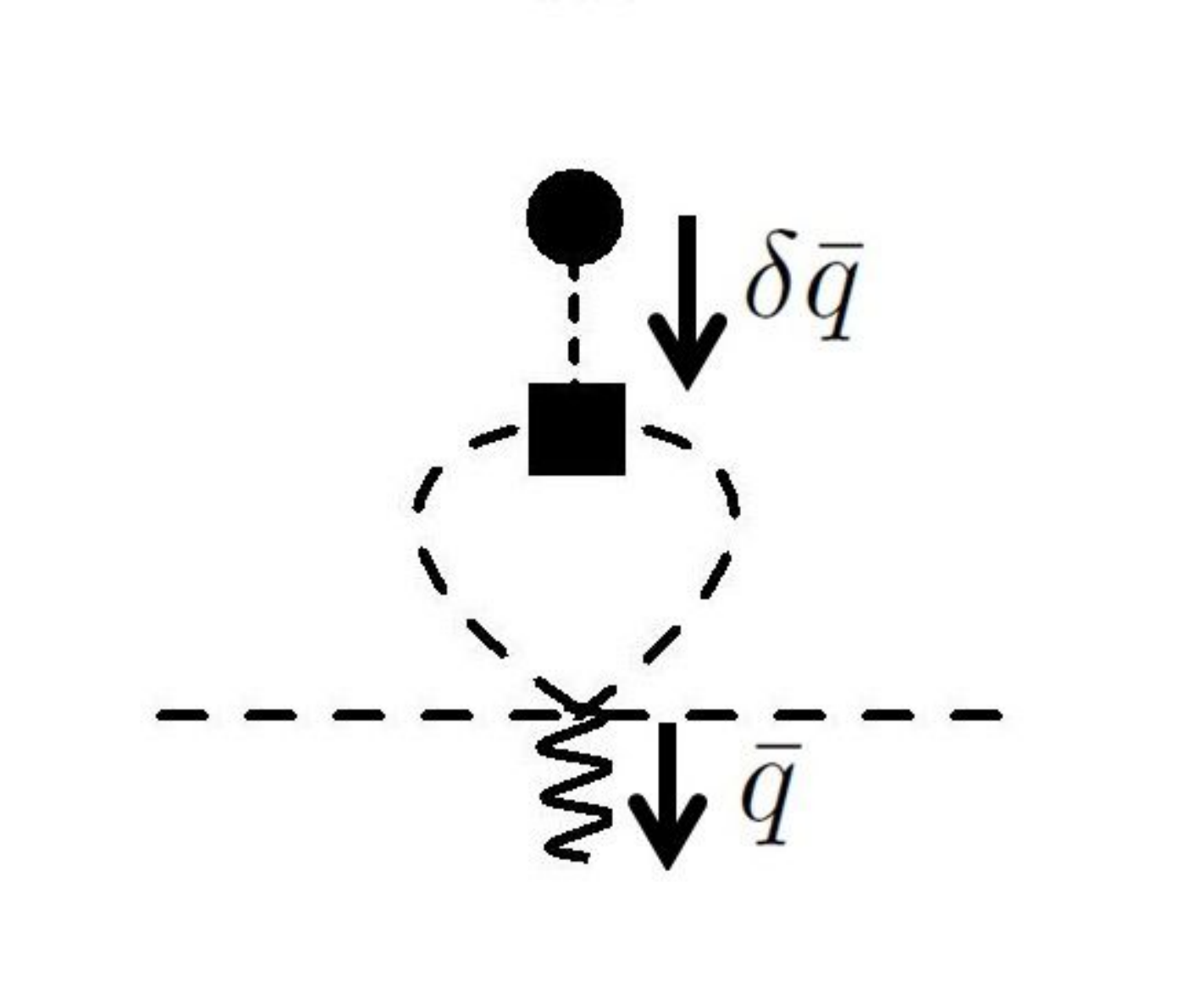}\hfill
		\includegraphics[scale=0.3]{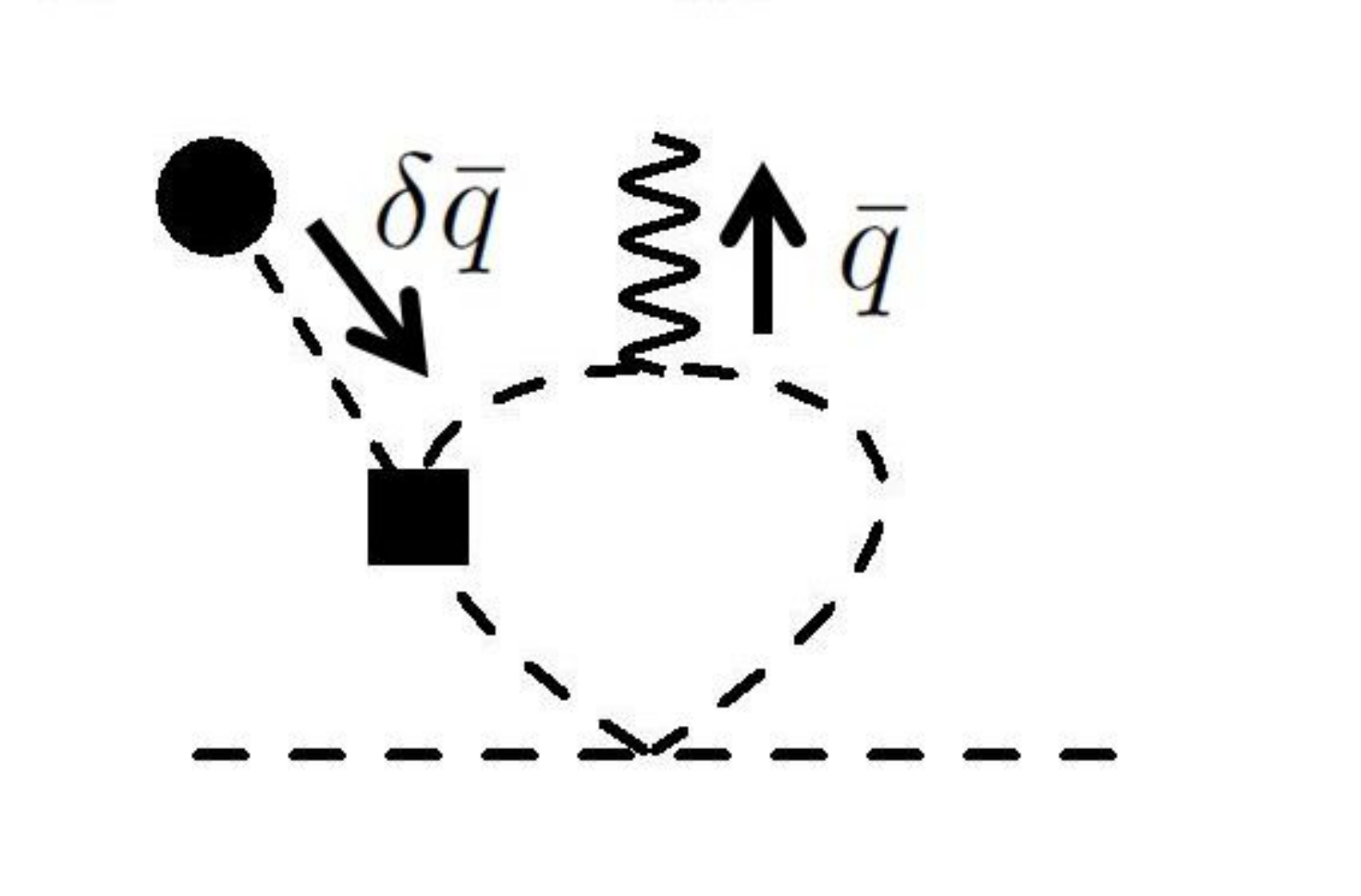}\hfill
		\includegraphics[scale=0.3]{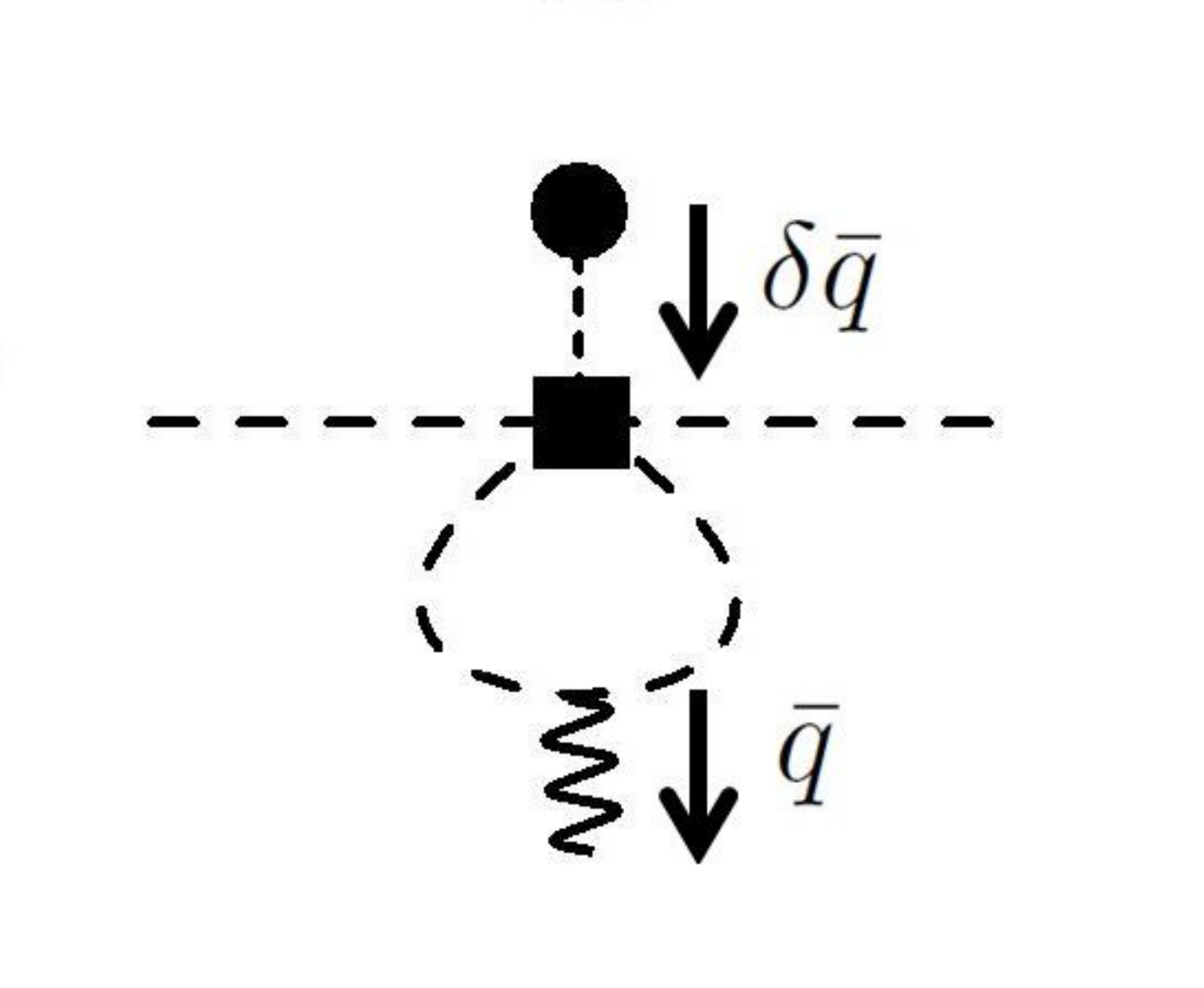}\hfill
		\includegraphics[scale=0.3]{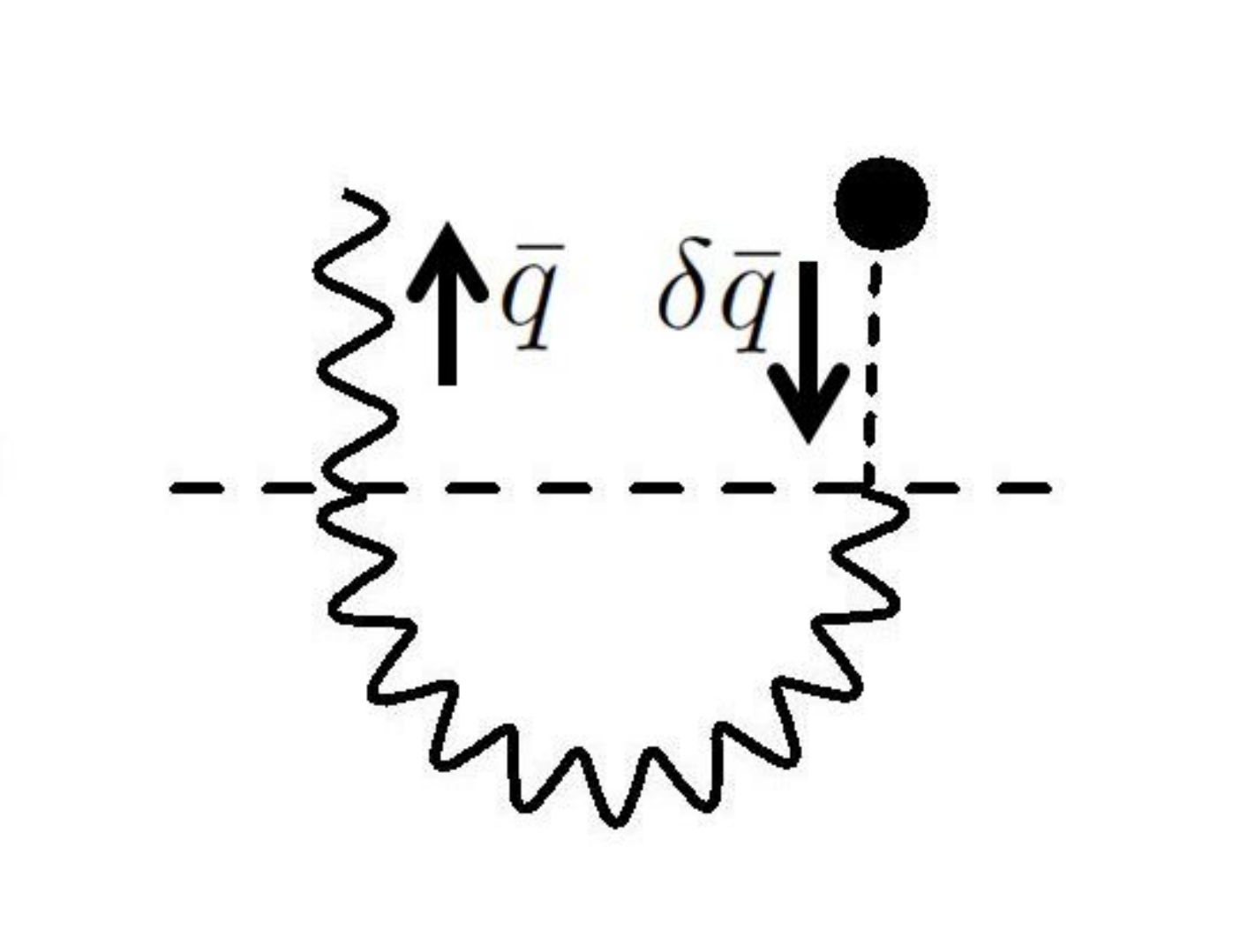}\hfill
		\includegraphics[scale=0.3]{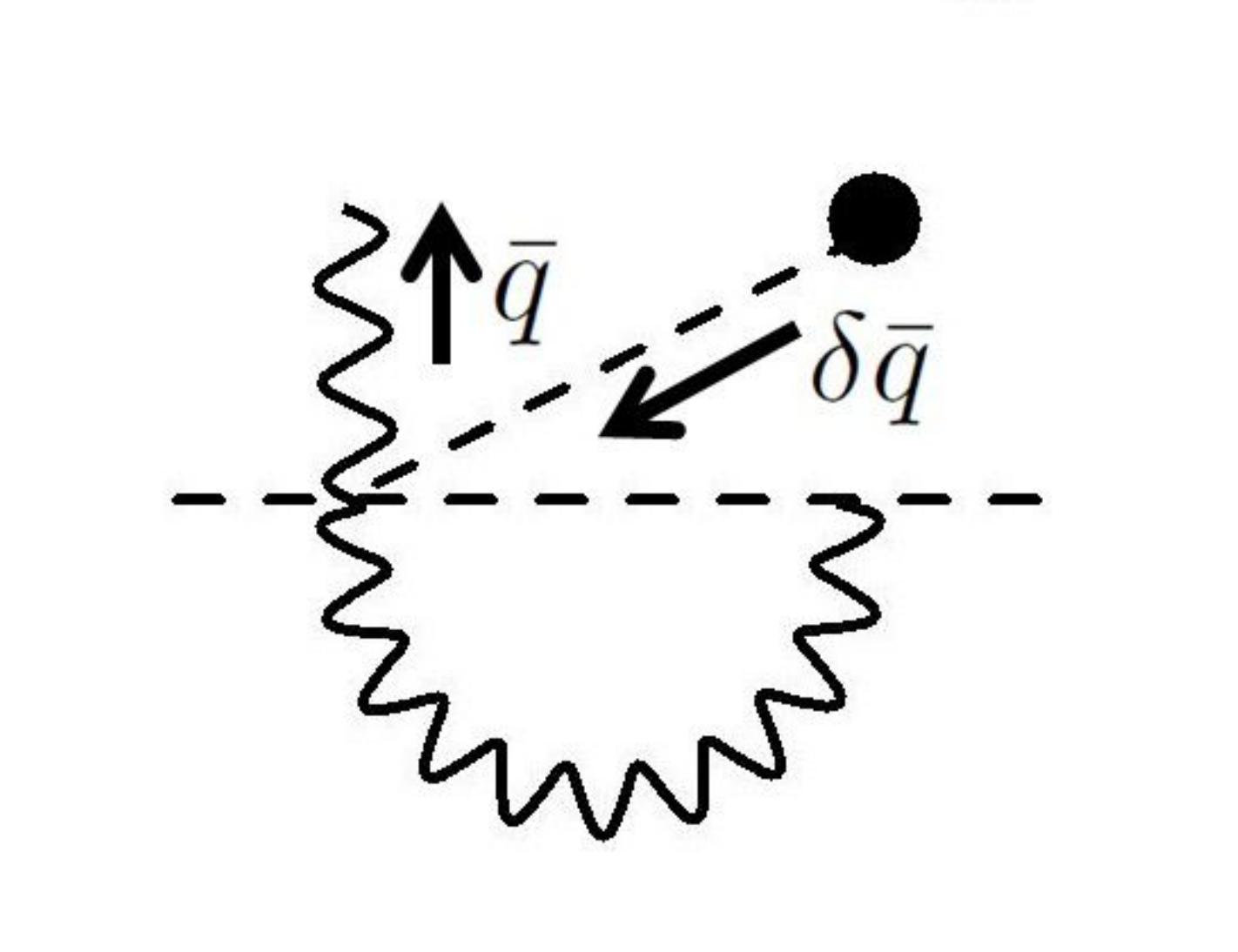}\hfill
		\includegraphics[scale=0.3]{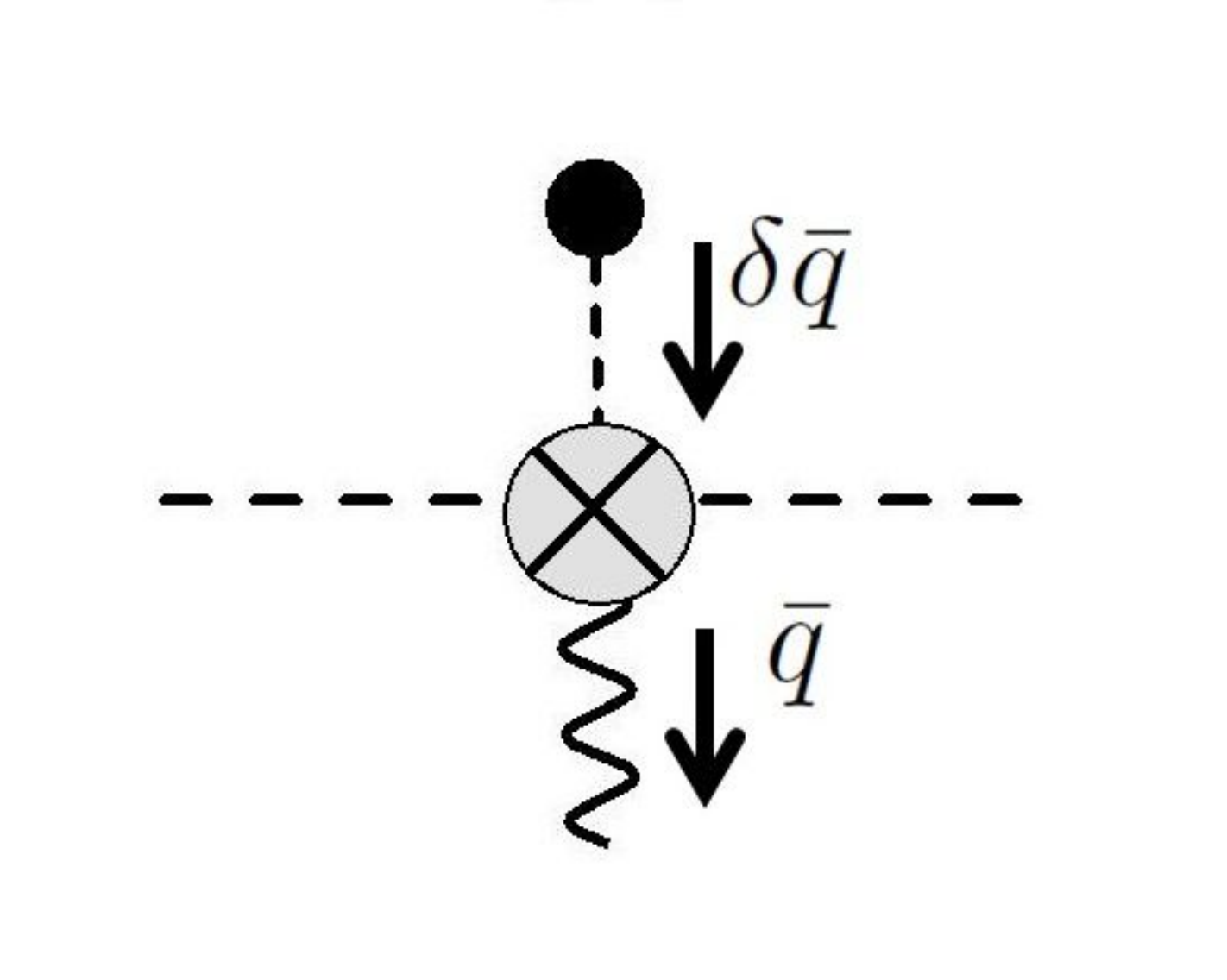}
		\par\end{centering}
	\caption{\label{fig:JdL1PI}Feynman diagrams up to $\mathcal{O}(e^{2}p^{2})$ for the 1PI correlation function with the simultaneous insertion of $(J_{W}^{\mu\dagger})_V$ (or $(\partial_{\mu} J^{\mu\dagger}_{W})_V$) and $i\delta\mathcal{L}$. The black box represent the vertex proportional to $Z$ whereas the grey crossed vertex represents $\mathcal{O}(e^{2}p^{2})$ counterterms.}
\end{figure}

\section{\label{sec:matching}Two-point and three-point functions of $K\pi$ form factors at $\mathcal{O}(e^{2}p^{2})$}

We are now in the position to split the $\mathcal{O}(e^{2}p^{2})$ ChPT corrections to $F_{K\pi}^{\mu}$ in Section~\ref{sec:ChPT} into two-point and three-point functions. We shall do it in terms of the invariant functions $\delta f_\pm^{K\pi}$, i.e.:
\begin{equation}
\delta f_\pm^{K\pi}(q^{2})=\delta f_{\pm,2}^{K\pi}(q^{2})+\delta f_{\pm,3}^{K\pi}(q^{2}).
\end{equation}

Our results are as follows: For $K_{l3}^{+}$,
\begin{eqnarray}
\delta f^{K^{+}\pi^{0}}_{+,2}(q^{2})&=&-\frac{\alpha}{16\sqrt{2}\pi}\left[-\ln\frac{M_{K}^{2}}{\mu^{2}}+\left(\frac{3M_{K}^{2}+M_{\pi}^{2}-q^{2}}{M_{K}^{2}}\right)\ln\frac{M_{K}^{2}}{M_{\gamma}^{2}}-\frac{11M_{K}^{2}+5M_{\pi}^{2}-5q^{2}}{2M_{K}^{2}}\right]\nonumber\\
&&-4\pi\sqrt{2}\alpha K_{12}^{r}\nonumber\\
\delta f^{K^{+}\pi^{0}}_{-,2}(q^{2})&=&\frac{\alpha}{16\sqrt{2}\pi}\left[-\left(\frac{3M_{K}^{2}+M_{\pi}^{2}-q^{2}}{M_{K}^{2}}\right)\ln\frac{M_{K}^{2}}{M_{\gamma}^{2}}+\frac{15M_{K}^{2}+5M_{\pi}^{2}-5q^{2}}{2M_{K}^{2}}\right],
\end{eqnarray}
\begin{eqnarray}
\delta f_{+,3}^{K^{+}\pi^{0}}(q^{2})&=&-\frac{\alpha}{16\sqrt{2}\pi}\left[\frac{M_{K}^{2}-M_{\pi}^{2}+q^{2}}{M_{K}^{2}}\right]\left[\ln\frac{M_{K}^{2}}{M_{\gamma}^{2}}-\frac{5}{2}\right]+\frac{Z\alpha}{2\sqrt{2}\pi}\frac{M_{K}^{2}}{M_{\eta}^{2}-M_{\pi}^{2}}\left[1+\ln\frac{M_{K}^{2}}{\mu^{2}}\right]\nonumber\\
&&-\frac{8\pi Z\alpha}{\sqrt{2}}\left[\frac{1}{2}\bar{h}_{K^{+}\pi^{0}}(q^{2})+\bar{h}_{K^{0}\pi^{-}}(q^{2})+\frac{3}{2}\bar{h}_{K^{+}\eta}(q^{2})\right]\nonumber\\
&&-\frac{4\pi\alpha}{\sqrt{2}}\left[-2K_{3}^{r}+K_{4}^{r}+\frac{2}{3}K_{5}^{r}+\frac{2}{3}K_{6}^{r}\right]-\frac{8\pi\alpha}{\sqrt{2}}\frac{M_{\pi}^{2}}{M_{\eta}^{2}-M_{\pi}^{2}}\left[-2K_{3}^{r}+K_{4}^{r}+\frac{2}{3}K_{5}^{r}\right.\nonumber\\
&&\left.+\frac{2}{3}K_{6}^{r}-\frac{2}{3}K_{9}^{r}-\frac{2}{3}K_{10}^{r}\right]\nonumber\\
\delta f_{-,3}^{K^{+}\pi^{0}}(q^{2})&=&\frac{\alpha}{16\sqrt{2}\pi}\left[-3\ln\frac{M_{K}^{2}}{\mu^{2}}+\left(\frac{3M_{K}^{2}+M_{\pi}^{2}-q^{2}}{M_{K}^{2}}\right)\ln\frac{M_{K}^{2}}{M_{\gamma}^{2}}-\frac{7M_{K}^{2}+5M_{\pi}^{2}-5q^{2}}{2M_{K}^{2}}\right]\nonumber\\
&&-\frac{Z\alpha}{4\sqrt{2}\pi}\left[1+\ln\frac{M_{K}^{2}}{\mu^{2}}\right]-\frac{3Z\alpha}{8\sqrt{2}\pi}\left[1+\ln\frac{M_{\pi}^{2}}{\mu^{2}}\right]\nonumber\\
&&+\frac{8\pi Z\alpha}{\sqrt{2}}\left[\frac{M_{K}^{2}-M_{\pi}^{2}+3q^{2}}{2q^{2}}K_{K\pi}(q^{2})+\frac{M_{K}^{2}-M_{\pi}^{2}-3q^{2}}{2q^{2}}K_{K\eta}(q^{2})\right.\nonumber\\
&&\left.\left.+\frac{3F_{0}^{2}}{2q^{2}}\left(H_{K\pi}(q^{2})+H_{K\eta}(q^{2})\right)\right]-\frac{8\pi Z\alpha F_{0}^{2}}{\sqrt{2}}\sum_{PQ}\left[\left(a_{PQ}+\frac{M_{P}^{2}-M_{Q}^{2}}{2q^{2}}b_{PQ}\right)\right.\right.\nonumber\\
&&\left.\bar{K}_{PQ}(q^{2})+b_{PQ}\frac{1}{q^{2}}\bar{h}_{PQ}(q^{2})\right]+\frac{4\pi\alpha}{\sqrt{2}}\left[-4K_{3}^{r}+2K_{4}^{r}+\frac{2}{3}K_{5}^{r}+\frac{2}{3}K_{6}^{r}\right],
\end{eqnarray}
and for $K_{l3}^{0}$,
\begin{eqnarray}
\delta f_{+,2}^{K^{0}\pi^{-}}(q^{2})&=&\frac{\alpha}{16\pi}\left[\ln\frac{M_{\pi}^{2}}{\mu^{2}}+\left(\frac{q^{2}-M_{K}^{2}-3M_{\pi}^{2}}{M_{\pi}^{2}}\right)\ln\frac{M_{\pi}^{2}}{M_{\gamma}^{2}}+\frac{5M_{K}^{2}+11M_{\pi}^{2}-5q^{2}}{2M_{\pi}^{2}}\right]-8\pi\alpha K_{12}^{r}\nonumber\\
\delta f_{-,2}^{K^{0}\pi^{-}}(q^{2})&=&-\frac{\alpha}{16\pi}\left[\left(\frac{q^{2}-M_{K}^{2}-3M_{\pi}^{2}}{M_{\pi}^{2}}\right)\ln\frac{M_{\pi}^{2}}{M_{\gamma}^{2}}+\frac{5M_{K}^{2}+15M_{\pi}^{2}-5q^{2}}{2M_{\pi}^{2}}\right],
\end{eqnarray}
\begin{eqnarray}
\delta f_{+,3}^{K^{0}\pi^{-}}(q^{2})&=&\frac{\alpha}{16\pi}\left[\frac{M_{K}^{2}-M_{\pi}^{2}-q^{2}}{M_{\pi}^{2}}\right]\left[\ln\frac{M_{\pi}^{2}}{M_{\gamma}^{2}}-\frac{5}{2}\right]-8\pi Z\alpha\left[\frac{1}{2}\bar{h}_{K^{+}\pi^{0}}(q^{2})+\bar{h}_{K^{0}\pi^{-}}(q^{2})\right.\nonumber\\
&&\left.+\frac{3}{2}\bar{h}_{K^{+}\eta}(q^{2})\right]\nonumber\\
\delta f_{-,3}^{K^{0}\pi^{-}}(q^{2})&=&\frac{\alpha}{16\pi}\left[3\ln\frac{M_{\pi}^{2}}{\mu^{2}}+\left(\frac{q^{2}-M_{K}^{2}-3M_{\pi}^{2}}{M_{\pi}^{2}}\right)\ln\frac{M_{\pi}^{2}}{M_{\gamma}^{2}}+\frac{5M_{K}^{2}+7M_{\pi}^{2}-5q^{2}}{2M_{\pi}^{2}}\right]\nonumber\\
&&-\frac{Z\alpha}{8\pi}\left[1+\ln\frac{M_{\pi}^{2}}{\mu^{2}}\right]-8\pi Z\alpha\left[\frac{M_{K}^{2}-M_{\pi}^{2}-q^{2}}{q^{2}}K_{K\pi}(q^{2})-\frac{M_{K}^{2}-M_{\pi}^{2}}{q^{2}}K_{K\eta}(q^{2})\right.\nonumber\\
&&\left.+\frac{3F_0^{2}}{2q^{2}}\left(H_{K\pi}(q^{2})+H_{K\eta}(q^{2})\right)\right]-8\pi Z\alpha F_0^{2}\sum_{PQ}\left[\left(c_{PQ}+d_{PQ}\frac{M_P^{2}-M_Q^{2}}{2q^{2}}\right)\right.\nonumber\\
&&\left.\bar{K}_{PQ}(q^{2})+\frac{1}{q^{2}}d_{PQ}\bar{h}_{PQ}(q^{2})\right]+\frac{8\pi\alpha}{3}\left(K_{5}^{r}+K_{6}^{r}\right).
\end{eqnarray}
Each term above is UV-finite and scale-independent. Another interesting feature is that the $Z$-dependent terms are totally contained in the three-point function.

We shall discuss the physical significance of the expressions above. First, we obtain an EFT description of the two-point function $\delta F_{K\pi,2}^{\mu}$ which depends on a poorly-constrained LEC $K_{12}^{r}$, however, we do not really need such an expression because we shall instead study it in its integral form, Eq.~\eqref{eq:twopt}, which involves the generalized Compton tensor $T^{\mu\nu}$. All the hadronic subtleties of this term as well as the $\gamma W$-box diagram which, in the EFT description, depends on a few more poorly-constrained LECs $\{X_i\}$, are fully contained in the tensor $T^{\mu\nu}$. At the same time, we also obtain an EFT description of the three-point function $\delta F_{K\pi,3}^{\mu}$ which depends on the LECs $\{K_{i}^{r}\}_{i\neq 12}$ and $L_{9}^{r}$, the former are poorly-constrained. However, they receive no short-distance enhancement so there is less ambiguity in their matching to UV-physics. 

To end this section, we point out a particularly interesting feature in the $K_{l3}^{0}$ channel, namely: the $\mathcal{O}(e^2p^2)$ counterterm contributions to the three-point function appear only in $\delta f_{-,3}^{K^{0}\pi^{-}}$, which effect on the $K_{l3}^{0}$ decay rate is suppressed by the factor $r_{l}$ as we discussed in Section~\ref{sec:tree}. Therefore, this channel possesses a significant practical advantage under our new formalism over the traditional ChPT treatment, that at the order $\mathcal{O}(e^{2}p^{2})$, instead of being scattered among various LECs $\{K_i^{r},X_i^{r}\}$, almost all the radiative correction uncertainties that go into the $K_{l3}^{0}$ decay rate come from the tensor $T^{\mu\nu}$. The latter can then be studied with the method outline at the end of Sec.~\ref{sec:logs}, which may lead to a better controlled theoretical uncertainty, and consequently a more precise extraction of $|V_{us}|$.

\section{\label{sec:conclusion}Conclusions}
 
Existing standard treatments of $K_{l3}$ electroweak radiative corrections are based on ChPT calculations: the short-distance effects are encoded in the universal factor $S_{\mathrm{EW}}$, and the long-distance electromagnetic effects are computed to $\mathcal{O}(e^2p^2)$. Within such a formalism, the theory uncertainty to the radiative corrections range from 0.22\% to 0.25\% depending on the actual decay channel. They come from (1) the neglected $\mathcal{O}(e^2p^4)$ terms which contributes to a universal uncertainty of 0.19\%, and (2) the associated uncertainties to the $\mathcal{O}(e^2p^2)$ LECs which range from 0.11\% to 0.16\%. The extracted value of $V_{us}$ from these results, combining with a recent update on $V_{ud}$, returns a 5$\sigma$ deviation of the first-row CKM unitarity. This triggered us to consider alternative approaches which may further reduce the uncertainties in electroweak radiative corrections in the $K_{l3}$ decays.

The new method we proposed is based on a hybridization of the current algebra formalism by Sirlin in the late 70ties and the modern ChPT approach. Under this framework, all the short-distance-enhanced terms in the $K_{l3}$ radiative corrections are described collectively by a single tensor $T^{\mu\nu}$, which could be studied using dispersion relations. Such a procedure effectively resums an important subset of $\mathcal{O}(e^2p^{2n})$ terms to all orders, thus we expect the existing 0.19\% uncertainty from the neglected $\mathcal{O}(e^2p^4)$ terms in the ChPT calculation to be significantly reduced. On the other hand, all the remaining, non-enhanced terms adopt a ChPT description in our formalism. We point out in particular that these terms receive only a minimal impact from the poorly-constrained LECs in the case of $K_{l3}^{0}$, therefore the effect of their associated uncertainties will also be reduced. As a conclusion, although further developments are still needed, we expect the new method to reduce the existing uncertainty as much as by a factor 2 depending on the actual decay channel. This may shed lights on various unanswered questions in kaon decay, including the $2\sigma$ discrepancy in the $V_{us}$ extraction between $K_{l2}$ and $K_{l3}$. 

\section*{Acknowledgements} 
The authors appreciate Vincenzo Cirigliano, Davide Giusti, Mikhail Gorchtein and Akaki Rusetsky for useful discussions. This work is supported in part by  the DFG (Grant No. TRR110)
and the NSFC (Grant No. 11621131001) through the funds provided
to the Sino-German CRC 110 ``Symmetries and the Emergence of
Structure in QCD", and also by the Alexander von Humboldt Foundation through the Humboldt Research Fellowship (CYS). The work of UGM was also supported by the Chinese Academy of Sciences (CAS) through a President's
International Fellowship Initiative (PIFI) (Grant No. 2018DM0034) and by the VolkswagenStiftung (Grant No. 93562).

\begin{appendix}
	
\section{\label{sec:loopfunction}Extra loop functions}	

In this appendix we define the extra loop functions $\{\bar{h}_{PQ},\bar{K}_{PQ}\}$ needed to isolate
the $\mathcal{O}(e^{2}p^{2})$ corrections to the $K\pi$ form factors from the chiral loops:
\begin{eqnarray}
\bar{h}_{K^{+}\pi^{0}}(q^{2})&=&\frac{1}{64\pi^{2}(M_{K}^{2}-M_{\pi}^{2})}\left[M_{\pi}^{2}-M_{K}^{2}+M_{\pi}^{2}\ln\frac{M_{K}^{2}}{M_{\pi}^{2}}\right]+\frac{M_{K}^{2}-M_{\pi}^{2}-q^{2}}{4q^{2}}\bar{J}_{K\pi}(q^{2})\nonumber\\
\bar{h}_{K^{0}\pi^{-}}(q^{2})&=&\frac{1}{64\pi^{2}(M_{K}^{2}-M_{\pi}^{2})}\left[M_{\pi}^{2}-M_{K}^{2}+M_{K}^{2}\ln\frac{M_{K}^{2}}{M_{\pi}^{2}}\right]+\frac{M_{\pi}^{2}-M_{K}^{2}-q^{2}}{4q^{2}}\bar{J}_{K\pi}(q^{2})\nonumber\\
\bar{h}_{K^{+}\eta}(q^{2})&=&\frac{1}{64\pi^{2}(M_{K}^{2}-M_{\eta}^{2})}\left[M_{\eta}^{2}-M_{K}^{2}+M_{\eta}^{2}\ln\frac{M_{K}^{2}}{M_{\eta}^{2}}\right]+\frac{M_{K}^{2}-M_{\eta}^{2}-q^{2}}{4q^{2}}\bar{J}_{K\eta}(q^{2})\nonumber\\
\bar{K}_{K^{+}\pi^{0}}(q^{2})&=&\frac{1}{2q^{2}}\left[\bar{J}_{K\pi}(q^{2})+(M_{K}^{2}-M_{\pi}^{2})\bar{J}_{K\pi}^{(1)}(q^{2})\right]\nonumber\\
\bar{K}_{K^{0}\pi^{-}}(q^{2})&=&\frac{1}{2q^{2}}\left[-\bar{J}_{K\pi}(q^{2})+(M_{K}^{2}-M_{\pi}^{2})\bar{J}_{K\pi}^{(2)}(q^{2})\right]\nonumber\\
\bar{K}_{K^{+}\eta}(q^{2})&=&\frac{1}{2q^{2}}\left[\bar{J}_{K\eta}(q^{2})+(M_{K}^{2}-M_{\eta}^{2})\bar{J}^{(1)}_{K\eta}(q^{2})\right]
\end{eqnarray}
where $\bar{J}_{PQ}(q^{2})$ is the standard mesonic loop function~\cite{Gasser:1984gg}
and
\begin{eqnarray}
\bar{J}^{(1)}_{PQ}(q^{2})&=&\frac{d}{dM_P^{2}}\bar{J}_{PQ}(q^{2})\nonumber\\
&&=\frac{1}{16\pi^{2}}\Biggl\{\frac{M_P^{2}-M_Q^{2}-M_Q^{2}\ln\frac{M_P^{2}}{M_Q^{2}}}{(M_P^{2}-M_Q^{2})^{2}}+\frac{1}{\lambda(q^{2},M_P^{2},M_Q^{2})}\biggl[q^{2}-M_P^{2}+M_Q^{2}\biggr.\Biggr.\nonumber\\
&&\Biggl.\biggl.+\frac{M_Q^{2}(M_P^{2}-M_Q^{2}+q^{2})}{M_P^{2}-M_Q^{2}}\ln\frac{M_P^{2}}{M_Q^{2}}\biggr]\Biggr\}+\frac{M_P^{2}-M_Q^{2}-q^{2}}{\lambda(q^{2},M_P^{2},M_Q^{2})}\bar{J}_{PQ}(q^{2})\nonumber\\
\bar{J}^{(2)}_{PQ}(q^{2})&=&\frac{d}{dM_Q^{2}}\bar{J}_{PQ}(q^{2})=\bar{J}^{(1)}_{QP}(q^{2}).
\end{eqnarray} 
Here $\lambda(a,b,c)=a^{2}+b^{2}+c^{2}-2ab-2bc-2ca$ is the K\"{a}ll\'{e}n function.

\section{\label{sec:Example}An explicit example of the calculation of two- and three-point functions}

\begin{figure}[tb]
	\begin{centering}
		\includegraphics[scale=0.5]{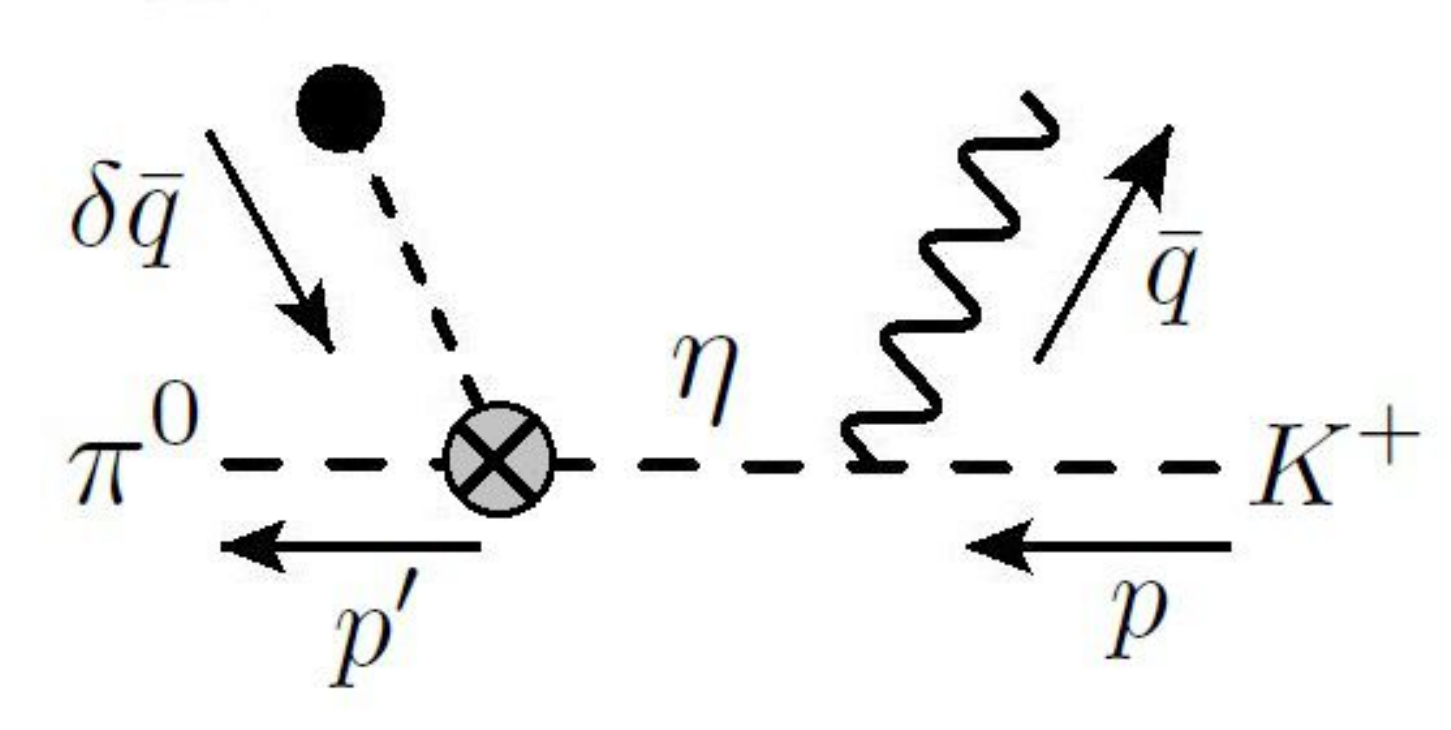}
		\par\end{centering}
	        \caption{\label{fig:example}The Feynman diagram for the calculation of $iT^{\mu}$ and $D$ in
                  Appendix~\ref{sec:Example}.}
\end{figure}

In this appendix, we use a simple example to demonstrate the steps to calculate the two-point and three-point function within the framework of an EFT. We shall consider the effect of the following $\pi^{0}-\eta$ mixing counterterm due to short-distance electromagnetic effects:
\begin{equation}
\delta\mathcal{L}=e^2\delta_{\pi\eta}\partial_\mu\eta \partial^{\mu}\pi^{0}
\end{equation}
to the $K_{l3}^{+}$ form factor. The answer is obvious from ordinary perturbation theory:
\begin{equation}
\delta F_{K^{+}\pi^{0}}^{\mu}(p',p)=\sqrt{\frac{3}{2}}e^2V_{us}^{*}\frac{M_{\pi}^{2}}{M_{\pi}^{2}-M_{\eta}^{2}}\delta_{\pi\eta}(p+p')^{\mu}.\label{eq:examplefull}
\end{equation}

Now we compute the quantities needed to evaluate the two-point and three-point functions:
\begin{eqnarray}
iT^{\mu}(\bar{q};p',p)&=&i\int d^{4} xe^{i\bar{q}\cdot x}\left\langle \pi^{0}(p')\right|T\left\{J_W^{\mu\dagger}(x)\delta\mathcal{L}(0)\right\}\left|K^{+}(p)\right\rangle\nonumber\\
D(\bar{q};p',p)&=&i\int d^{4}xe^{i\bar{q}\cdot x}\left\langle \pi^{0}(p')\right|T\left\{\partial_{\mu} J_{W}^{\mu\dagger}(x)\delta\mathcal{L}(0)\right\}\left|K^{+}(p)\right\rangle,
\end{eqnarray}
where one utilizes the ChPT representation of the charged weak current in Eq.~\eqref{eq:JWEFT}. Notice also that there is no one-particle singularity in this case so the $B^{\mu}$ term does not exist. As we discussed in Section~\ref{sec:diagrammatic}, there are two ways to calculate $iT^{\mu}$: (1) using covariant perturbation theory (see Fig.~\ref{fig:example}), and (2) using the definition of the time-ordered product in Eq.~\eqref{eq:timeordered}; the outcomes are named $iT^{\mu}_{1}$ and $iT^{\mu}_{2}$ respectively. On the other hand, one should calculate $D$ using covariant perturbation theory, but with the EOM-simplified version of $\partial_{\mu}J_W^{\mu\dagger}$ as in Eq.~\eqref{eq:JWEOM}. The results are:
\begin{eqnarray}
iT_{1}^{\mu}&=&\sqrt{\frac{3}{2}}e^{2}V_{us}^{*}\delta_{\pi\eta}\frac{p'\cdot(p-\bar{q})}{(p-\bar{q})^{2}-M_{\eta}^{2}}(2p-\bar{q})^{\mu}\nonumber\\
iT_{2}^{\mu}&=&iT_{1}^{\mu}-\sqrt{\frac{3}{2}}e^{2}V_{us}^{*}\delta_{\pi\eta}g^{\mu 0}p_{0}^{\prime}\nonumber\\
D&=&-i\sqrt{\frac{3}{2}}e^{2}V_{us}^{*}\delta_{\pi\eta}\frac{M_{K}^{2}-M_{\eta}^{2}}{(p-\bar{q})^{2}-M_{\eta}^{2}}p'\cdot(p-\bar{q}).
\end{eqnarray}
In particular, the difference between $iT_{1}^{\mu}$ and $iT_{2}^{\mu}$ will later enter the two-point function. 

The three-point function is therefore given by:
\begin{equation}
\delta F_{K^{+}\pi^{0},3}^{\mu}(p',p)=\lim_{\bar{q}\rightarrow q}\left[-\bar{q}_{\nu}\frac{\partial}{\partial\bar{q}_{\mu}}iT_{1}^{\nu}+i\frac{\partial}{\partial\bar{q}_{\mu}}D\right]=\sqrt{\frac{3}{2}}e^{2}V_{us}^{*}\delta_{\pi\eta}\left[\frac{M_{\pi}^{2}}{M_{\pi}^{2}-M_{\eta}^{2}}(p+p')^{\mu}-p^{\prime\mu}\right].
\end{equation}
Whereas, the two-point function is given by:
\begin{eqnarray}
\delta F^{\mu}_{K^{+}\pi^{0},2}(p',p)&=&-\lim_{\bar{q}\rightarrow q}\frac{\partial}{\partial\bar{q}_{\mu}}\int d^{4}xe^{i\bar{q}\cdot x}\delta(x_0)\left\langle \pi^{0}(p')\right|\left[J_{W}^{0\dagger}(x),\delta\mathcal{L}(0)\right]\left|K^{+}(p)\right\rangle\nonumber\\ &&+\sqrt{\frac{3}{2}}e^{2}V_{us}^{*}\delta_{\pi\eta}g^{\mu 0}p_{0}^{\prime}.
\end{eqnarray}
To compute the first term, one utilizes the canonical commutation relation for free real scalar fields:
\begin{equation}
\delta(x_{0}-y_{0})[\dot{\phi}(x),\partial_{\mu}\phi(y)]=g_{\mu i}i\partial_{x}^{i}\delta^{4}(x-y).
\end{equation}
This gives:
\begin{equation}
\delta F^{\mu}_{K^{+}\pi^{0},2}(p',p)=\sqrt{\frac{3}{2}}e^{2}V_{us}^{*}\delta_{\pi\eta}g^{\mu i}p_{i}^{\prime}+\sqrt{\frac{3}{2}}e^{2}V_{us}^{*}\delta_{\pi\eta}g^{\mu 0}p_{0}^{\prime}=\sqrt{\frac{3}{2}}e^{2}V_{us}^{*}\delta_{\pi\eta}p^{\prime\mu}.
\end{equation}
With this we have calculated both the two-point and three-point function, and they add up to give the full result, Eq.~\eqref{eq:examplefull}, as expected. 

\end{appendix}


\begin{thebibliography}{10}
	\expandafter\ifx\csname url\endcsname\relax
	\def\url#1{\texttt{#1}}\fi
	\expandafter\ifx\csname urlprefix\endcsname\relax\def\urlprefix{URL }\fi
	\expandafter\ifx\csname href\endcsname\relax
	\def\href#1#2{#2} \def\path#1{#1}\fi
	
	\bibitem{Cabibbo:1963yz}
	N.~Cabibbo, {\it {Unitary Symmetry and Leptonic Decays}},  {\em Phys. Rev.
		Lett.} {\bf 10} (1963) 531--533.
	
	\bibitem{Kobayashi:1973fv}
	M.~Kobayashi and T.~Maskawa, {\it {CP Violation in the Renormalizable Theory of
			Weak Interaction}},  {\em Prog. Theor. Phys.} {\bf 49} (1973) 652--657.
	
	\bibitem{Tanabashi:2018oca}
	{\bf Particle Data Group} Collaboration, M.~Tanabashi {\em et.~al.}, {\it
		{Review of Particle Physics}},  {\em Phys. Rev.} {\bf D98} (2018), no.~3
	030001.
	
	\bibitem{Bauman:2012fx}
	S.~Bauman, J.~Erler and M.~Ramsey-Musolf, {\it {Charged Current Universality
			and the MSSM}},  {\em Phys. Rev.} {\bf D87} (2013), no.~3 035012
	[\href{http://arXiv.org/abs/1204.0035}{{\tt 1204.0035}}].
	
	\bibitem{Gonzalez-Alonso:2018omy}
	M.~Gonzalez-Alonso, O.~Naviliat-Cuncic and N.~Severijns, {\it {New physics
			searches in nuclear and neutron $\beta$ decay}},  {\em Prog. Part. Nucl.
		Phys.} {\bf 104} (2019) 165--223 [\href{http://arXiv.org/abs/1803.08732}{{\tt
			1803.08732}}].
	
	\bibitem{Cirgiliano:2019nyn}
	V.~Cirigliano, A.~Garcia, D.~Gazit, O.~Naviliat-Cuncic, G.~Savard and A.~Young,
	{\it {Precision Beta Decay as a Probe of New Physics}},
	\href{http://arXiv.org/abs/1907.02164}{{\tt 1907.02164}}.
	
	\bibitem{Seng:2018yzq}
	C.-Y. Seng, M.~Gorchtein, H.~H. Patel and M.~J. Ramsey-Musolf, {\it {Reduced
			Hadronic Uncertainty in the Determination of $V_{ud}$}},  {\em Phys. Rev.
		Lett.} {\bf 121} (2018), no.~24 241804
	[\href{http://arXiv.org/abs/1807.10197}{{\tt 1807.10197}}].
	
	\bibitem{Marciano:2005ec}
	W.~J. Marciano and A.~Sirlin, {\it {Improved calculation of electroweak
			radiative corrections and the value of V(ud)}},  {\em Phys. Rev. Lett.} {\bf
		96} (2006) 032002 [\href{http://arXiv.org/abs/hep-ph/0510099}{{\tt
			hep-ph/0510099}}].
	
	\bibitem{Seng:2018qru}
	C.~Y. Seng, M.~Gorchtein and M.~J. Ramsey-Musolf, {\it {Dispersive evaluation
			of the inner radiative correction in neutron and nuclear $\beta$ decay}},
	{\em Phys. Rev.} {\bf D100} (2019), no.~1 013001
	[\href{http://arXiv.org/abs/1812.03352}{{\tt 1812.03352}}].
	
	\bibitem{Gorchtein:2018fxl}
	M.~Gorchtein, {\it {$\gamma W$ Box Inside Out: Nuclear Polarizabilities Distort
			the Beta Decay Spectrum}},  {\em Phys. Rev. Lett.} {\bf 123} (2019), no.~4
	042503 [\href{http://arXiv.org/abs/1812.04229}{{\tt 1812.04229}}].
	
	\bibitem{POETIC2019}
	R.~Petti, 2019.
	\newblock in the \textit{9th International Conference on Physics Opportunities
		at an ElecTron-Ion-Collider}.
	
	\bibitem{Seng:2019plg}
	C.-Y. Seng and U.-G. Mei\ss{}ner, {\it {Toward a First-Principles Calculation
			of Electroweak Box Diagrams}},  {\em Phys. Rev. Lett.} {\bf 122} (2019),
	no.~21 211802 [\href{http://arXiv.org/abs/1903.07969}{{\tt 1903.07969}}].
	
	\bibitem{Czarnecki:2019mwq}
	A.~Czarnecki, W.~J. Marciano and A.~Sirlin, {\it {Radiative Corrections to
			Neutron and Nuclear Beta Decays Revisited}},
	{\em Phys. Rev.} {\bf D100} (2019), no.~7 073008
	[\href{http://arXiv.org/abs/1907.06737}{{\tt 1907.06737}}].
	
	
	\bibitem{Bazavov:2018kjg}
	{\bf Fermilab Lattice, MILC} Collaboration, A.~Bazavov {\em et.~al.}, {\it
		{$|V_{us}|$ from $K_{\ell 3}$ decay and four-flavor lattice QCD}},  {\em
		Phys. Rev.} {\bf D99} (2019), no.~11 114509
	[\href{http://arXiv.org/abs/1809.02827}{{\tt 1809.02827}}].
	
	\bibitem{Cirigliano:2011ny}
	V.~Cirigliano, G.~Ecker, H.~Neufeld, A.~Pich and J.~Portoles, {\it {Kaon Decays
			in the Standard Model}},  {\em Rev. Mod. Phys.} {\bf 84} (2012) 399
	[\href{http://arXiv.org/abs/1107.6001}{{\tt 1107.6001}}].
	
	\bibitem{Cirigliano:2001mk}
	V.~Cirigliano, M.~Knecht, H.~Neufeld, H.~Rupertsberger and P.~Talavera, {\it
		{Radiative corrections to K(l3) decays}},  {\em Eur. Phys. J.} {\bf C23}
	(2002) 121--133 [\href{http://arXiv.org/abs/hep-ph/0110153}{{\tt
			hep-ph/0110153}}].
	
	\bibitem{Cirigliano:2008wn}
	V.~Cirigliano, M.~Giannotti and H.~Neufeld, {\it {Electromagnetic effects in
			K(l3) decays}},  {\em JHEP} {\bf 11} (2008) 006
	[\href{http://arXiv.org/abs/0807.4507}{{\tt 0807.4507}}].
	
	\bibitem{Sirlin:1977sv}
	A.~Sirlin, {\it {Current Algebra Formulation of Radiative Corrections in Gauge
			Theories and the Universality of the Weak Interactions}},  {\em Rev. Mod.
		Phys.} {\bf 50} (1978) 573. [Erratum: Rev. Mod. Phys.50,905(1978)].
	
	\bibitem{Kinoshita:1958ru}
	T.~Kinoshita and A.~Sirlin, {\it {Radiative corrections to Fermi
			interactions}},  {\em Phys. Rev.} {\bf 113} (1959) 1652--1660.
	
	\bibitem{Pauli:1949zm}
	W.~Pauli and F.~Villars, {\it {On the Invariant regularization in relativistic
			quantum theory}},  {\em Rev. Mod. Phys.} {\bf 21} (1949) 434--444.
	
	\bibitem{DescotesGenon:2005pw}
	S.~Descotes-Genon and B.~Moussallam, {\it {Radiative corrections in weak
			semi-leptonic processes at low energy: A Two-step matching determination}},
	{\em Eur. Phys. J.} {\bf C42} (2005) 403--417
	[\href{http://arXiv.org/abs/hep-ph/0505077}{{\tt hep-ph/0505077}}].
	
	\bibitem{Ananthanarayan:2004qk}
	B.~Ananthanarayan and B.~Moussallam, {\it {Four-point correlator constraints on
			electromagnetic chiral parameters and resonance effective Lagrangians}},
	{\em JHEP} {\bf 06} (2004) 047
	[\href{http://arXiv.org/abs/hep-ph/0405206}{{\tt hep-ph/0405206}}].
	
	\bibitem{Ecker:1988te}
	G.~Ecker, J.~Gasser, A.~Pich and E.~de~Rafael, {\it {The Role of Resonances in
			Chiral Perturbation Theory}},  {\em Nucl. Phys.} {\bf B321} (1989) 311--342.
	
	\bibitem{Gasser:1984gg}
	J.~Gasser and H.~Leutwyler, {\it {Chiral Perturbation Theory: Expansions in the
			Mass of the Strange Quark}},  {\em Nucl. Phys.} {\bf B250} (1985) 465--516.
	
	\bibitem{Urech:1994hd}
	R.~Urech, {\it {Virtual photons in chiral perturbation theory}},  {\em Nucl.
		Phys.} {\bf B433} (1995) 234--254
	[\href{http://arXiv.org/abs/hep-ph/9405341}{{\tt hep-ph/9405341}}].
	
	\bibitem{Knecht:1999ag}
	M.~Knecht, H.~Neufeld, H.~Rupertsberger and P.~Talavera, {\it {Chiral
			perturbation theory with virtual photons and leptons}},  {\em Eur. Phys. J.}
	{\bf C12} (2000) 469--478 [\href{http://arXiv.org/abs/hep-ph/9909284}{{\tt
			hep-ph/9909284}}].
	
	\bibitem{Bernard:2006gy}
	V.~Bernard, M.~Oertel, E.~Passemar and J.~Stern, {\it {K(mu3)**L decay: A
			Stringent test of right-handed quark currents}},  {\em Phys. Lett.} {\bf
		B638} (2006) 480--486 [\href{http://arXiv.org/abs/hep-ph/0603202}{{\tt
			hep-ph/0603202}}].
	
	\bibitem{Bernard:2007tk}
	V.~Bernard and E.~Passemar, {\it {Matching chiral perturbation theory and the
			dispersive representation of the scalar K pi form-factor}},  {\em Phys.
		Lett.} {\bf B661} (2008) 95--102 [\href{http://arXiv.org/abs/0711.3450}{{\tt
			0711.3450}}].
	
	\bibitem{Bernard:2009zm}
	V.~Bernard, M.~Oertel, E.~Passemar and J.~Stern, {\it {Dispersive
			representation and shape of the K(l3) form factors: Robustness}},  {\em Phys.
		Rev.} {\bf D80} (2009) 034034 [\href{http://arXiv.org/abs/0903.1654}{{\tt
			0903.1654}}].
	
	\bibitem{Bernard:2011ae}
	V.~Bernard, D.~R. Boito and E.~Passemar, {\it {Dispersive representation of the
			scalar and vector Kpi form factors for $\tau\to K\pi \nu_\tau$ and $K_{l3}$
			decays}},  {\em Nucl. Phys. Proc. Suppl.} {\bf 218} (2011) 140--145
	[\href{http://arXiv.org/abs/1103.4855}{{\tt 1103.4855}}].
	
	\bibitem{Bernard:2013jxa}
	V.~Bernard, {\it {First determination of $f_+(0) |V_{us}|$ from a combined
			analysis of $\tau\to K\pi \nu_\tau$ decay and $\pi K$ scattering with
			constraints from $K_{\ell3}$ decays}},  {\em JHEP} {\bf 06} (2014) 082
	[\href{http://arXiv.org/abs/1311.2569}{{\tt 1311.2569}}].
	
	\bibitem{Giusti:2017dwk}
	D.~Giusti, V.~Lubicz, G.~Martinelli, C.~T. Sachrajda, F.~Sanfilippo, S.~Simula,
	N.~Tantalo and C.~Tarantino, {\it {First lattice calculation of the QED
			corrections to leptonic decay rates}},  {\em Phys. Rev. Lett.} {\bf 120}
	(2018), no.~7 072001 [\href{http://arXiv.org/abs/1711.06537}{{\tt
			1711.06537}}].
	
	\bibitem{Cirigliano:2019jig}
	{\bf USQCD} Collaboration, V.~Cirigliano, Z.~Davoudi, T.~Bhattacharya,
	T.~Izubuchi, P.~E. Shanahan, S.~Syritsyn and M.~L. Wagman, {\it {The Role of
			Lattice QCD in Searches for Violations of Fundamental Symmetries and Signals
			for New Physics}},  \href{http://arXiv.org/abs/1904.09704}{{\tt 1904.09704}}.
	
	\bibitem{DiCarlo:2019thl}
	M.~Di~Carlo, D.~Giusti, V.~Lubicz, G.~Martinelli, C.~T. Sachrajda,
	F.~Sanfilippo, S.~Simula and N.~Tantalo, {\it {Light-meson leptonic decay
			rates in lattice QCD+QED}},  {\em Phys. Rev.} {\bf D100} (2019), no.~3 034514
	[\href{http://arXiv.org/abs/1904.08731}{{\tt 1904.08731}}].
	
	\bibitem{Baur:1996ya}
	R.~Baur and R.~Urech, {\it {Resonance contributions to the electromagnetic
			low-energy constants of chiral perturbation theory}},  {\em Nucl. Phys.} {\bf
		B499} (1997) 319--348 [\href{http://arXiv.org/abs/hep-ph/9612328}{{\tt
			hep-ph/9612328}}].
	
	\bibitem{Bijnens:1996kk}
	J.~Bijnens and J.~Prades, {\it {Electromagnetic corrections for pions and
			kaons: Masses and polarizabilities}},  {\em Nucl. Phys.} {\bf B490} (1997)
	239--271 [\href{http://arXiv.org/abs/hep-ph/9610360}{{\tt hep-ph/9610360}}].
	
	\bibitem{Brown:1970dd}
	L.~S. Brown, {\it {Perturbation theory and selfmass insertions}},  {\em Phys.
		Rev.} {\bf 187} (1969) 2260--2265.
	
	\bibitem{Sirlin:1981ie}
	A.~Sirlin, {\it {Large m(W), m(Z) Behavior of the O(alpha) Corrections to
			Semileptonic Processes Mediated by W}},  {\em Nucl. Phys.} {\bf B196} (1982)
	83--92.
	
	
	\bibitem{Bjorken:1966jh}
	J.~D. Bjorken, {\it {Applications of the Chiral U(6) x (6) Algebra of Current
			Densities}},  {\em Phys. Rev.} {\bf 148} (1966) 1467--1478.
	
	\bibitem{Johnson:1966se}
	K.~Johnson and F.~E. Low, {\it {Current algebras in a simple model}},  {\em
		Prog. Theor. Phys. Suppl.} {\bf 37} (1966) 74--93.
	
	
\end{thebibliography}


\end{document}